\documentclass{emulateapj}
\usepackage{lscape}

\newcommand{\scr}{\scriptsize}
\newcommand{\lan}{\langle}
\newcommand{\ran}{\rangle}

\newcommand{\B}[1]{$B_{\rm{#1}}$}
\newcommand{\R}[1]{$R_{\rm{#1}}$}
\newcommand{\Bz}[2]{$B^{#2}_{\rm{#1}}$}
\newcommand{\Rz}[2]{$R^{#2}_{\rm{#1}}$}
\newcommand{\gunnr}{Gunn $r$ }
\newcommand{\Log}{{\rm \log\,}}
\newcommand{\Ie} {{\lan I_e \ran}}
\newcommand{\mue}{{\lan \mu_e \ran}}
\newcommand{\devauc}{de Vaucouleurs }
\newcommand{\ml}{{\sl M/L }}

\newcommand{\dcdr}{$\Delta(B-R)/\Delta\Log r$}

\slugcomment{Published in ApJ. A typo fixed}

\shorttitle{E+As and Early Type Galaxies}
\shortauthors{Yang et al.}

\begin{document}

\title{The Detailed Evolution of E+A Galaxies into Early Types\altaffilmark{1}}
                                                                                          
\author{Yujin Yang, Ann I. Zabludoff, and Dennis Zaritsky}
\affil{Steward Observatory, University of Arizona, Tucson, AZ 85721}
\email{yyang, azabludoff, dzaritsky@as.arizona.edu}

\author{J. Christopher Mihos}
\affil{Department of Astronomy, Case Western Reserve University, 
       10900 Euclid Ave, Cleveland, OH 44106}
\email{mihos@case.edu}

\altaffiltext{1} 
{Based on observations with the NASA/ESA Hubble Space Telescope obtained
at the Space Telescope Science Institute, which is operated by the
Association of Universities for Research in Astronomy, Incorporated,
under NASA contract NAS5-26555.}

\begin{abstract}

Post-starburst, or E+A galaxies, are the best candidates for galaxies
in transition from being gas-rich and star-forming to gas-poor and
passively-evolving as a result of galaxy-galaxy interactions.  To focus
on what E+A galaxies become after their young stellar populations fade
away, we present the detailed morphologies of 21 E+A galaxies using
high resolution {\sl HST}/{\sl ACS} and {\sl WFPC2} images.  Most of
these galaxies lie in the field, well outside of rich clusters, and
at least 11 (55\%) have dramatic tidal features indicative of mergers.
Our sample includes one binary E+A system, in which both E+As are tidally
disturbed and interacting with each other.
Our E+As are similar to early types in that they have large bulge-to-total
light ratios (median $B/T$ = 0.59), high S\'ersic indices, ($n \gtrsim
4$), and high concentration indices ($C \gtrsim 4.3$), but they have
considerably larger asymmetry indices ($A \gtrsim 0.04$) than ellipticals,
presumably due to the disturbances within a few $r_e$ caused by the
starburst and/or the galaxy-galaxy interaction.  We conclude that E+As
will be morphologically classified as early-type galaxies once these
disturbances and the low surface brightness tidal features fade.
The color morphologies are diverse, including six E+As with compact
(0.4\,--\,1.4 kpc) blue cores, which might be local analogs of high-$z$
ellipticals with blue-cores.  The large fraction (70\%) of E+As with
positive color gradients indicates that the young stellar populations
are more concentrated than the old.  These positive color gradients
(i.e., bluer nuclei) could evolve into the negative gradients typical
in E/S0s if the central parts of these galaxies are metal enhanced.
Our E+As stand apart from the E/S0s in the edge-on projection of the
Fundamental Plane (FP), implying that their stellar populations differ
from those of E/S0s and that E+As have, on average, a \ml that is 3.8
times smaller.  The tilt of the E+A FP indicates that the variation among
their stellar populations is closely tied to the structural parameters,
i.e., E+As follow their own scaling relationships such that smaller or
less massive galaxies have smaller {\sl M/L}.  
We find a population of unresolved compact sources in nine E+As (45\%),
all of which have merger signatures.  In the four E+As with suitable
color data, the compact sources have colors and luminosities consistent
with newly-formed star clusters.  The bright end of the cluster LF is
fainter in redder E+A's, suggesting that the young star clusters fade
or are disrupted as the merger remnant ages.
In summary, the morphologies, color profiles, scaling relations, and
cluster populations are all consistent with E+As evolving ultimately
into early-types, making the study of E+As critical to understanding
the origin of the red sequence of galaxies.

\end{abstract}

\keywords{
 galaxies: evolution ---
 galaxies: interactions ---
 galaxies: starburst ---
 galaxies: star clusters ---
 galaxies: stellar content
}


\section{Introduction}

If some galaxies evolve from star-forming, gas-rich, disk-dominated
galaxies (late-types) into quiescent, gas-poor, spheroid-dominated
galaxies (early-types), we should find objects caught in the midst of
this transformation.  The best candidates are the so-called ``E+A'',
``K+A'', or "post-starburst" galaxies \citep{Dressler83,Couch87}
due to their combination of late- and early-type characteristics,
including both a significant young stellar population (age $\lesssim$
1 Gyr) and a lack of on-going star formation.  These galaxies have been
spectroscopically identified by their strong Balmer absorption lines and
absence of emission lines (e.g., [\ion{O}{2}] and H$\alpha$) in various
environments and at all redshifts \citep{Zabludoff96, Poggianti99,
Goto03, Blake04, Tran03, Tran04}.

While the cause of the abrupt end of their star formation is poorly
understood, there is strong evidence that galaxy-galaxy tidal interactions
or mergers trigger the starburst in many cases.  First, most E+A galaxies
reside in low-density environments, such as poor groups, that are similar
to those of star-forming galaxies \citep{Zabludoff96, Quintero04, Blake04,
Balogh05, Goto05, Hogg06, Yan08}. Therefore, many E+As must arise from
a process common in the field, such as galaxy-galaxy interactions,
instead of a mechanism limited to denser, hotter environments,
such as ram pressure stripping \citep{Gunn&Gott72} or strangulation
\citep{Balogh00}.  Second, a significant fraction of E+As have tidal
features \citep{Zabludoff96, Yang04, Blake04, Tran03, Tran04, Goto05}.
Third, optical and {\sl NIR} colors show that the spectral signatures
of E+As require enhanced recent star formation, rather than simply a
truncation of star formation in a normal spiral galaxies \citep{Balogh05}.

What will E+A galaxies become? In general, E+As are bulge-dominated,
highly-concentrated \citep{Quintero04, Tran04, Blake04, Goto05, Balogh05},
relatively gas-poor \citep{Chang01, Buyle06}, and kinematically hot
systems \citep{Norton01}. Therefore, in a statistical sense, E+As are
likely to become E/S0 galaxies.  However, due to a lack of spatial
resolution in previous studies, we do not know whether the detailed
properties of individual E+As, e.g., their bulge fractions, color
gradients, internal kinematics, and newly formed stellar clusters,
are consistent with their presumed evolution into early type galaxies.
Using {\sl HST/WFPC2}, \citet{Yang04} showed that their morphological
features are consistent with a  transition from late to early types,
but their sample contained only the five bluest E+As galaxies from
the Las Campanas Redshift Survey (LCRS; \cite{Zabludoff96}) and thus
was not representative.  As a result, we still do not know whether the
entire population of E+As will evolve into E/S0s, whether there is a
distinguishable subclass of E+As that evolves into E/S0s, or whether
E+As evolve into typical E/S0s.

To answer these questions, we must understand how well E+As match the
{\it full} range of E/S0 properties.
Most fundamentally, are the global morphologies (e.g., bulge-to-total
light ratios and concentration) of the whole LCRS E+A sample consistent
with those of E/S0s?
Second, E/S0s in the local universe become redder toward their center;
these negative color gradients originate from metallicity gradients
\citep{Peletier90}.  In contrast, E+As exhibit a wide range of color
morphologies \citep{Yang04, Yamauchi05}.  Can the color profiles of E+As
evolve into those of the typical early types?
Third, the number of globular clusters per unit luminosity is higher in
early types than in late types \citep{Harris&vandenBergh81}. If E+As
are in transition from late to early types, one should find new star
clusters formed during the starburst. Are there such clusters, and, if
so, do their colors and numbers coincide with the expected evolution of
the globular cluster systems of present-day E/S0s?
Fourth, early-type galaxies lie on the Fundamental Plane (FP), an
empirical scaling relation between the effective radius, the central
velocity dispersion, and the mean surface brightness, with remarkably
small scatter \citep{Djorgovski87, Dressler87}.  Will E+As lie on the
same FP once they evolve?

To determine whether E+As evolve into objects that are indistinguishable
from the bulk of E/S0s, we present {\sl HST/ACS} observations of the 15
remaining E+A galaxies from the \citet{Zabludoff96} sample.  We combine
these with the previous {\sl HST/WFPC2} observations of five blue E+As
\citep{Yang04} and of one serendipitously discovered E+A \citep{Yang06}.
The resulting high resolution imaging of the 21 confirmed E+A galaxies
in the LCRS sample enables us to study the detailed color morphologies
and the properties of the newly formed star cluster candidates at the
sub-kpc scale.  The detailed morphologies, color profiles, and cluster
populations of many E+As are consistent with the galaxy-galaxy interaction
scenario.  We discuss how these features will evolve and then relate this
evolution to the properties of E/S0s.  Using existing kinematic data
\citep{Norton01} and our {\sl HST} photometry, we also address whether
or not the various scaling relations of E+As are consistent with those
of E/S0s.

This paper is organized as follows.  We describe our E+A galaxy sample
and the {\sl HST/ACS} data reduction in \S \ref{sec:observation}. In \S
\ref{sec:morphology}, we examine the general morphology of E+As, including
a discussion of tidal features (\S \ref{sec:qualitative_morphology}),
surface brightness profiles (\S \ref{sec:fitting}), structural
parameters (\S\ref{sec:profile}), and concentration/asymmetry
measures (\S \ref{sec:ca}).  The color profiles, including a class
of E+As with luminous blue cores \citep{Yang06}, are presented in \S
\ref{sec:color_profile}.  We compare the scaling relations of E+As with
those of E/S0s in \S \ref{sec:fp}.  We present the properties of the
newly-formed young star clusters in \S \ref{sec:cluster}.  We summarize
in \S\ref{sec:conclusion}.



\begin{figure*}
\epsscale{0.9}
\plotone{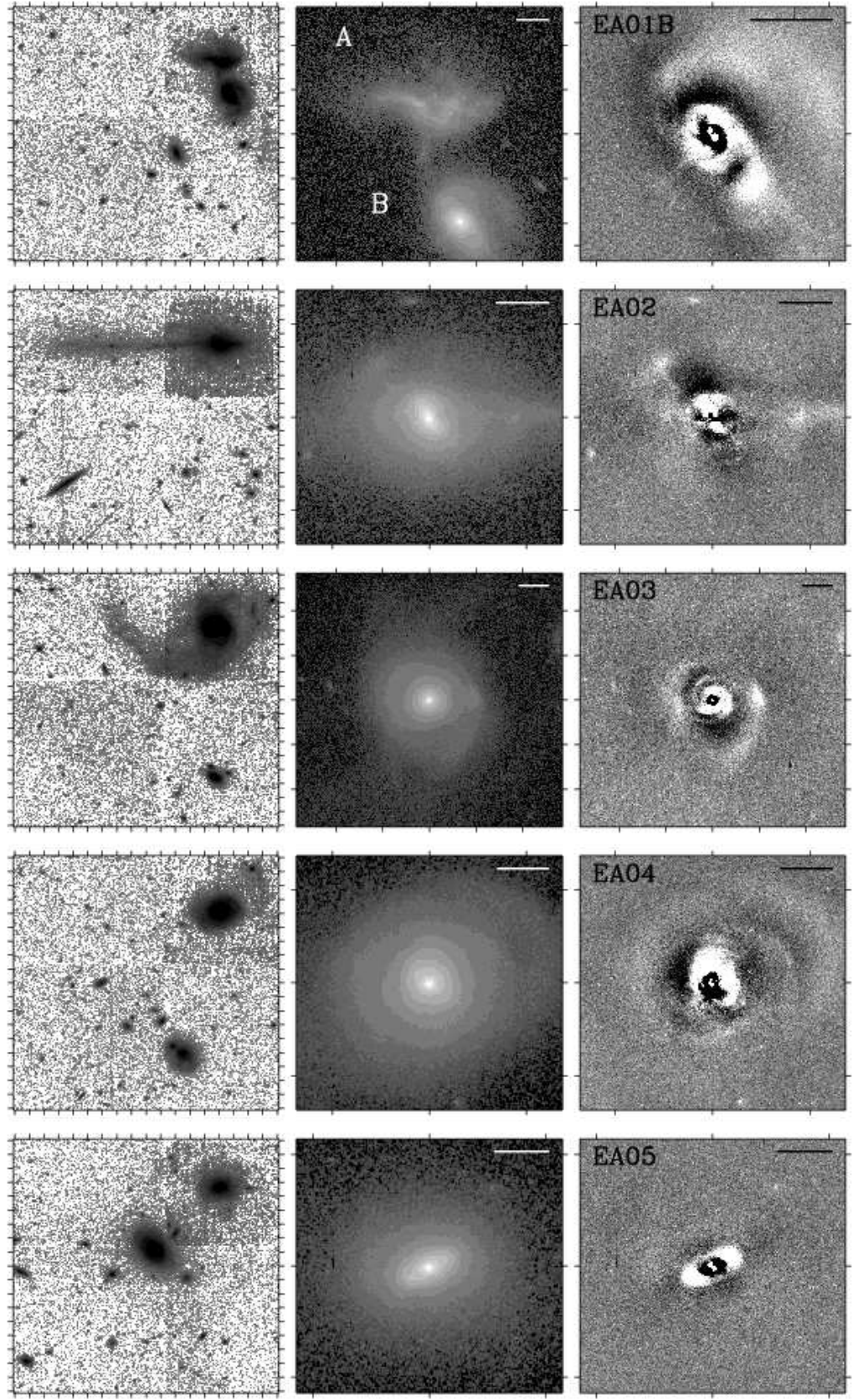} 
\caption{
({\it Left}) High-contrast $R$ band (\R{702}) images show the low surface
brightness tidal features.  ({\it Middle})  $R$ band images for the {\sl
WFPC2} sample (EA01AB -- EA05).
({\it Right}) Residual $R$ band images subtracted from the smooth
symmetric model components.  We bound each image with 4 arcsec tickmarks
and include a 4 kpc horizontal scalebar.  Note the diverse morphologies of
E+A galaxies: tidal and disturbed features, dusty galaxies, blue-cores,
bars, and even compact star clusters.  Because EA01A is too disturbed
to be modeled by axisymmetric models, we restrict our analysis to EA01B
and show the residual image only for EA01B in the top panel.
\label{fig:images_wfpc2}}
\end{figure*}

\begin{figure*}
\epsscale{0.9}
\plotone{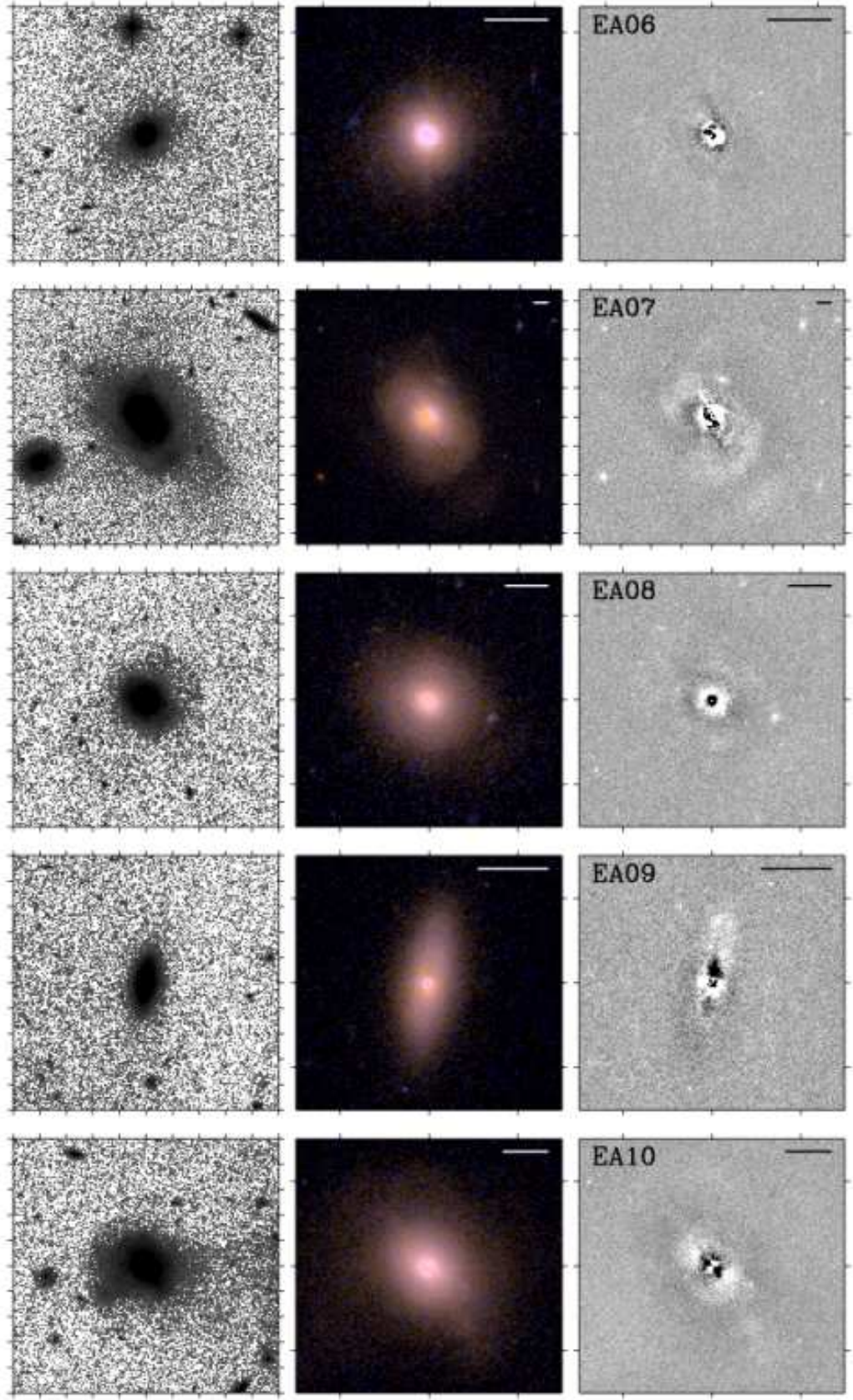} 
\caption{
({\it Left})   Same as for Fig \ref{fig:images_wfpc2}, except these are {\sl ACS} images
of EA06--20.
({\it Middle}) Two-color composite images from the $B$ and $R$ bands.
({\it Right})  Same as for Fig \ref{fig:images_wfpc2}, except for the {\sl ACS} images.
\label{fig:images_acs}
}
\end{figure*}

\begin{figure*}
\addtocounter{figure}{-1}
\epsscale{0.9}
\plotone{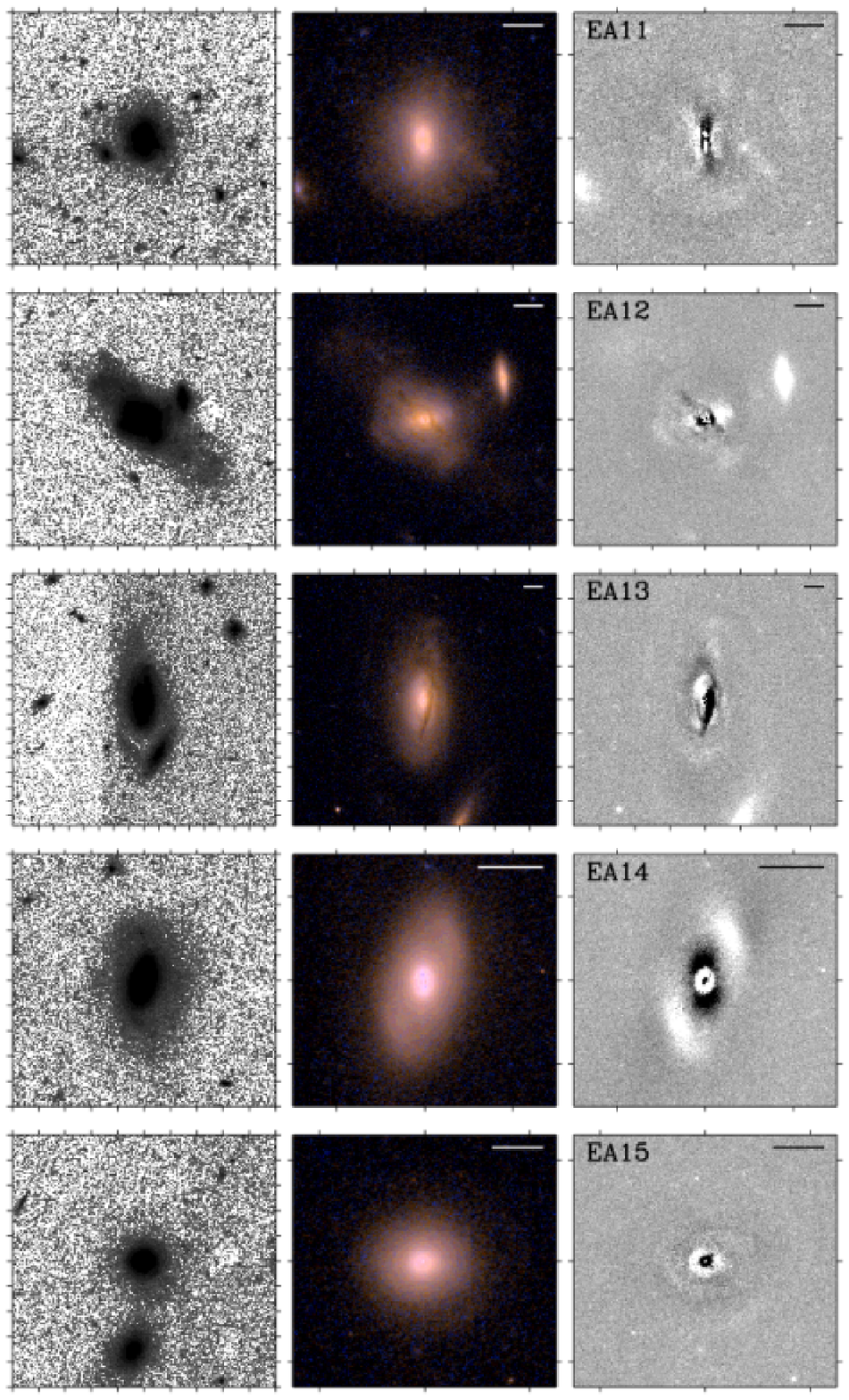}
\caption{Continued.}
\end{figure*}

\begin{figure*}
\addtocounter{figure}{-1}
\epsscale{0.9}
\plotone{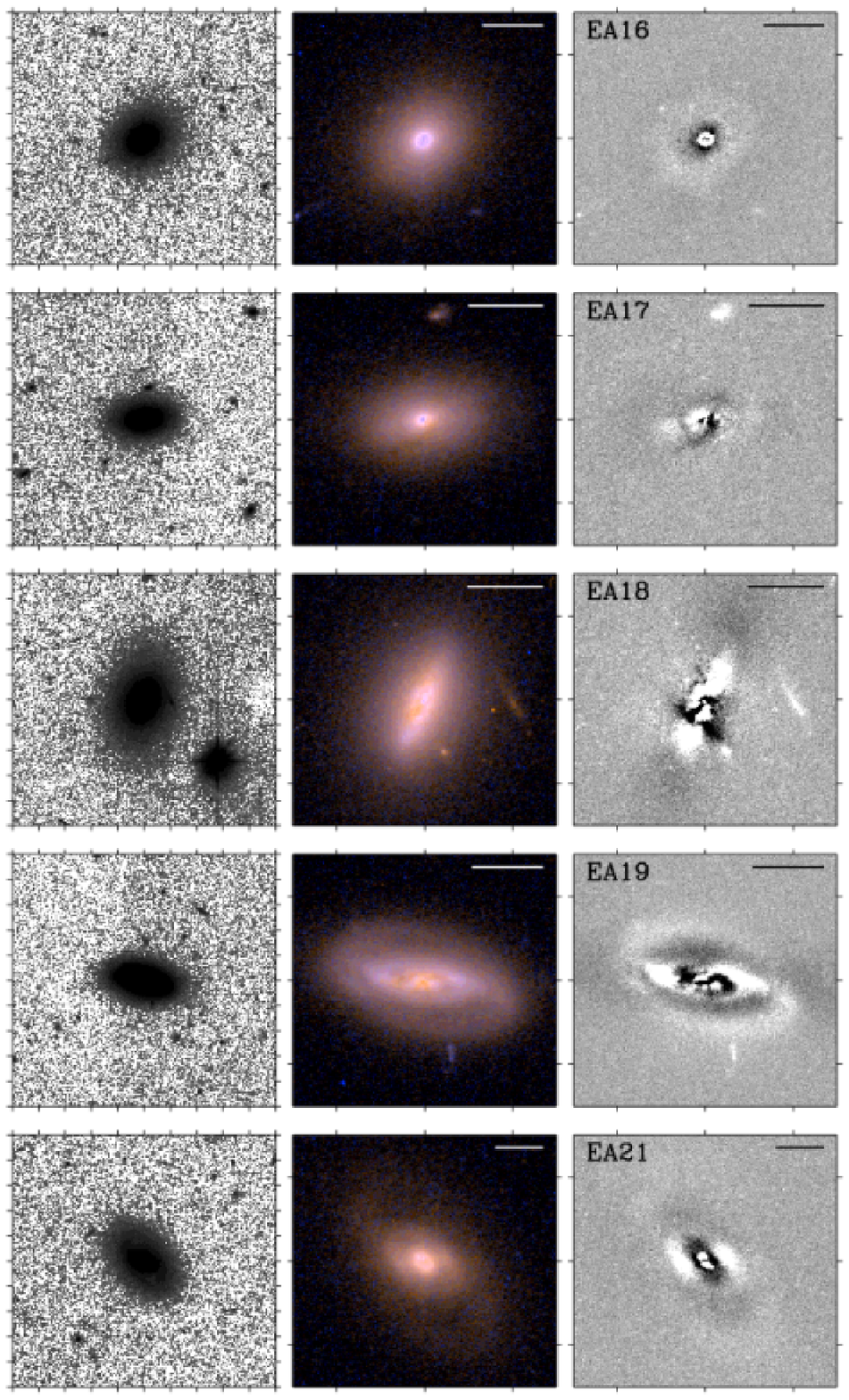}
\caption{Continued.}
\end{figure*}

\section{Observations and Data Reduction}
\label{sec:observation}
\subsection{LCRS E+A Sample}

Our {\sl HST} imaging sample consists of 21 nearby E+A galaxies
spectroscopically identified from the 11,113 galaxy spectra of the
Las Campanas Redshift Survey \cite[LCRS;][]{Shectman96} with redshifts
between 0.07 and 0.18 \citep{Zabludoff96}.  These E+As are required to
have strong Balmer absorption features (average equivalent width $\lan
H\beta\gamma\delta \ran$ $>$ 5.5\AA) and little if any [\ion{O}{2}]
emission (EW[\ion{O}{2}] $<$ 2.5 \AA).  Three-quarters of this sample are
in the field, well outside rich cluster environments.  The number of each
E+A (e.g., EA01) is adopted from \citet{Zabludoff96} and increases with
increasing 4000 {\AA} break ($D_{4000}$) strength.  $D_{4000}$ is related
to the galaxy's color --- bluer galaxies have smaller $D_{4000}$.  For the
remainder of this paper, we refer to each galaxy by its assigned number.

There have been two changes to the original sample since published.
One of the original galaxies (EA20) was misclassified as an E+A due to
noise in the region of the spectral line diagnostics \citep{Norton01}. On
the other hand, the morphologically disturbed companion galaxy of EA01
lies at the same redshift and also has an E+A spectrum \citep{Yang06}.
We include it in our sample and refer to it as EA01B, and to the original
EA01 as EA01A.  The EA01AB system is the first known binary E+A system
and provides additional evidence that the E+A phase of galaxy evolution
can be triggered by galaxy-galaxy interactions.

This LCRS sample is the most extensively studied E+A sample, with data
that includes measurements of internal kinematics \citep{Norton01},
\ion{H}{1} content \citep{Chang01, Buyle06}, radio-continuum emission
\citep{Chang01, Miller_Owen01}, {\sl NIR} fluxes and morphologies
\citep{Galaz00}, {\sl HST} optical morphologies \citep{Yang04},
and spectral diagnostics of nuclear activity \citep{Yang06}. Table
\ref{tab:basic} summarizes the basic data and inferred properties from
the various studies.  Throughout this paper, we adopt $H_0=70\ \mathrm{km\
s^{-1}\ Mpc^{-1}}$, $\Omega_\mathrm{M}=0.3$, and $\Omega_{\Lambda}=0.7$.

\subsection{{\sl HST ACS} Observations}

By combining the observations of six E+As with {\sl HST/WFPC2}, EA01AB
through EA05 \citep{Yang04,Yang06}, with the new {\sl ACS/WFC} imaging
of the remaining 15 E+As, we now have data for the complete LCRS sample.
The {\sl ACS/WFC} imaging is in the F435W and F625W bands, which closely
match the Johnson $B$ and the SDSS $r$ bands, respectively. We chose
these filters to match the two {\sl WFPC2} filters (F439W and F702W)
used by \cite{Yang04}. Hereafter, we will refer to magnitudes in each
of these filters as \B{435}, \R{625}, \B{439}, and \R{702}, respectively.

The F625W band images consist of three cosmic ray split exposures, one
short (200s) and two long (350s) exposures, while the F435W band images
consist of only two exposures.  The exposure time for each split in F435W
ranges from 500 to 550 sec depending on the galaxy's location on the sky.
These uneven exposure times per pointing and the number of splits were
adopted to obtain both \B{435} and \R{625} band exposures within an orbit,
thereby maximizing the efficiency of our {\sl HST} program by avoiding
unnecessary readout overheads.

The images are bias-subtracted and flat-fielded with the standard
{\sl ACS} calibration pipeline. We determine the offsets and rotations
between the exposures by cross-correlating the bright field stars and
then ``drizzle'' the individual frames into the geometrically corrected
output frame using the Pyraf/Multidrizzle package.  We use an output
pixel scale of 0.05\arcsec per pixel and the square interpolation kernel
for the drizzle.

We reject cosmic-rays using the standard Multidrizzle routine for
the \R{625} band images, which have three exposures, and the LACOSMIC
algorithms \citep{vanDokkum01} and Multidrizzle together for the \B{435}
band images. Because there are only two images and each exposure is
moderately deep, our \B{435} band images do have coincident cosmic
ray hits in the exposures. To address this problem, we identify cosmic
rays in the individual images using  LACOSMIC, fix the flagged pixels
by interpolating across neighboring pixels, and then apply the standard
drizzle process.  We apply LACOSMIC in a conservative way to retain real
objects (e.g., star clusters) and visually check the rejected cosmic
rays within a 20\arcsec\ radius from each E+A.

We use the Vega photometric system calibrated with the {\sl ACS WFC} zero
points, 25.779 (\B{435}) and 25.731 (\R{625}) from \citet{Sirianni05},
and correct for Galactic extinction using the reddening laws
\citep{Sirianni05} $A_{F625W} = 2.633\, E(B-V)$ and $A_{F435W} = 4.103\,
E(B-V)$ for an Sc type spectral energy distribution (SED).  The color
excess, $E(B-V)$, at each position is obtained from the \citet{Schlegel98}
maps. Because the LCRS sample is at high galactic latitude, the extinction
correction is almost negligible for our sample.  We do not include the
errors from the extinction correction in our error budget.

\subsection{Magnitude Transformation and K-correction}

To compare our photometric measurements with our previous {\sl WFPC2}
imaging, and with the color gradients (Johnson $B$ and Cousins $R$)
and the Fundamental Plane data (Johnson $B$ and Gunn $r$) of early type
galaxies in the literature, we establish the magnitude transformation
relation between the ground, {\sl WFPC2}, and {\sl ACS/WFC} photometric
systems using synthetic photometry of K+A type templates as outlined
by \citet{Sirianni05}.  Because the spectra of post-starburst galaxies
depend strongly on burst strength and time elapsed since the burst, we
build a library of K+A type spectra using stellar population synthesis
models \cite[][BC03]{BC03}.  The model stellar populations are composed
of a young burst population (A--type) and an old underlying population
(K--type). We assume that the old stellar population started forming
5 or 10 Gyr ago and has since experienced an exponentially declining
star formation rate ($\tau$ = 2 Gyr). Superposed is a young stellar
population with burst mass fraction, $f_{\rm burst}$, which represents
the fraction by mass of all stars formed in a brief, $\Delta t_{\rm
burst}$ = 20--300 Myr, burst. We allow the burst mass fractions to range
from 0.02 to 1.0. We also vary the metallicities of the old and young
stellar populations between $Z=0.004$ and $Z=0.02$.  The purpose of this
analysis is \emph{not} to reconstruct the star formation history (SFH)
of our galaxies, but to generate a grid of SEDs that covers the wide
range of observed spectral parameters and broad-band colors.

Using this grid of SEDs and the IRAF/Synphot routine, we measure
the Johnson $B$, \B{435}, \B{439}, Cousins $R$, Gunn $r$, \R{702}
and \R{625} magnitudes over time until 2 Gyr after the end of burst
(well beyond when the galaxy would be classified as an E+A).  We derive
transformation equations by fitting to the data across the model grid
using the following linear relations,
\begin{equation}
\label{eq:transform}
{\rm TCOL} = c_0 + c_1\ {\rm SCOL} + c_2 \ {\rm SCOL}^2
\end{equation}
where TCOL and SCOL represent the target and source color (e.g.,
TCOL = \B{435} $-$ \R{625} and SCOL = \B{439} $-$ \R{702}).  In Table
\ref{tab:transform}, we list the conversion parameters for the WFPC2,
ACS, Landolt (i.e., Johnson $B$ + Cousins $R$) and Thuan-Gunn systems.

To compare the properties of E+As across a range of redshift ($z =
0-0.15$), one should apply {\sl K}-corrections. In principle, the {\sl
K}-correction can be derived from the known SED of the object for an
arbitrary redshift and filter. If SEDs are not available, but there are
enough measurements of the broadband magnitudes, one can reconstruct
the galaxy SED and calculate the {\sl K}-correction \cite[e.g.,
][]{Blanton03b}. However, neither flux-calibrated spectra nor multiple
broadband colors are available for our sample.

Therefore, we adopt an alternative approach using the synthetic photometry
of the described model K+A SED library. The procedure is similar to what
we use to derive the magnitude transformation equations. Using the grid of
SEDs, we redshift the rest frame spectra to the appropriate E+A redshift,
$z$, measure the observed \Bz{435}{z} and \Rz{625}{z} magnitudes, and
calculate the differences from the corresponding rest frame magnitudes
(i.e., {\sl K}-corrections) as a function of \Bz{435}{z}$-$\Rz{625}{z}
color.  For each E+A, we determine the {\sl K}-correction at the measured
(\Bz{435}{z}$-$\Rz{625}{z}) color and use the range of  model values as
an error estimate for the {\sl K}-correction.



\begin{figure*}
\epsscale{0.85}
\plotone{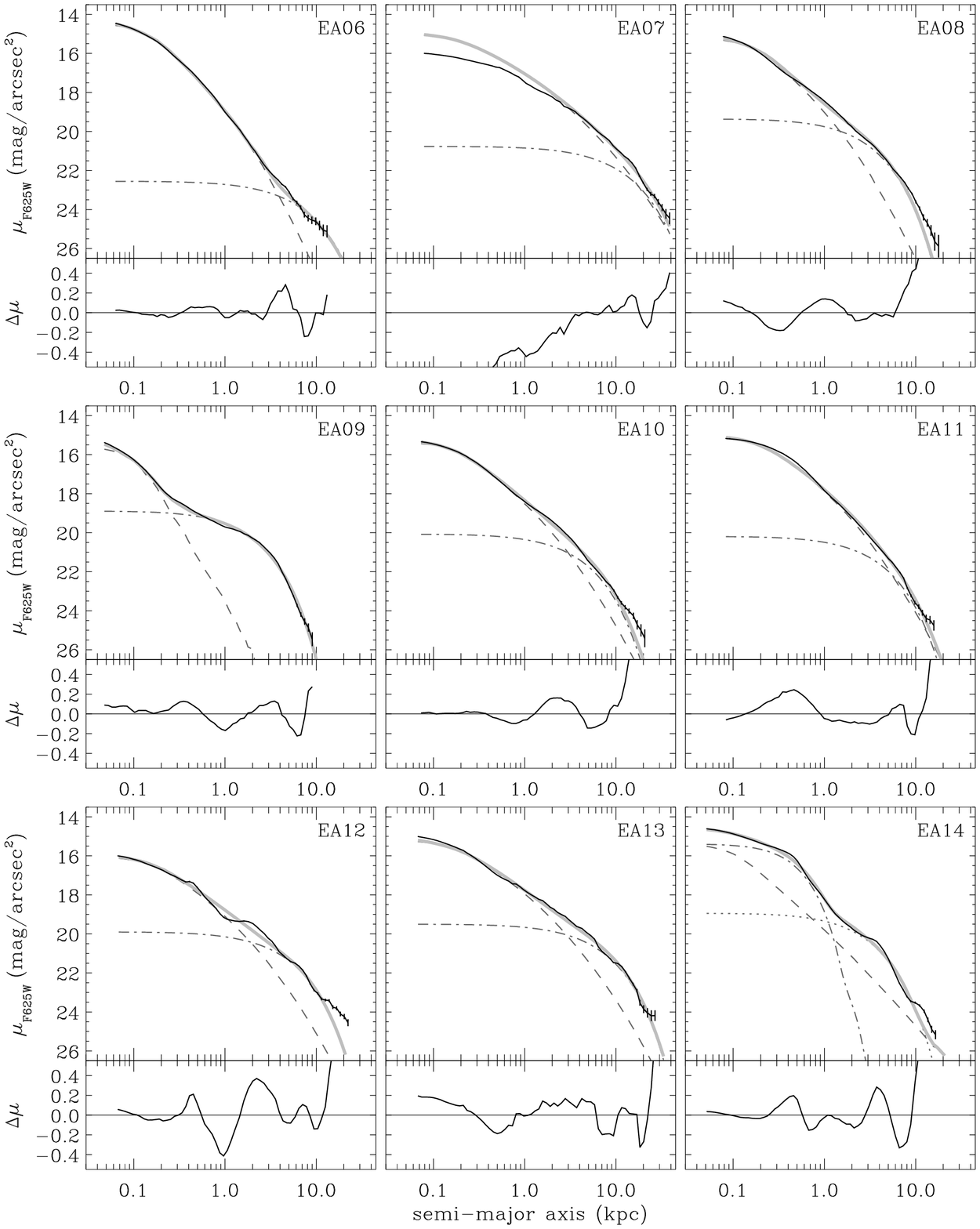} 
\caption{
E+A surface brightness profiles. The data are extracted using the
ELLIPSE task (dark solid line) and the model profiles with GALFIT
(gray solid line).  The dashed and dot-dashed lines are the bulge and
disk components from GALFIT, respectively. For the barred galaxies (EA14
and 21), we show the three components fits (S\'ersic + S\'ersic + disk;
Appendix \ref{apdx:qualitative_morphology}).  The differences between the
data and the model are shown in the bottom panels.  The large excess of
light at large radii ($r \geqslant 10$kpc) in some E+As (EA08, 10, 11,
12, 18) is mainly due to low surface brightness tidal features.
\label{fig:profile}}
\end{figure*}

\begin{figure*}
\addtocounter{figure}{-1}
\epsscale{0.85}
\plotone{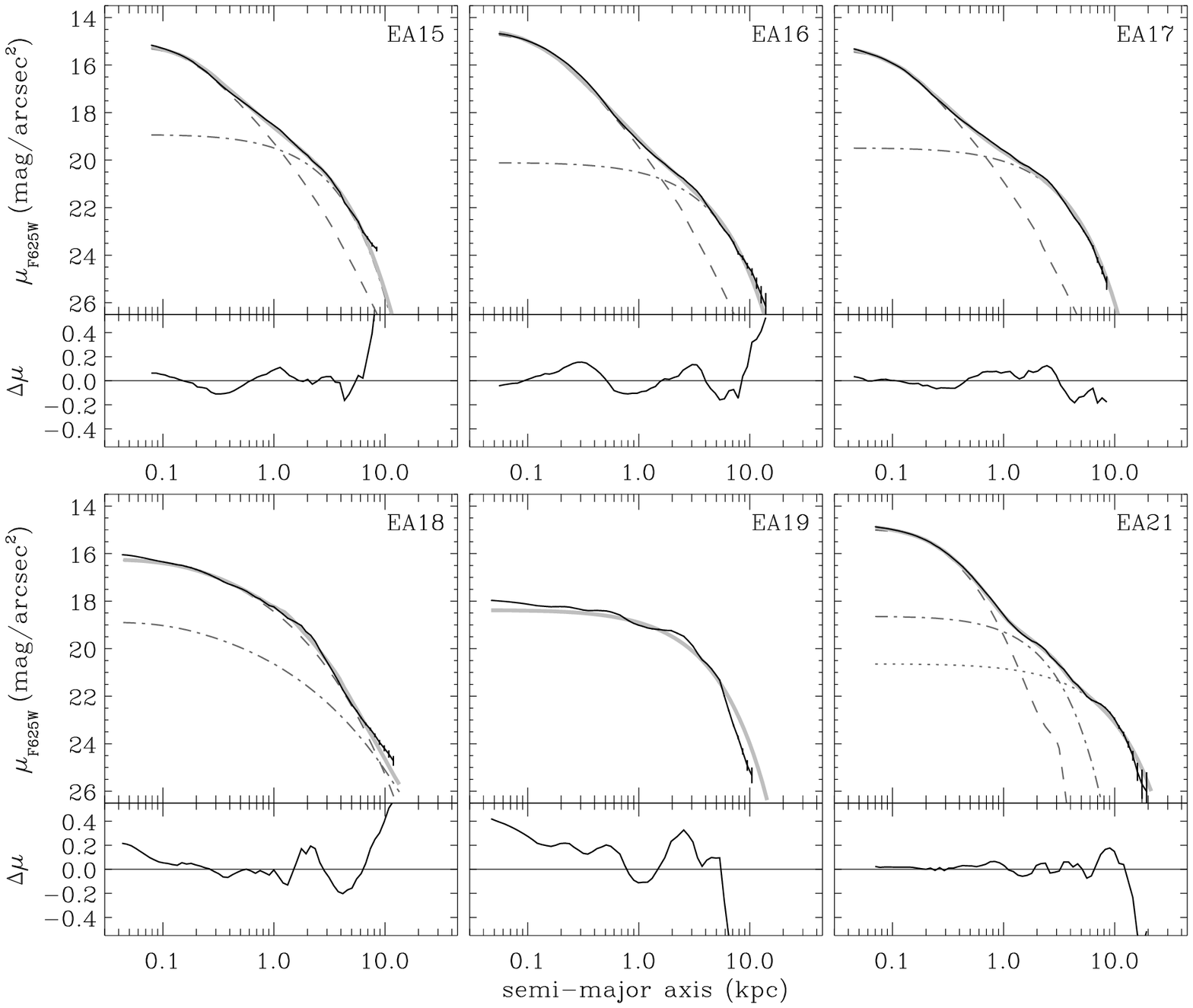}
\caption{Continued.}
\end{figure*}

\section{Morphology}
\label{sec:morphology}

\subsection{Qualitative Description}
\label{sec:qualitative_morphology}

We show images of the six {\sl WFPC2} E+A galaxies in Figure
\ref{fig:images_wfpc2}.  The leftmost panels are high contrast images
in $R$ and reveal low surface brightness features.  The middle panels
are the $R$ band images (the $B$ band images are far more shallow).
Figure \ref{fig:images_acs} shows the {\sl ACS} data for the remaining 15
E+As. The leftmost panels are the same as Figure \ref{fig:images_wfpc2}.
The middle panels here are the two-color composite images from the $B$
and $R$ bands. We adopt a color scheme employing the arcsinh stretch from
\citet{Lupton04}. We bound each image with 4 arcsec tickmarks and include
a 4 kpc horizontal scalebar.  The right panels of both figures are the $R$
band residual images resulting from the subtraction of best-fit models
(see \S \ref{sec:fitting}).

High resolution {\sl HST} images enable us to identify a wealth of
small and large scale features. Readers are referred to Appendix
\ref{apdx:qualitative_morphology} for detailed descriptions of the
morphological features. Here we briefly describe our findings.  First,
the morphologies of E+As are extremely diverse, including train-wrecks,
barred galaxies, blue-cores, and relaxed-looking disky galaxies.  Given
that our E+A sample was selected using uniform spectroscopic criteria, it
is striking that the morphologies are so varied.  Second, more than half
(55\%) of the E+As have tidal and/or disturbed features brighter than our
detection limit ($\mu_R < 25.1\pm0.5$ mag arcsec$^{-2}$), supporting the
picture that galaxy-galaxy interactions/mergers are responsible for E+A
phase. Five E+As also have apparent companion galaxies within $\sim$
30 kpc that appear to be interacting with the E+A.  Third, six E+As
(30\%) exhibit distinct compact blue cores with a characteristic size
of $\sim$ 0.5\arcsec.  Fourth, seven E+As show dust features, such
as lanes and filamentary structures, in the two-color composites or
the residual images. Only three of those have color profiles that are
seriously affected by these dust lanes (\S\ref{sec:color_profile}).

\subsection{Surface Photometry and Model Fitting}
\label{sec:fitting}

To compare E+A morphologies with those of early- and late-type galaxies
in a quantitative way, we fit models to the surface photometry. First,
to obtain the surface brightness profiles (and the color profiles
in \S\ref{sec:color_profile}), we extract the one-dimensional
azimuthally-averaged light profiles of the galaxies using the IRAF/ELLIPSE
task. Because a large fraction of E+As are disturbed and have asymmetric
features, we allow the model to follow the light distribution as closely
as possible. Thus, we allow the center, major axis position angle, and
ellipticity of each ellipse to change freely, but we force ellipses to not
overlap by fitting over a limited radial range along the major axis. The
only difference from our analysis of the previous WFPC2 imaging is that
we forego point-spread function deconvolution because the errors in the
drizzled ACS images are correlated.



To obtain global photometric parameters, such as effective radius $r_e$,
effective surface brightness $\mu_e$, and S\'ersic index $n$, we use
the two-dimensional image fitting algorithm GALFIT \citep{Peng02}.
GALFIT assumes a two-dimensional model profile for the galaxy with
the following free parameters: the $(x,y)$ position of the center,
$\mathrm{M}_{tot}$ (the total magnitude of the component), $r_e$ (the
effective radius), $n$ (the S\'ersic index), $q$ (the axis ratio defined
as $b/a$), the major axis position angle, and $c$ (the diskiness/boxiness
index, where $c > 0$ indicates boxy).  This index $c$ plays the same role
as the $\cos 4\theta$ Fourier coefficient term used often in isophote
analysis \citep{Jedrzejewski87}.  As GALFIT explores parameter space,
it convolves the model image with a point-spread function (PSF) and
compares it to the data for each parameter set.

\begin{figure*}
\epsscale{1.1}
\plotone{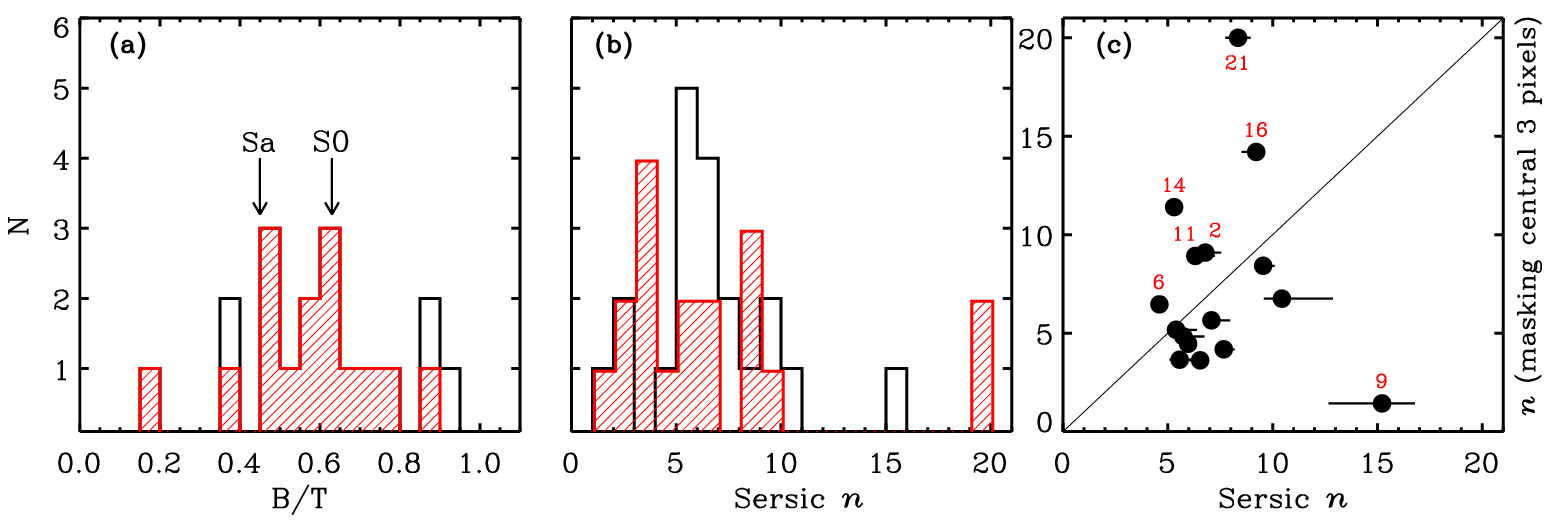} 
\caption{
($a$) Distribution of $R$ band bulge fractions ($B/T$) of 15
E+As where the bulge-disk decomposition is reliable (shaded histogram).
Except for two disky E+As (EA09 and EA17), most E+As (13/15) have $B/T$'s
greater than the median for Sa galaxy bulges (0.45). The median $B/T$
of E+As (0.59) is roughly consistent with that of S0 galaxies (0.63).
($b$) Distribution of S\'ersic indices $n$ in the $R$ band for 20
E+As (except EA01A) using a single S\'ersic fit (solid) and S\'ersic +
disk fit (shaded).  Most E+As (17/20) have very high S\'ersic index
($n\gtrsim5$), even higher than the $n=4$ of \devauc profiles of normal
ellipticals, indicating that the galaxy luminosity is highly concentrated.
($c$) S\'ersic indices before and after masking the galaxy center.
Most galaxies (65\%) have a smaller $n$ after masking, suggesting that
the central pixels play a role in the large derived $n$ values.
\label{fig:hist_bulge}}
\end{figure*}

We use three types of model profiles with a pre-determined fixed sky
background:
1) $r^{1/4}$-law bulge plus an exponential disk, 
2) single \devauc $r^{1/4}$-law profile, and 
3) single S\'ersic $r^{1/n}$-law profile.
Therefore, for each galaxy, we have three sets of structural parameters,
unless bulge/disk decomposition proves to be impossible (e.g., EA01A
and 19).  Bulge-disk decomposition (model 1) is useful for determining
whether there is a disk and for measuring a bulge fraction ($B/T$).  Model 2
is used mostly to compare the morphologies of E+As with those of normal
ellipticals or S0 galaxies, and we use the derived parameters to place
E+As on  the Fundamental Plane (\S\ref{sec:fp}).  Model 3 is flexible
enough to distinguish early type galaxies ($n>3$) from disk-dominated
late type galaxies ($n\sim1$).  We determine the adopted sky value using
median values from 40$\times$40 pixel regions that appear free from
tidal features or other large scale sources.  The GALFIT input PSFs are
evaluated using unsaturated stars in the ACS field with S/N higher than
100 as identified by the IRAF/DAOPHOT package.  Because the number of
available stars is often too small ($N<5$) in each E+A field, we generate
the time- and spatially-averaged PSFs using all of the stars ($N\sim90$)
in our \emph{{\sl HST}} observations.  We fit the galaxy images with all
three profiles and determine which profiles are the best fits based on the
$\chi^2$ values and visual inspection of the residual images.

We  plot the light profiles derived from the ELLIPSE fit and the
best fitting model profiles from GALFIT in Figure \ref{fig:profile}.
As expected, the 1--D and 2--D fits do not always agree, especially when
there are a significant changes in PA or ellipticity, but generally they
are consistent to within $\sim$0.2 mag.
Because systematic errors, such as that introduced by the fixed sky value,
dominate, we run GALFIT with various plausible fitting input parameters
and take the full range of fitting results as representative of the
final errors. Therefore, the errors given in Table \ref{tab:devauc}
and \ref{tab:decomp} are conservative estimates of the true
uncertainties.

\subsection{Light Profiles}
\label{sec:profile}

\subsubsection{Bulge-disk Decomposition}

In Figure \ref{fig:hist_bulge}$a$, we show the distribution of the $R$
band bulge fractions ($B/T$) of the 15 E+As for which the bulge-disk
decomposition is reliable (red shaded histogram).  The
decompositions are not reliable in other cases, for example, in those galaxies
(EA07, 18, 19) where a large portion of the bulge is masked by dust
lanes and in strongly disturbed systems (EA01A, EA02, EA12) where the
distinction between bulge and disk is meaningless.  Except for two
disky E+As (EA09 and EA17), most (13/15) of the remaining E+As have
$B/T$'s greater than the median for Sa's \cite[0.45;][]{Kent85}. The
median $B/T$ of E+As (0.59) is consistent with that of S0 galaxies
\cite[0.63;][]{Kent85}. E+A galaxies are mostly bulge-dominated
systems, confirming previous studies using bulge-disk decomposition
\cite[e.g.,][]{Yang04,Blake04,Balogh05}.

\subsubsection{S\'ersic Profile}

The S\'ersic index $n$ is often used to quantify galaxy morphology
\cite[e.g.,][]{Blanton03a} because $r^{1/n}$ profiles are flexible
enough to represent a wide range of profiles, from those of disk galaxies
($n=1$) to those of spheroidals ($n=4$). E+A galaxies in the SDSS have
large S\'ersic indices (median $n = 3$), and therefore their luminosity
is centrally concentrated \citep{Quintero04}. We fit single S\'ersic
profiles to all of our E+A galaxies, and present the results in  Table
\ref{tab:devauc}. This model generally results in larger $\chi^2_{\nu}$
values than those obtained with bulge+disk decomposition, but both sets
of models are statistically acceptable.

In Figure \ref{fig:hist_bulge}$b$, we plot the distribution of the
S\'ersic index $n$ in the $R$ band for 20 E+As,  
the entire sample except for EA01A, which is too disturbed to fit.  Most E+As
(17/20) have very high S\'ersic index, $n\gtrsim5$, even higher than the
$n=4$ \devauc profile of normal ellipticals, indicating that the galaxy's
luminosity is highly concentrated, presumably due to recent centralized
star formation.  The exceptions are the three dusty galaxies (EA07, EA18,
EA19) with $n=1-3$, where we hypothesize that the dust lanes partially
mask the bulge component.  High central concentrations do not necessarily
mean that the galaxies are bulge-dominated systems. For example, two
of our most disky systems, EA09 and EA17, with $B/T = $ 0.17 and 0.35,
respectively, have unusually high S\'ersic indices ($n>10$).

What causes the S\'ersic index of E+As to be even larger than those of
early-types \cite[e.g.,][]{Kelson00a, Graham01, Graham02}?  We consider
three possibilities: 1) we are being misled by the comparison 
across different studies because the derived S\'ersic
indices are strongly dependent on the fitted radial range, the choice
of fitting method (1D vs. 2D), and the depth of the images, 2) the
index is dominated by the presence of an additional, highly concentrated,
central source, and 3) misidentification of a disk or tidal feature
as the shallow wing of the S\'ersic profile is creating the apparently
large central concentration.  For the remainder of this discussion, we
presume that the first possibility is not the origin of these results,
although a more thorough cross-comparison among samples is warranted.

To test the degree to which the central regions of E+As affect the
S\'ersic fits, we mask the centers with a 3 pixel radius mask and repeat
the fitting.  In Figure \ref{fig:hist_bulge}$c$, we plot the S\'ersic
indices before and after masking. Most (65\%) of the galaxies have a smaller
$n$ after masking, which suggests that a central concentration does
play a role in causing the high $n$'s. This conclusion is supported
by positive central residuals in the model subtracted images
(Fig. \ref{fig:images_wfpc2} and \ref{fig:images_acs}). EA09 is a
special case because the bulge is so small that this masking leaves
only the disk component. However, S\'ersic indices in six E+As actually
increase after the masking. The E+As with a larger resulting $n$ after
masking generally have ring-like substructures in the residual images
and negative central residuals.  We conclude that substructures near the
center, e.g., bright nuclei, bars, and rings, play an important role in
generating the highly concentrated light of most E+As.

The last of our three possibilities is that disk or tidal features are
fitted as the shallow wings of the S\'ersic profiles. To examine the
S\'ersic index of the bulge component alone, we fit a S\'ersic+disk
model. In Figure \ref{fig:hist_bulge}$b$, we show the resulting
distribution of S\'ersic indices for the bulge components (red shaded
histogram) compared with the original, single-component values.
The S\'ersic index of the bulge itself is more consistent with that
of normal bulges ($n\simeq 3-5$); however, some bulges still have
large S\'ersic indices ($n > 5$). We conclude that both the central
substructures and disk-like features are responsible for the unusually
high S\'ersic indices in E+As. In any case, our E+As are more consistent
with $n\sim4$ than $n\sim1$.

\subsection{Concentration and Asymmetry, Residuals}
\label{sec:ca}

So far, we have derived structural parameters by fitting smooth
symmetric models to the images, even though asymmetric features are
quite common in the residual images (Fig. \ref{fig:images_wfpc2}
and \ref{fig:images_acs}). To quantify the asymmetric features
and measure morphologies in a model-independent way, we calculate
nonparametric measures of the galaxy morphologies: the concentration
index $C$ \citep{Abraham94} and the rotational asymmetry index $A$
\citep{Schade95}.  In previous work \citep{Yang04}, we used the
azimuthal Fourier decomposition method \cite[e.g.,][]{Rix95} to
investigate asymmetric features of the disk-like components, such as
the lopsideness in two face-on disky E+As (EA03 and EA04). We use $CA$
indices in this paper, because a large comparison sample for the azimuthal
Fourier decomposition is not available.

To determine the $C$ and $A$ indices (collectively referred to as
$CA$), we follow the methodology described by \citet{Conselice00} and
\citet{Bershady00}.  First, the Petrosian radius $r_p$ is determined
from the growth curve for circular apertures requiring that $\eta \equiv
{I(r_p)}/{\lan I(r_p) \ran} = 0.2$, where $I(r)$ is the surface brightness
at the radius $r$ and ${\lan I(r) \ran}$ is the mean surface brightness
within $r$ \citep{Petrosian76}.  We calculate the $CA$ indices for image
pixels within the ``total'' aperture defined to be twice the Petrosian
radius.  We measure the asymmetry index $A$ by subtracting the galaxy
image rotated by 180\degr\ from the original image and correcting for
the contribution to the asymmetry arising from background noise:
\begin{equation}
\label{eq:A_def}
A \equiv min \bigg(\frac{\sum |I_0 - I_{180}|}{\sum |I_0|}\bigg)
                 - \frac{\sum |B_0 - B_{180}|}{\sum |I_0|},
\end{equation}
where $I$ and $B$ represent the image pixels within a circular aperture
of radius $2 r_p$ and the background region, respectively. We choose a
rotational center that minimizes $A$ by searching over sub-pixel grids
with 0.1 pixel resolution.
The concentration index $C$ is defined by 
\begin{equation}
C \equiv 5\,\Log \bigg(\frac{r_{80}}{r_{20}}\bigg), 
\end{equation} 
where $r_{20}$ and $r_{80}$ are the radii containing 20\% and 80\% of
the total flux, respectively.  Galaxies with a \devauc profile ($n=4$)
or a pure exponential disk profile have  $C = 4.50$ or 2.79, respectively.

\begin{figure}
\epsscale{1.0}
\plotone{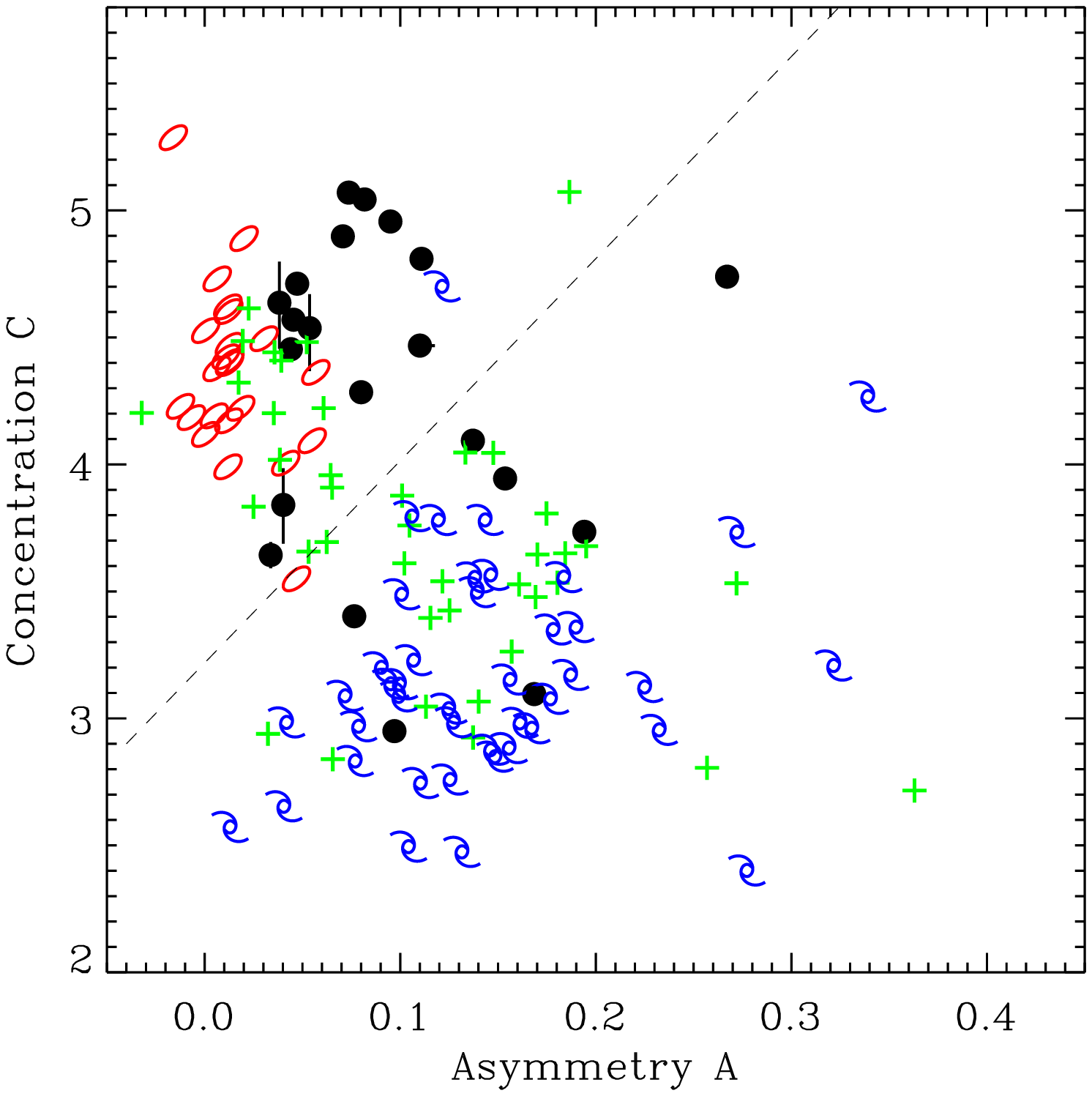} 
\caption{
$CA$ classification diagram for our 21 E+As as well as 113 local
galaxies with various Hubble types drawn from the \citet{Frei96} catalog.
The ellipses, crosses and spirals represent early-types ($T\leqslant0$),
intermediate spirals, and late-type spirals ($T\geqslant5$), respectively.
E+As are represented with filled circles.  We overlay a dashed line
providing a rough division of the $C-A$ plane into regions of early-
and late-types.  E+As populate the $CA$ plane widely, as suggested by
their morphological diversity.  Two-thirds of E+As occupy a unique part
of the plane:  high concentration ($C$ $\gtrsim$ 4.3) and moderately
large asymmetry ($A$ $\gtrsim$ 0.04).  These E+As have concentrations
consistent with those of early type galaxies ($\lan C \ran = 4.4 \pm
0.3$) or even higher, as expected from their $B/T$ fractions and high
S\'ersic indices $n$.   However, they have considerably larger asymmetry
than ellipticals ($\lan A \ran = 0.02\pm0.02$) due to disturbances
in their centers, presumably arising from the starburst and/or recent
galaxy-galaxy interaction.
\label{fig:ca}}
\end{figure}

In Figure \ref{fig:ca}, we show the $CA$ classification diagram, which
is frequently used for the morphological classification of high-$z$ galaxies
\cite[e.g.][]{Abraham96}, for our 21 E+As and for 113 local galaxies with various
Hubble types drawn from the \citet{Frei96} catalog. The $CA$ indices
of the Frei sample were measured in exactly the same way as those of
the E+As. Although the $CA$ diagram is not an ideal classification tool
\cite[e.g.,  see][]{Conselice00}, it is quite useful in distinguishing
early from late type galaxies.  In Figure \ref{fig:ca},
intermediate spirals occupy a wide space
in the $CA$ plane, but early-types are segregated from late-types.
The $CA$ indices have low sensitivity to S/N and
spatial resolution, within certain limits. For example, \citet{Lotz04}
show that $CA$ indices can be robustly measured within $\Delta C \lesssim
0.1$ and $\Delta A \lesssim 0.05$ for images with an average $\lan S/N
\ran$ $\gtrsim$ 5 and a spatial resolution of $\sim$ 500 pc per pixel
or better \cite[see also][]{Conselice00}.  Because our {\sl HST} images
satisfy these criteria ($\lan S/N \ran$ $\gtrsim$ 7 within $r_p$ and
a spatial resolution $\lesssim$ 100 pc per pixel), a direct comparison
between the E+As at $z\sim0.1$ with the local Frei sample is valid.

E+As populate a wide range in the $CA$ plane,  as already suggested
by their morphological diversity. We overlay a dashed line providing a
rough division of the $CA$ plane into regions of early- and late-types.
Most of the dusty E+As (EA07, 12, 13, 18, 19) and the train-wreck E+A
(EA01A) lie below the dividing line and have large asymmetry and/or low
concentration, presumably due to extinction.  The remaining E+As are
located above the line, but populate a unique part of the $CA$ plane
where there is both high concentration $C$ $\gtrsim$ 4.3 and moderately
large asymmetry $A$ $\gtrsim$ 0.04.  These E+As have concentration
indices greater than or equal to those of early types ($\lan C \ran =
4.4 \pm 0.3$), as expected from their $B/T$ fractions and high S\'ersic
index $n$'s, but they have a considerably larger $A$ than ellipticals
($\lan A \ran = 0.02\pm0.02$).


Where in radius does the largest contribution to the $A$ parameter
come from?  Faint tidal features do not contribute significantly to
$A$ because 1) the total aperture size is not large enough to enclose
features at such large radii and 2) the asymmetry is normalized using
the total galaxy luminosity (eq. \ref{eq:A_def}), so little weight is
placed on low surface brightness tidal features.  In the local galaxy
sample, small scale features such as spiral arms and flocculence in the
disk are the main contributors to $A$ \citep{Conselice00}.  However,
such features are not visible in the disks of our E+As (except for in
EA03 and EA19), which are smooth \cite[see also the {\sl NIR} imaging
from][]{Balogh05}.   Therefore, the large asymmetry of E+As is probably
due to relic structures within a few effective radii arising from the
starburst and/or recent galaxy-galaxy interaction.  \citet{Yamauchi05}
reached a similar conclusion from their analysis of 22 SDSS E+As.

To further investigate the nature of the asymmetry, we compare the
asymmetry profiles, i.e., the asymmetry index interior to radius $r$,
$A(<r)$, of E+As and late-type spirals ($T\geqslant5$).  Except for
three E+A galaxies (EA01B, 03, 11), E+As have decreasing or flat $A(r)$
profiles.  Conversely, the late-type galaxies tend to have increasing
asymmetry profiles, because spiral arms and flocculence in the disk are
increasingly included as the aperture increases.  We conclude that the
asymmetry in E+As originates mostly in the central parts (within a few
$r_e$), presumably due to unrelaxed structures or residual dust from
the starburst and/or recent merger.  These E+As would be classified as
``normal'' elliptical galaxies or early-type spirals if they were observed
at high redshift, unless the disturbances in their innermost regions
(within a few kpc) or the low surface brightness features with $\mu \simeq
24-25$ could be clearly identified.  In Table \ref{tab:cas}, we list the
aperture sizes ($r_p$), $CA$ indices, and relevant aperture magnitudes.

\section{Color Profiles}
\label{sec:color_profile}


\begin{figure*}
\epsscale{0.95}
\plotone{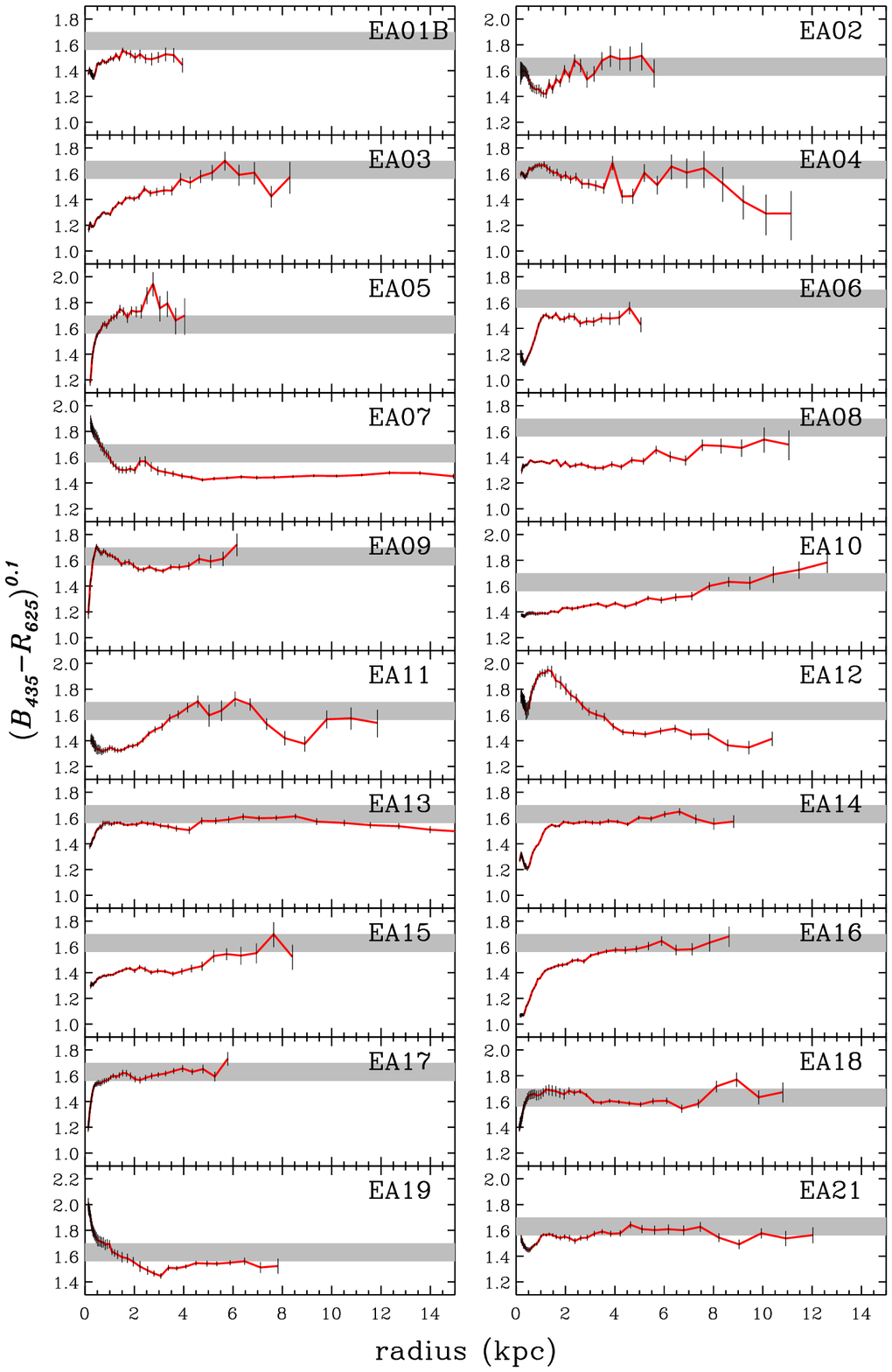} 
\caption{
Redshifted radial (\B{435}$-$\R{625})$^{0.1}$ color profiles of
the 20 E+As.  All the colors are {\sl K}-corrected to $z = 0.1$.
The color profiles are as diverse as the morphologies of E+As, ranging
from negative color gradients (EA02, 04, 07, 12, 19) to positive
color gradients including compact blue cores (EA05, 06, 09, 14, 16,
17). The large fraction (70\%) of positive color gradients suggests
that many E+As had a central burst of star formation and therefore that
the E+A phase is triggered by galaxy interactions/mergers that funnel
gas toward the center of the remnant.  The shaded region indicates
the (\B{435}$-$\R{625})$^{0.1}$ colors of a 5--10 Gyr old SSP with
metallicity $Z=0.004$ (1/5$Z_{\sun}$).  The color of an SSP with solar
metallicity is redder and varies from 1.86 to 2.04 between 5 and 10 Gyr.
Recent star formation is not limited to the central region, as evidenced
by the fact that the color is bluer than the shaded region at most radii.
\label{fig:color_profile}}
\end{figure*}

The radial color profile of a galaxy depends on its dust content and
the spatial distributions, ages, and metallicities of its stellar
populations, which in turn depend on the evolutionary history of the
galaxy.  For example, if galaxy-galaxy interactions are responsible for
an E+A's recent starburst, the young stellar population is expected to
be centrally concentrated \cite[e.g.,][]{Mihos&Hernquist94a,Bekki05}
and to produce a positive color gradient, i.e., a redder color with
increasing radius. On the other hand, if the E+A arises from a truncation
of continuous star formation, due to a mechanism such as ram pressure
stripping in the cluster environment, then it may have a uniform color
profile \cite[see the extensive discussions in][]{Caldwell99,Rose01}.

Current observational evidence supports the model of centralized star
formation in E+A galaxies. Using long slit spectroscopy, \citet{Norton01}
find that young stars are more centrally concentrated than older ones
in our E+As. Even for E+As in clusters, \citet{Bartholomew01} show
that on average they become slightly bluer toward the center than normal
early-type galaxies. Recently, \citet{Yamauchi05} found that a substantial
fraction of SDSS E+As have such a positive color gradient. However,
these studies are somewhat limited due to the galaxies' small angular
extent and/or the effects of ground-based seeing. In this section, we
present high resolution internal color distributions of our E+As and
show that indeed a significant fraction have positive color gradients
and sometimes distinct blue cores. Then, we investigate whether the
color profiles of E+As can evolve into those of early-type galaxies.


\begin{figure}
\epsscale{1.10}
\plotone{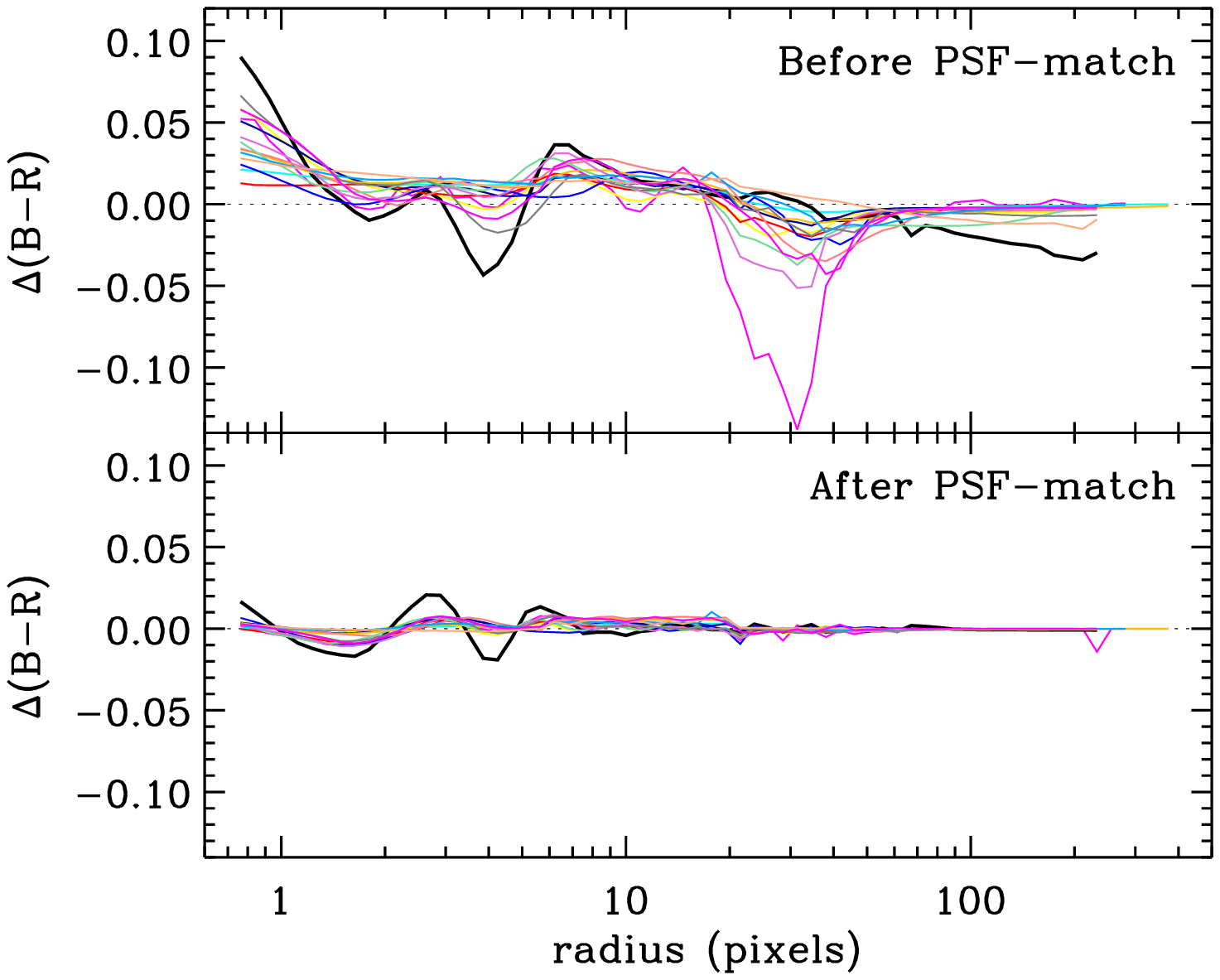} 
\caption{
Effect of the PSF correction. ({\it Top}) Measured color profiles for
model galaxies with a uniform color distribution, i.e., a zero color
gradient, without convolving the PSF-matching kernels with the $R$
band images. The non-zero color gradients arising due to differences
in the PSF function are clear over the entire range of radii depending
on the model galaxy. ({\it Bottom}) Color profiles extracted after the
PSF-matching. The measured color profiles deviate from the flat profiles
by no more than $\sim$ 0.02 magnitudes. Our color gradient measurement
is valid even in the innermost region.
\label{fig:psf_match}}
\end{figure}

\subsection{Positive Color Gradients in E+A Galaxies}

We show the (\B{435}$-$\R{625}) radial color profiles of 20 E+As in Figure
\ref{fig:color_profile}, excluding EA01A because it is so irregular.
To enable direct comparison between the profiles, the (\B{439}$-$\R{702})
colors were transformed into (\B{435}$-$\R{625}), and all colors were {\sl
K}-corrected with respect to $z = 0.1$ to minimize the {\sl K}-correction
uncertainties. We refer to these colors as $(B-R)^{0.1}$.  Note that we do
not apply different {\sl K}-corrections along the color gradients because
these corrections are small (typically 0.02 mag) compared to the overall
{\sl K}-correction values.  We measure the radial color profiles using
the surface brightness profiles obtained with the ELLIPSE task. Before
fitting ellipses to the isophotes, we correct for the different point
spread functions in the $B$ and $R$ bands by convolving the sharper
$R$ band images to match the $B$ band images.  The smoothing kernel is
determined using unsaturated stars in both images.  We extract surface
brightness profiles in the $R$ band as described in \S \ref{sec:profile},
and then extract the $B$ band profiles using the same isophotes as in
the $R$ band analysis.


Note that this PSF-matching is critical in order to investigate
the innermost color profiles of E+A galaxies and to identify color
structures such as blue cores that can be as small as a kpc (see also
\citet{Menanteau04} for a comprehensive treatment of PSF issues).  Figure
\ref{fig:psf_match} shows the effect of this PSF correction.  Assuming
an intrinsically uniform color distribution, i.e., a zero color gradient, we
generate model $B$ and $R$ images with the best fitting GALFIT models for
our 20 E+As, and then convolve these with the appropriate $B$ and $R$
PSF's to generate the ``observed'' $B$ and $R$ images.  In the upper
panel, we plot the observed color profiles when we do not match PSFs.
We plot the measured color profiles after PSF-matching in the lower
panel. The improvement is obvious and the measured color profiles deviate
from the intrinsic flat profile by no more than $\sim$ 0.02 magnitudes.
The extracted profiles tend to be biased toward redder cores, but the
effect is negligible.

We find that all E+As have globally blue colors, except within dusty
regions, and so conclude that recent star formation extends across the
face of these galaxies \cite[consistent with the results of][]{Franx93,
Caldwell96, Norton01}. In Figure \ref{fig:color_profile}, we overlay the
model $(B-R)^{0.1}$ colors of 5 to 10 Gyr-old simple stellar populations
which are consistent with those of early type galaxies for comparison.

While E+As are relatively blue overall, their internal color variations
are surprisingly diverse, as suggested by their morphological diversity
(see Table \ref{tab:color_gradient}). Twelve E+As (57\%) have positive
color gradients, i.e., they become redder with increasing radius, while
five E+As (20\%; EA02, 04, 07, 12, 19) have negative color gradients. The
remaining five E+As have relatively flat color profiles or a mixture of
positive and negative color gradients.  Half of the E+As with positive
gradients exhibit blue cores, which we discuss below.  This diversity
echoes the variation in H$\delta$ absorption line strength profiles
observed in cluster E+As \citep{Pracy05}.  We list the overall color
morphologies of E+A galaxies in Table \ref{tab:color_gradient} within
broad categories: positive (including blue core), negative, or flat
color gradient.

Among the E+As with negative color gradients, three (EA07, 12, 19)
show clear dust signatures, such as dust lanes and irregular filamentary
structures in the two-color composite images.  The clear detection of dust
features in the {\sl ACS} sample is mainly due to the excellent $B$ band
sensitivity, and so we cannot exclude the possibility that similar dust
exists in the {\sl WFPC2} sample galaxies with negative color gradients
(EA02 and 04).  Therefore, although there are other possible explanations,
including a truncated IMF during the starburst (see \citet{Bekki05}),
we speculate that the red cores in these galaxies arise from increasing
dust extinction toward the center.

To quantify the color gradients, \dcdr, we perform linear least-square fits
to the radial color profiles. Because of the highly variable color
profiles and compact blue cores in some E+As, a single color gradient
is obviously not the best representation.
Therefore, we measure the color gradients with two piecewise linear
fits, allowing the break radius $R_{\rm break}$ between the inner and
outer slope to change freely. Our only constraint 
is $R_{\rm break}\gtrsim0.3$ kpc ($\sim$ 4 pixels) to prevent
measuring break radii that might be seriously affected by the PSF.
In Table \ref{tab:color_gradient}, we list the color gradients and break
radii from the piecewise linear fits, as well as the color gradient
derived from the single line fit. The color gradients are measured in
(\B{435}$-$\R{625}) at $z=0.1$ to minimize the systematic errors that may
arise from the {\sl K}-corrections and color transformations. For EA01-05,
the WFPC2 sample, we measure the color gradient in (\B{439}$-$\R{702})
at $z=0.1$ and convert it to (\B{435}$-$\R{625}) at
$z=0.1$ using the relation $\Delta$(\B{435}$-$\R{625}) $\simeq c_1$
$\Delta$(\B{439}$-$\R{702}) in Table \ref{tab:transform}. We find that
the broken linear fits significantly improve the fits ($\chi^2$s) in
all but three E+As (EA01B, 03, 15).

\begin{figure}
\epsscale{1.15}
\plotone{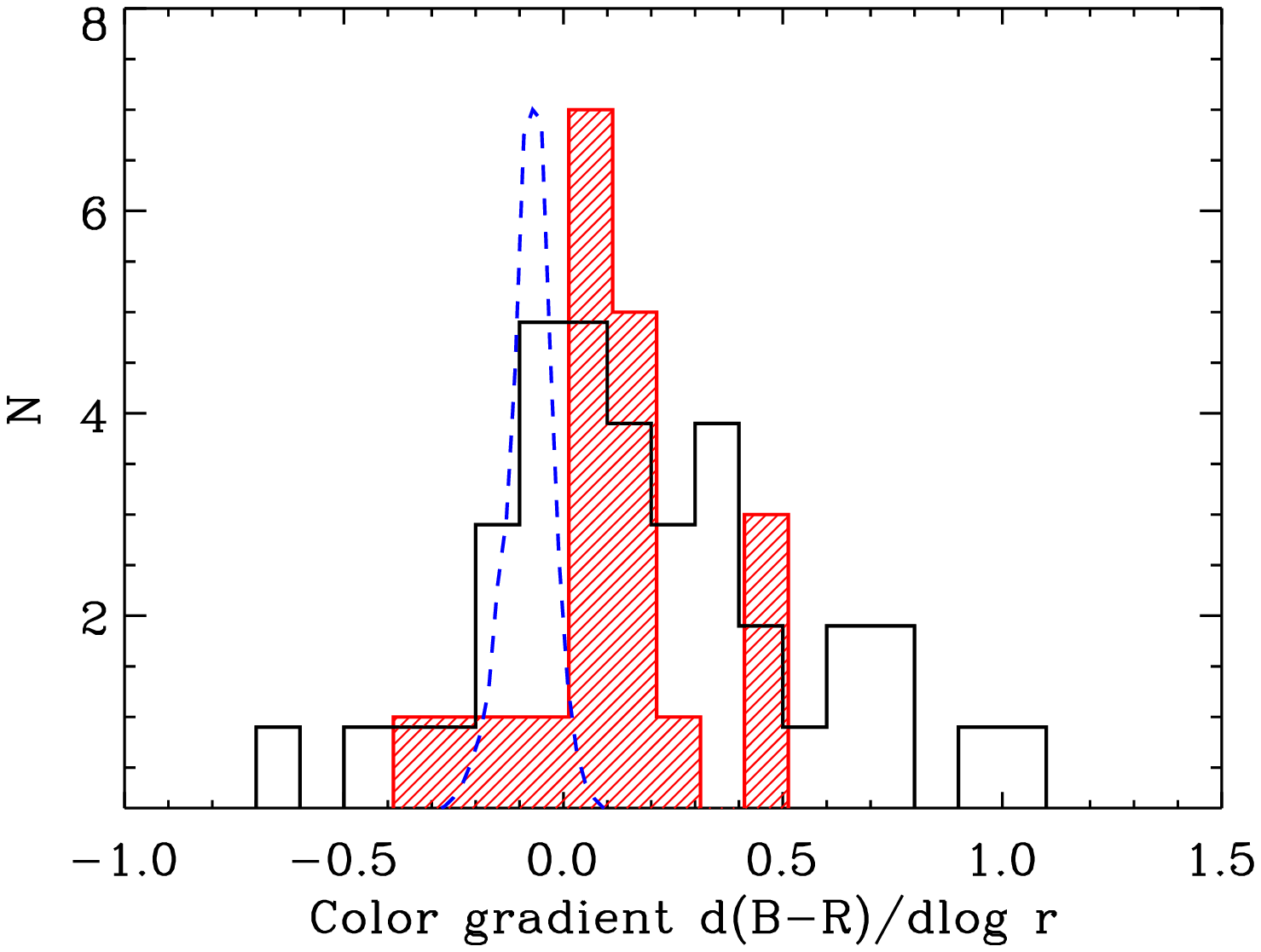} 
\caption{
Distribution of the E+A color gradients \dcdr. The solid and shaded
histograms represent the gradients measured from the two-line fits and
single line fits, respectively. Most E+A galaxies have positive color
gradients, i.e., the color becomes redder as the radius increases. Note
that there are some very steep [\dcdr\ $>$ 0.6] color gradients due to
the blue cores. The dashed curve represents the distribution of the
color gradients of local E/S0s \citep{Wise&Silva96} with arbitrary
normalization.  E/S0s have a narrow range of color gradients [\dcdr\
= $-0.09\pm0.06$]. These color gradients are interpreted as being the
result of metallicity gradients.
\label{fig:hist_gradient}}
\end{figure}

The distribution of color gradients is shown in Figure
\ref{fig:hist_gradient}, where the solid and hatched histograms indicate
the color gradients for single and double line fits, respectively.
The range of color gradients for early-type galaxies \cite[E and S0s
from][]{Franx89,Peletier90} is shown for comparison (dashed curve). The
majority of E+As ($\sim$ 70\%) have a positive color gradient that
deviates from the slightly negative profile typical of early-type
galaxies.  We conclude that the young stellar populations are more
concentrated than the underlying old populations.  This conclusion is
independent of whether the single or double line fits are used and is
qualitatively consistent with that of \citet{Yamauchi05}.


\begin{figure*}
\epsscale{0.91}
\plotone{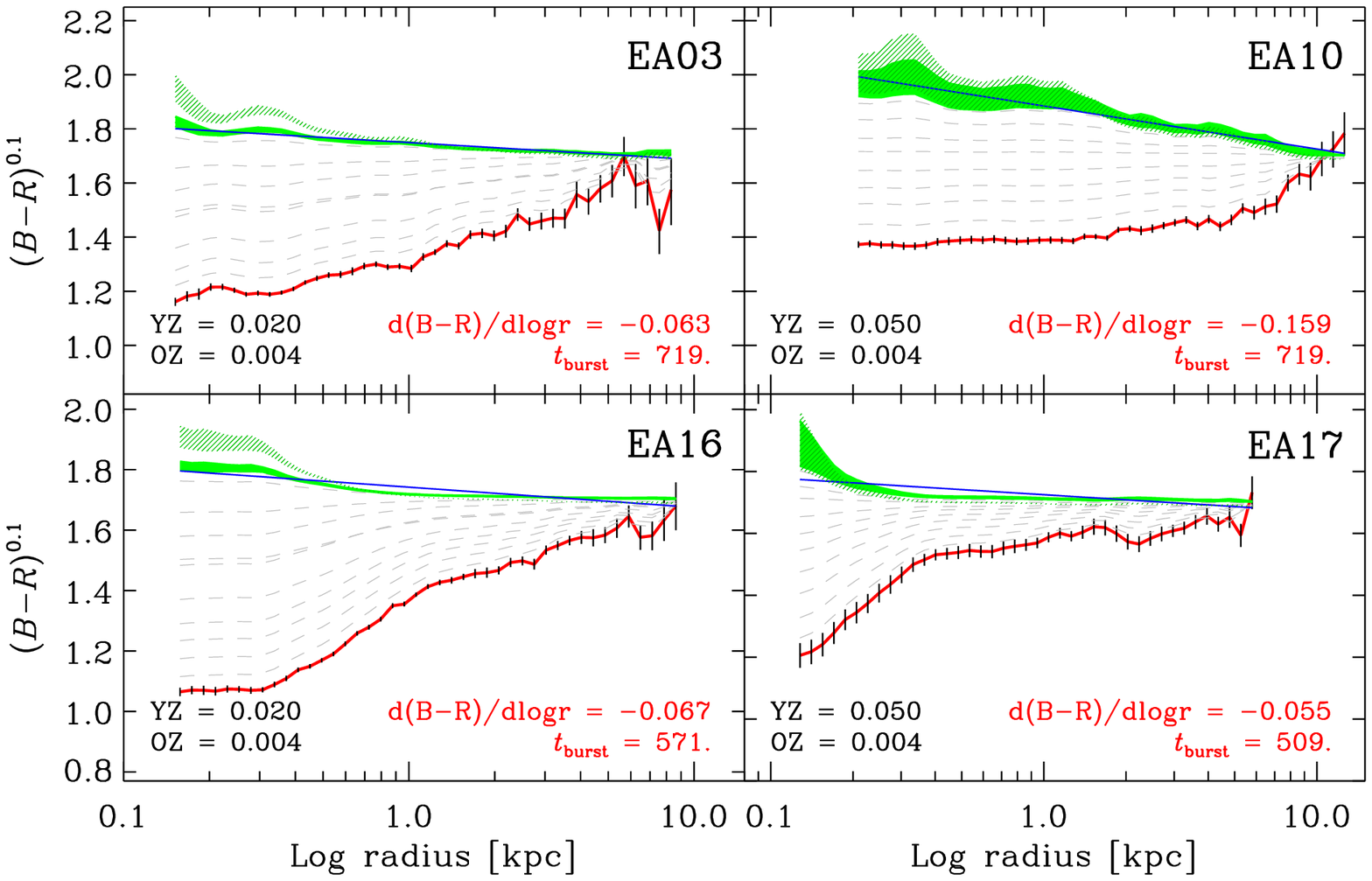} 
\caption{
Examples of color gradient evolution for two E+As with positive gradients
(EA3 and 10) and for two blue-core E+As (EA16 and 17). The solid and
dashed lines represent the current color profiles and the subsequent
evolution, respectively. The ranges of color profiles after 5 and 10 Gyr
are represented with solid and hatched regions, respectively. We adopt
$Z_{\rm young} = 0.02$ for EA03 and EA16 ({\it left}), and $Z_{\rm young}
= 0.05$ for EA10 and EA17 ({\it right}).  The chosen burst ages ($t_{\rm
burst}$) can produce the whole range of colors in each galaxy. The
positive color gradients in E+As can be inverted within $\sim$ 5 Gyr,
evolving into a range [$-0.05< $ \dcdr\ $ <-0.16$] consistent with that
observed for E/S0s.
\label{fig:color_grad_evol}}
\end{figure*}

\subsection{Blue Cores}

The most interesting morphological feature in certain E+As is the compact,
almost stellar-like, blue core that is distinct from the other parts
of the galaxy. We define blue core galaxies as those that have a color
variation, $\delta(B-R)$, within the inner 2 kpc region that is greater
than 0.3 magnitudes. This definition is comparable to the original
criteria adopted by \citet{Menanteau01a}, $\delta(V-I)_{\rm obs} \gtrsim
0.2$, for morphologically selected early-type galaxies. Six E+As (EA05,
06, 09, 14, 16, 17) satisfy our definition and have characteristic core
sizes of $R_{\rm break}$ = 0.4 -- 1.4 kpc.  Within these break radii,
the color gradients of blue core galaxies are very steep, \dcdr $ > 0.6$.
The slopes become relatively flat at radii larger than the break radii.
Note that we classified EA01B as one of the blue core galaxies based on
its {\sl WFPC2} $(B-R)$ colors in previous work \citep{Yang06}; however,
EA01B is not counted as a blue core galaxy in this paper because its
color profile does not have a break radius.



Though the origin of blue cores is not fully understood, they are
common in early-type galaxies at higher redshifts ($z$ $\gtrsim$ 0.5),
when field spheroids are presumably still assembling. For example,
30\% of the morphologically-selected elliptical galaxies in the Hubble
Deep Field North have color inhomogeneities, mostly due to blue cores
\citep{Menanteau01a}. \citet{Treu05} find that $\sim$ 8\% (14/165)
of early-type spheroidals in the Great Observatories Origins Deep
Survey North (GOODS-N) have blue cores. \citet{LeeJH06} also find
``blue clumps'' in almost half of the early-type galaxies with blue
colors in the GOODS North and South fields. The blue-core E+As in our
sample have relaxed morphologies and bulge fractions of $0.2-0.6$.
Given that the $CA$ classification scheme identifies these galaxies
as early type galaxies, the blue-core E+As would be classified as
early-types if they were observed at high redshift. Therefore, these
blue core E+A galaxies might be the local analog of the blue core
spheroids often found at high redshift.  The importance of blue core
E+As for better understanding galaxy-galaxy mergers and the origin of
the black hole mass-bulge velocity dispersion relation for galaxies
\cite[$M_\bullet-\sigma_B$;][]{Ferrarese00,Gebhardt00} is described
by \cite{Yang06}.

\subsection{Evolution of Color Gradient}

It is now well established that E/S0s in the local Universe have
negative color gradients [\dcdr\ = $-0.09 \pm 0.06$ mag dex$^{-1}$ in
Fig. \ref{fig:hist_gradient}], which originate from their metallicity
gradients; their stellar populations become more metal rich and redder
toward the center \cite[e.g.,][]{Tamura00}.  If E+As evolve into E/S0s,
how can their diverse color morphologies converge into such a narrow range
of color gradients within a few Gyr? In particular, is it possible that
the observed positive color gradients invert into the negative color
gradients typical of early-types?

For those E+As whose negative color gradients arise from dust, it
is difficult to determine the underlying color gradients without
incorporating radiative transfer calculations \citep{Wise&Silva96}.
Nor do we know how the dust content will evolve. However, were the
dust in these E+As to disperse somehow, the negative color gradients
[\dcdr\ $\lesssim$ $-0.3$] would flatten and be more consistent with
those of E/S0s.

For those E+As with positive color gradients --- which are most of the
sample --- the centralized starburst will naturally lead to metallicity
and color gradients if the young stellar populations are more metal rich
than the underlying old populations. Although violent mixing during the
merger could dilute any previously established metallicity gradients,
numerical simulations suggest that metallicity gradients might be
regenerated by centralized star formation \citep{Mihos&Hernquist94b,
Kobayashi04}.

To investigate the evolution of the color gradients of E+As
quantitatively, we take the current color gradients and evolve the
stellar populations using simple assumptions.  We assume that the
current color gradients are due solely to burst-strength gradients and
that the young populations are coeval, i.e., that the last starburst
was instantaneous. We adopt a uniform metallicity ($Z_{\rm old} =
0.004$) for the old population, and consider $Z_{\rm young} = 0.02$
and $0.05$ for the young population. Then, for each galaxy, we choose
a single post-burst age (the time since the starburst ended) that
produces the whole range of colors observed in that galaxy, and derive
the burst strength profile that reproduces the color profile.  As older
post-burst ages are selected, the burst strength gradient increases for a
given color, and therefore larger metallicity gradients are introduced.
We evolve these stellar populations passively for 5 and 10 Gyr ignoring
further dynamical evolution, and measure the resulting color gradients.

In Figure \ref{fig:color_grad_evol}, we show the color gradient evolution
for two E+As with positive gradients (EA03 and 10) and for two with
blue-cores (EA16 and 17). The solid and the dashed lines represent the
current color profiles and their subsequent evolution, respectively.
Taking into account  measurement errors, we show the range of color
profiles after 5 and 10 Gyr with the solid and hatched regions,
respectively.  We adopt $Z_{\rm young} = 0.02$ for EA03 and EA16 (left
panels), and $Z_{\rm young} = 0.05$ for EA10 and EA17 (right panels).

In these examples, the positive color gradients can invert within
$\sim$ 5 Gyr and evolve to a color gradient range [$-0.05< $ \dcdr\ $
<-0.16$] consistent with that of E/S0s.  Larger metallicity differences
between the old and young stellar populations and older post-burst ages
will result in steeper metallicity and color profiles after 5--10 Gyr.
While it is hard to predict the true color gradient evolution of the E+As
without exact knowledge of the post-burst ages and metallicities, our
calculation shows that the current color profiles, most likely arising
from a centralized starburst during a galaxy-galaxy tidal interaction
or merger, can evolve into those of early-type galaxies.


\begin{figure*}
\epsscale{0.9}
\plotone{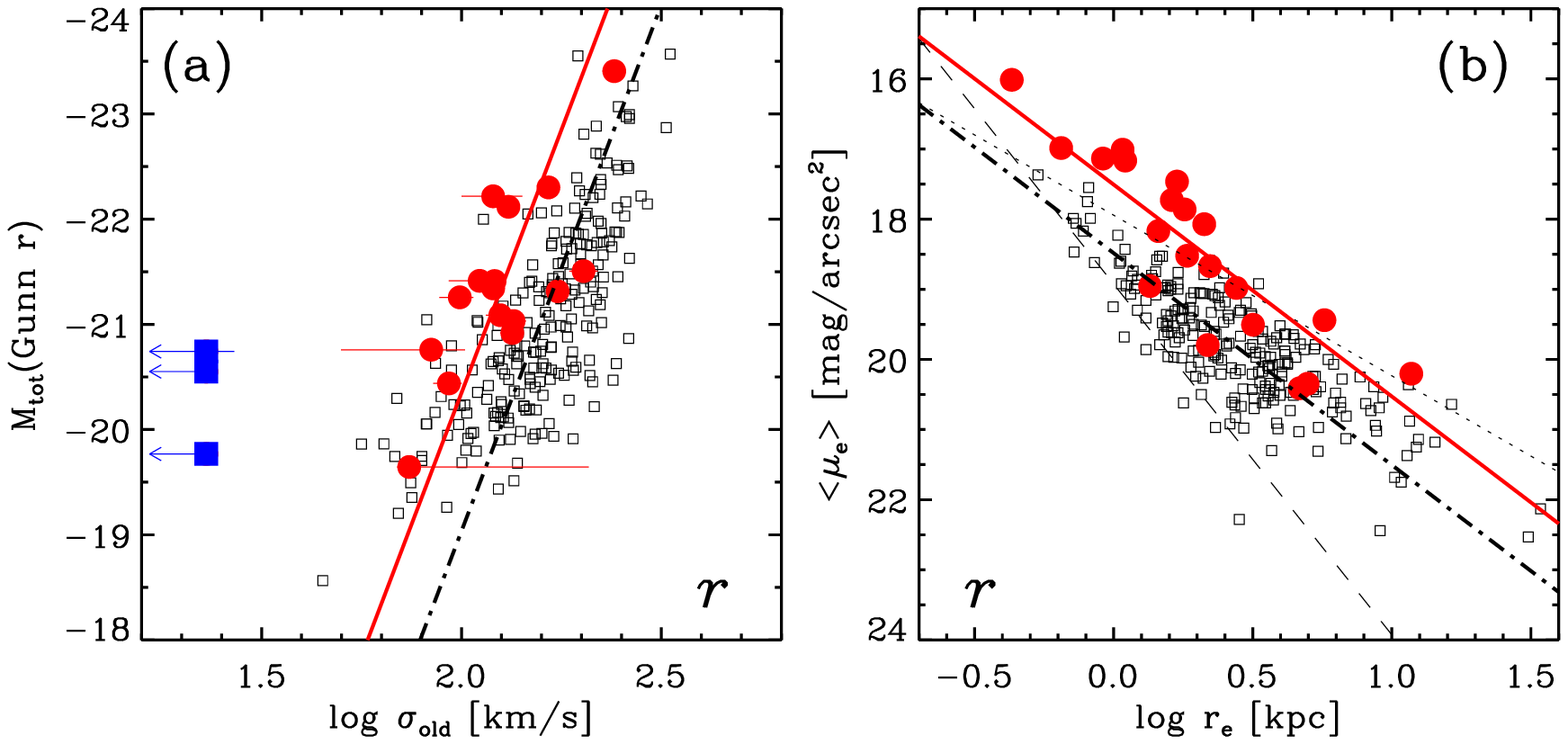}
\plotone{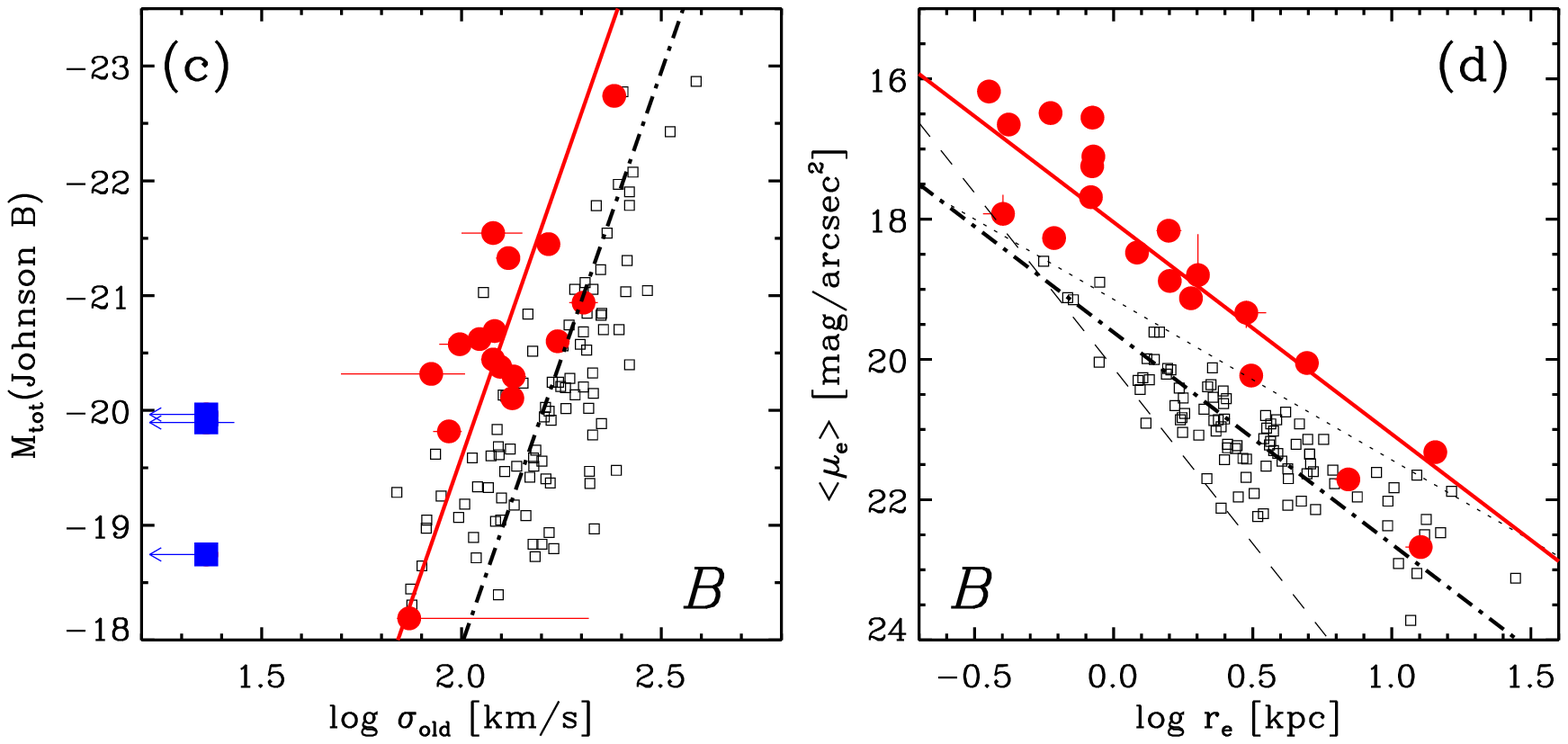}
\caption{
Faber-Jackson and Kormendy relations for E+A galaxies in the $B$
({\it bottom}) and \gunnr bands ({\it top}). The small squares and
large circles represent cluster E/S0s in the J96 sample and our E+As,
respectively. The large filled squares are the E+As  for which we have
only an upper limit on the velocity dispersion of their old stellar
population, $\sigma_{\rm old}$ from N01.  In the left panels, the solid
and dot-dashed lines represent the best-fit Faber-Jackson relations
($L \propto \sigma^4$) for E+As and E/S0s, respectively.  On average,
E+A galaxies are 1.31 ($\pm0.73$) and 1.64 ($\pm0.71$) mag brighter than
E/S0s in the \gunnr and $B$ bands.  In the right panels, the solid and
dot-dashed lines are the Kormendy relations with fixed slope for E+As
and E/S0s, respectively.  The average offsets between E+As and E/S0s are
$0.98 \pm 0.45$ and $1.57 \pm 0.52$ mag in $r$ and $B$, respectively. The
dashed line is the boundary set by the limiting magnitude.  The dotted
line is the sharp physical boundary that divides the plane into the
E/S0s and the so-called \emph{exclusion zone} where dynamically relaxed
systems are not allowed \citep{Bender92}.
\label{fig:FJR}}
\end{figure*}

\section{E/S0 Scaling Relationships and E+As}
\label{sec:fp}

In previous sections, we have shown that some aspects of E+A morphology
($B/T$, concentration $C$, S\'ersic index $n$) are consistent with
those of current early-type galaxies. We have also presented models or
arguments in which other aspects now inconsistent with current early-type
properties (blue cores, tidal features, and positive color gradients)
could, via evolution, become consistent with the properties of current
early-type galaxies.  To further investigate whether E+A galaxies
truly evolve into present-day early-types after a few Gyr, when the
discrepant morphological features might disappear, we test whether E+As
will eventually lie on the various scaling relationships of early-type
galaxies.  For example, \citet{Tacconi02} and \citet{Rothberg06} show that
late-stage ultra-luminous infrared galaxy (ULIRG) mergers and optically
selected merger remnants (other possible progenitors of present-day E/S0s)
are located very close to or on the {\sl K}-band Fundamental Plane (FP)
of early type galaxies.  We start with the simplest forms of the scaling
relations, i.e., the Faber-Jackson relation \cite[FJR;][]{Faber_Jackson}
and the Kormendy relation \cite[KR;][]{Kormendy77}.


\begin{figure*}
\epsscale{1.0}
\plotone{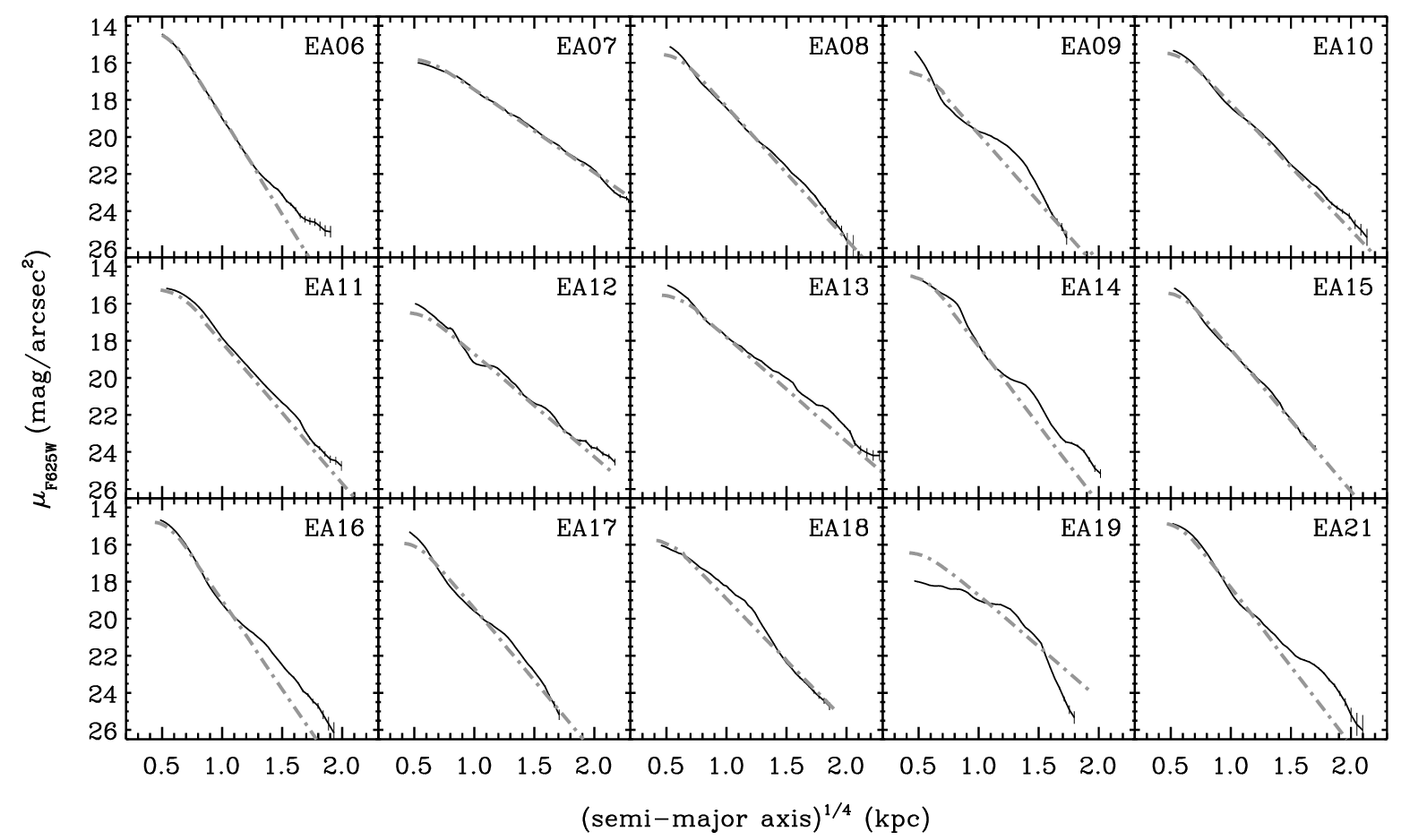}
\caption{
Light profiles of E+As with a single $r^{1/4}$-law fit (dark lines).
The thick dot-dashed lines are the best fits. The structural parameters
($r_e$ and $\mue$) from the single \devauc profiles fits are used to
construct the FP of E+A galaxies.
\label{fig:profile_devauc}}
\end{figure*}

\subsection{Faber-Jackson and Kormendy Relations}

\begin{figure*}
\epsscale{0.9}
\plotone{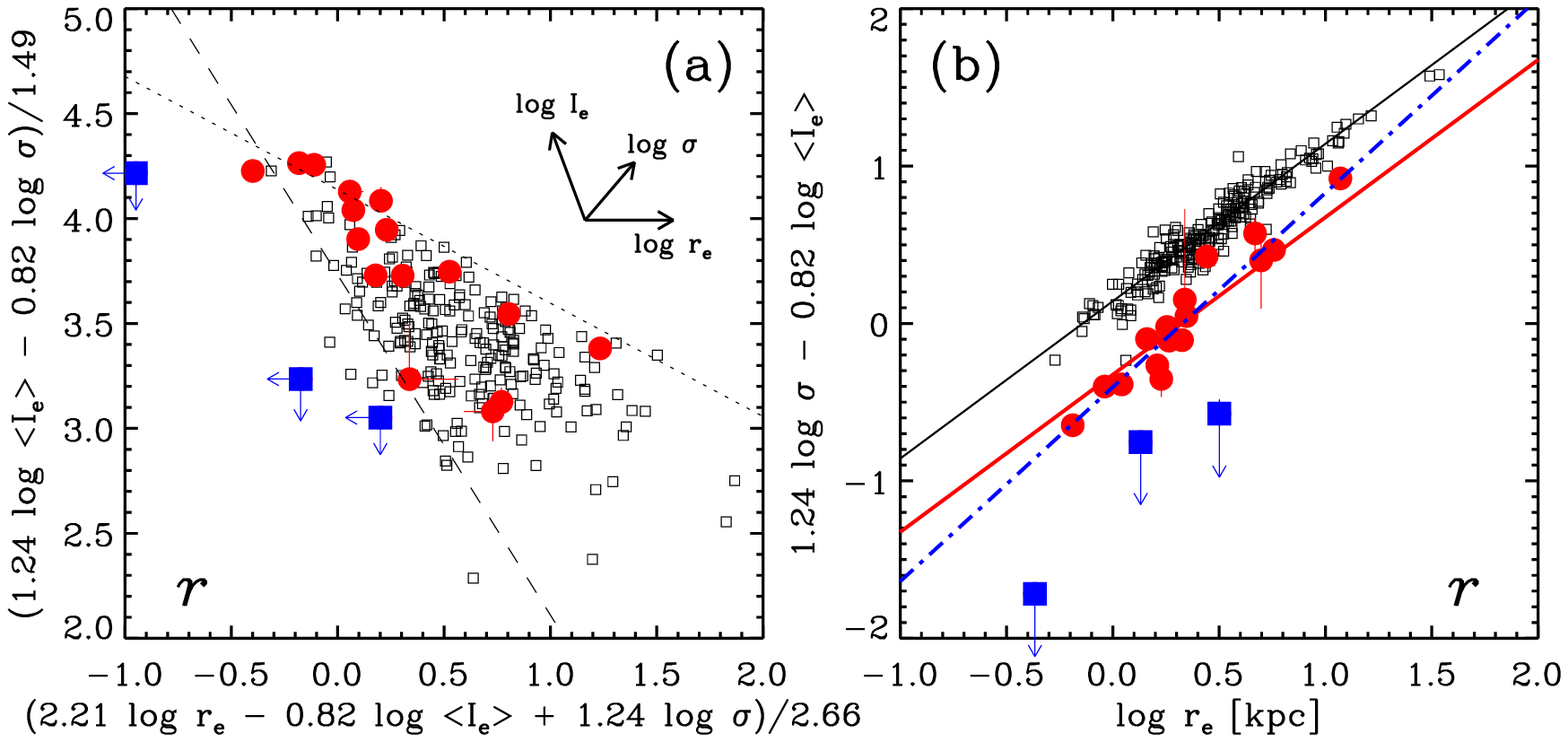}
\plotone{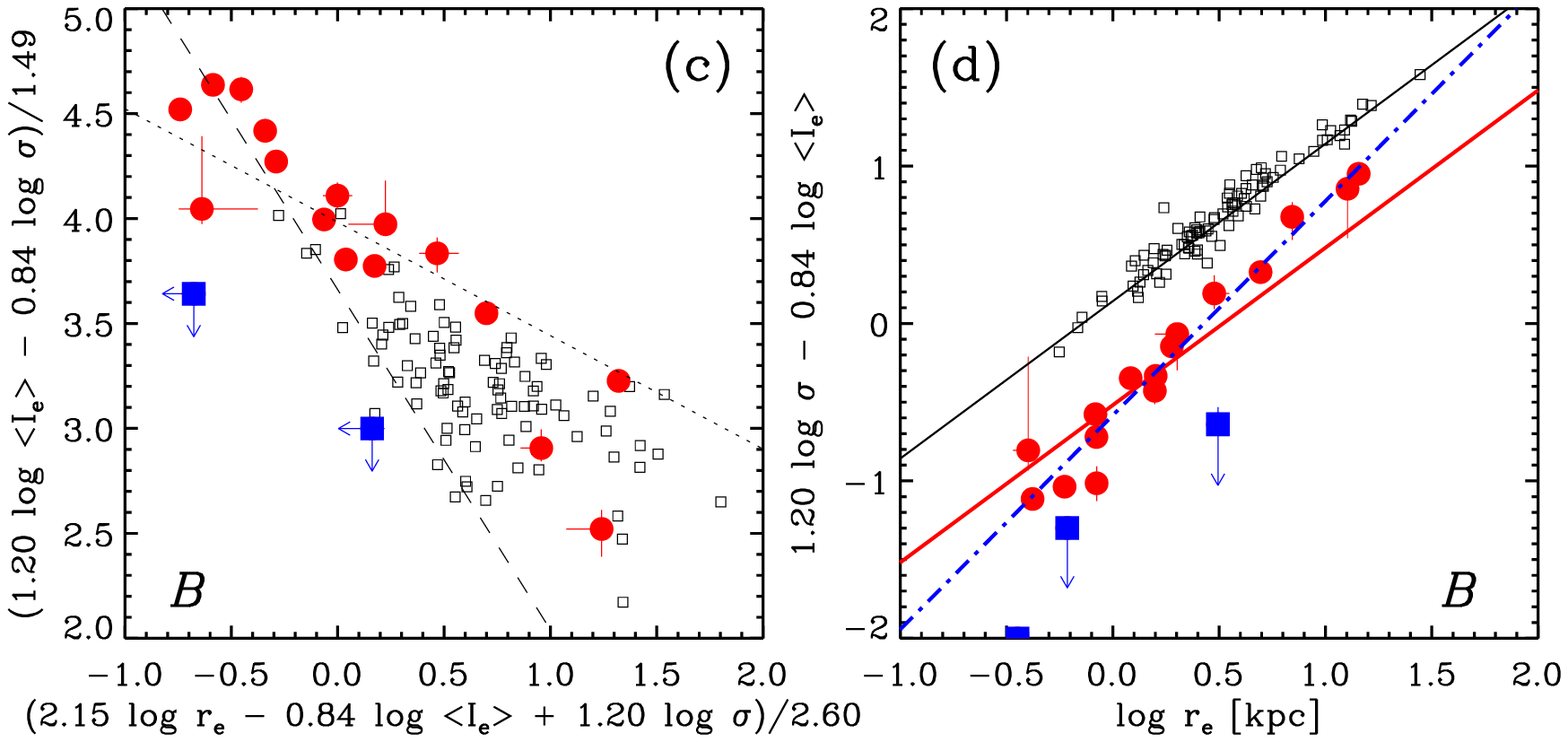}
\caption{
Face-on and edge-on projections of the fundamental plane in the $B$
({\it bottom}) and $r$ ({\it top}) bands.  The small squares and
large circles represent the E/S0s in the J96 sample and our E+As,
respectively. The large filled squares are the three E+As (EA06, EA17,
and EA18) that have only upper limits on $\sigma_{\rm old}$ from N01.
Of all the sample galaxies, these three E+As have the largest light
fraction contributed by the young A-type population ($f_{A}$ in N01).
($a$)($c$) Face-on projection of the FP.  The dashed line is the boundary
set by the limiting magnitude.  The dotted line is the sharp physical
boundary that divides the plane into the E/S0 FP and the so-called
\emph{exclusion zone} where dynamically relaxed systems are not allowed
\citep{Bender92}.
($b$)($d$) Edge-on view of the FP.  The solid and dot-dashed lines
represent the best-fit planes for E+As with and without fixed slopes
($\alpha$, $\beta$), respectively. E+As follow their own scaling relations
and have significant offsets from the E/S0s.
\label{fig:FP}}
\end{figure*}

In Figure \ref{fig:FJR}, we show the Faber-Jackson relation ($L-\sigma^4$)
and the Kormendy relation ($r_e - \mue$) of E/S0s in comparison to those
of E+As. We draw the E/S0 comparison sample from \citet{Jorgensen96},
because their photometric bands (\gunnr and $B$ band) closely match our
{{\sl HST}} {\sl WFPC2} and {\sl WFC} filters.  For the E+As, we adopt
the velocity dispersions of the old components ($\sigma_{\rm old}$)
from \citet{Norton01}.  To disentangle the kinematics of the young and
old stellar populations, \citet{Norton01} simultaneously fit the velocity
profile and the relative contributions of different stellar components
using longslit (1.5\arcsec $\times$\, 6.6\arcsec aperture) spectra.
For three E+As (EA06, EA17, and EA18), we have only upper limits on
$\sigma_{\rm old}$, because a large fraction of the light is contributed
by a young A-type population ($f_{A}$ in N01) and thus the velocity
dispersion of the old K-type population is not measured reliably. Although
we include these galaxies in Figures \ref{fig:FJR} and \ref{fig:FP},
we exclude them from the following analysis.
For the structural parameters, we adopt $r_e$, $\mue$ and
the total magnitude $M_{\rm tot}$ from our \devauc fit (Figure
\ref{fig:profile_devauc}).  While a single \devauc profile is not an ideal
fit/model for complex morphologies (e.g., EA02) and/or for disky systems
(EA09 and EA17), we justify the choice of $r_e$ from the $r^{1/4}$ fits
as a simple measurement of the half light radius of the galaxy. Indeed,
we find that the luminosity within $r_e$ given by the $r^{1/4}$ GALFIT
fit agrees with a circular aperture measurement of the half light radius
to within 10--20\%.

We reproduce \citet{Norton01}'s result that there is a strong correlation
between E+A magnitude and velocity dispersion, and find that it is
roughly parallel to the E/S0 Faber-Jackson relation with some offset
(left panels in Figure \ref{fig:FJR}).  This relation implies that E+A
galaxies already harbor dynamically relaxed old populations and that they
are pressure-supported systems.  While \citet{Norton01} had to perform
several steps to transform the LCRS Gunn/Kron-Cousins magnitude into
\gunnr magnitude to compare the properties of E+As with those of E/S0s,
we determine more directly the relative offsets between the E+As and
E/S0s with our {\sl HST} photometry.  We measure the offset by fitting
the $L\propto\sigma^4$ (i.e., $M_{\rm tot}\propto10\log \sigma$) relation
to the E+As and the J96 sample.

On average, E+A galaxies are 1.31 ($\pm0.73$) and 1.64 ($\pm0.71$)
mag brighter than the E/S0s in the \gunnr and $B$ bands, respectively.
This offset in \gunnr is almost twice as large ($\sim$ 0.6 mag) as that
derived by \citet{Norton01}. This difference arises not only because we
use improved photometry and magnitude transformations, but also because
the functional forms used to calculate the offset are different.


%
%

These offsets can be interpreted as either that E+As have an elevated
luminosity for a given velocity dispersion or that they have an unusually
low velocity dispersion at a given luminosity.  To investigate which
interpretation is more correct, we compare E+As to E/S0s on the purely
photometric scaling relation, i.e., the \citet{Kormendy77} relation
(hereafter KR), between the effective radius ($r_e$) and the mean surface
brightness within $r_e$.  The right panels in Figure \ref{fig:FJR}
show the KRs for the E+As and E/S0s, the latter of which is corrected
from the original KR \citet{Kormendy77} by assuming an average color,
$B-r$ $\simeq$ 1.20 \citep{Fukugita95}.  The slope of the relation is
known to be insensitive to the band.  For example, \citet{Hamabe87}
find a slope of 2.98 in the $V$ band compared to 3.02 in the $B$ band.

In the KR plane, most of E+As, except for a few disky ones, have a higher
surface brightness than E/S0s at a given effective radius.  The average
offset between E+As and E/S0s is 0.98 $\pm 0.45$ and 1.57 $\pm 0.52$
mag in $r$ and $B$, respectively.  The offsets appear to increase as the
galaxies become smaller; roughly half of the E+As do not overlap with the
$B$ and $r$ band E/S0 loci.  The lower boundary (dashed line) of the E/S0
region is set by the limiting magnitude of $M_r \simeq -20.45$. The upper
boundary is not due to selection effects (see \S\ref{sec:fp_location}).
Because the surface brightness offsets in the KR are consistent with
the magnitude offsets obtained from the FJR, we conclude that elevated
luminosities, probably due to the recent star formation, are responsible
for the offsets of the E+A scaling relations, rather than unusually low
velocity dispersions.  While the FJR and KR suggest that roughly one
magnitude of fading may bring E+As down to the E/S0 scaling relations,
the large scatter makes such an analysis difficult.

\subsection{Fundamental Plane}

The fundamental plane, hereafter FP, is an empirical scaling relation
between the effective (or half-light) radius $r_e$, the central velocity
dispersion $\sigma$, and the mean surface brightness $\Ie$ within $r_e$
for early type galaxies \citep{Djorgovski87, Dressler87}.  Although the
physical origin of the FP is not fully understood, both cluster and
field early-type galaxies follow this relation with remarkably small
scatter (e.g., $\sim$ 0.1 dex in $\Log r_e$ in the \gunnr band; J96),
and an extension of this formalism appears to fit all spheroids ranging
from Galactic dSph galaxies to the intracluster stellar component of
galaxy clusters \citep{ZGZ}.

To examine the FP of E+As and compare it with that of E/S0s, we adopt
the definition and methodologies of \citet{Jorgensen96} \citep[see
also][]{Kelson00c}. The fundamental plane is defined as
\begin{equation}
\label{eq:fp_def}
\Log r_e = \alpha\, \Log \sigma + \beta\, \Log \Ie + \gamma \,,
\end{equation} 
where $r_e$, $\sigma$, and $\Ie$ are the effective or half-light radius
in kpc, central velocity dispersion in km s$^{-1}$, and mean
surface brightness within $r_e$ in units of $L_\sun {\rm pc^{-2}}$.
For the E+A structural parameters, we adopt $r_e$ and $\lan \mu_e
\ran$ from the single \devauc fits to our {\sl HST} images (Figure
\ref{fig:profile_devauc}).  
We transform the mean surface brightness in our {\sl HST} bands into 
\gunnr and Johnson $B$ using our transformation
relations (eq.\ref{eq:transform}) and the coefficients in Table
\ref{tab:transform}.  We convert to units of $L_\sun {\rm pc^{-2}}$ using
the relation $\Ie = -0.4 (\mue - {\rm constant})$, where the constants
are 26.4 and 27.0 for \gunnr and Johnson $B$ magnitudes, respectively.
We do not apply a color-dependent correction to $r_e$, because
the E/S0 FP is only weakly sensitive to color \citep{Pahre98,Bernardi03}.


\subsection{Location of E+As on the FP}
\label{sec:fp_location}

We show the location of our E+As on various projections of the J96 FP in
the \gunnr and $B$ bands in Figure \ref{fig:FP}.  Figures \ref{fig:FP}$a$
and \ref{fig:FP}$c$ present the face-on view of the E/S0 FP in $r$ and
$B$, respectively.  The dashed line is the boundary set by the limiting
magnitude of $M_r \simeq -20.45$, which corresponds to the constant
magnitude line (dashed line) in \ref{fig:FJR}$b$ and \ref{fig:FJR}$d$.
The boundary set by the dotted line divides the plane into the E/S0 FP and
the so-called \emph{zone of exclusion} (ZOE), where dynamically relaxed
systems do not reside \citep{Bender92}.  The physical origin of this
boundary is not fully understood, although it might correspond to the
upper limit of the velocity dispersion function of early-type galaxies
\citep[$\sigma\lesssim350$ km s$^{-1}$;][]{Treu06}.  In the face-on
projection, E+As span the same region defined by the normal early-type
galaxies in the $r$ band; however, half of our E+As violate the zone of
exclusion in the $B$ band.  We plot vectors parallel to the FP parameters
to find out which parameters drive this shift in the $r$ band. 
E+As are shifted from the E/S0s along the direction of increasing
$\Ie$, as we concluded earlier from the Kormendy relation. Given that
E+As already populate the same region as E/S0s in the $r$ band and that
the $\Ie$ offset is larger in $B$, this violation of the ZOE must be
due to the temporary increase of the surface brightness caused by the
recent starburst.


Figures \ref{fig:FP}$b$ and \ref{fig:FP}$d$ present the edge-on projection
of the E/S0 FP, defined as $x = \Log r_e$ and $y = \alpha \Log \sigma
+ \beta \Log \Ie$, in $r$ and $B$, respectively. In this projection,
E+As are distinct from E/S0s.  The existence of a tight correlation
among E+As indicates that they follow their own scaling relation. The
large offset between the E/S0s and E+As suggests that the E+As' current
stellar content is different.  Although the mechanisms that trigger the
E+A phase and stop star formation are not fully understood, any model
must be able to explain the relative tilt and offset of the E+A FP.
In the following section, we discuss how the observable properties
are related to physical ones, such as mass-to-light ratio {\sl M/L},
and possible interpretations of the tilt and offset.

\subsubsection{FP Offsets: Evolution of the FP Zero Points}

E+A galaxies stand apart from the E/S0 FP in the edge-on projection
(Figures \ref{fig:FP}$b$ and \ref{fig:FP}$d$).  The difficulty in
interpreting this difference is that any, or all, of the three plotted
parameters or the implicit parameter, {\sl M/L}, may be evolving as the
E+A evolves.  The simplest option is to assume that only \ml evolves. In
this case, the offset between the two populations can be directly related
to a difference in \ml,
  \begin{equation}
  \label{eq:fp_ml}
  \Delta \Log \frac{M}{L} = \frac{\Delta \gamma}{\beta},
  \end{equation}
where the FP offset $\Delta \gamma$ is measured from the difference
between the intercepts of two parallel lines in Figures \ref{fig:FP}$b$
and \ref{fig:FP}$d$.
We find $\Delta \gamma = -0.47 \pm 0.12$ in the Gunn $r$ band, which
implies $\Delta \Log (M/L)$ = $-0.57 \pm 0.14$.  In other words, E+As
have, on average, a \ml that is 3.8 times smaller than that
of E/S0s.  Note that this average offset has a smaller error than
the offsets measured from the FJR and KR.  However, as we have shown
previously in \S\ref{sec:color_profile}, the structural properties are
affected by the E+A phase (the color gradients are altered, some E+As have
blue cores) and so assuming that only \ml evolves cannot be completely
correct. Our expectation is that $r_e$ will be smaller in E+As because
of the centrally concentrated star formation, and hence that $r_e$ will
increase (and $\langle I_e \rangle$ decrease) as the E+A evolves, leading
to a complex evolution on the FP.  Therefore, the \ml difference discussed
above is the minimum amount of fading that E+As should experience to
settle onto the E/S0 locus.  By examining the tilt of the FP, and other
projections of the scaling laws, we explore whether we can proceed beyond
this simple model and the potential complexity of evolution on the FP.

\subsubsection{FP Tilts: Variation of \ml along the FP}
\label{sec:fp_tilt}

Given the range of possible evolutionary paths on the FP, it is rather
surprising that there is a tight relation for E+As on the E/S0 FP edge-on
projection.  Furthermore, the difference appears to be both a relative
shift and a tilt. To measure the relative tilt of E+As with respect to
E/S0s, we find the best-fit plane in the form of equation \ref{eq:fp_def}
through the E+As using an orthogonal fitting method that minimizes the
normal distances to the plane. We adopt this fitting method to enable
direct comparison with the fits from J96.
J96 find
$\alpha = 1.24 $$\pm 0.07$, $\beta=-0.82 $$\pm 0.02$ for \gunnr and 
$\alpha = 1.20 $$\pm 0.06$, $\beta=-0.83 $$\pm 0.02$ for Johnson $B$ 
using 226 and 91 cluster galaxies, respectively.  Using 16 E+A
galaxies\footnote{This excludes EA01AB, 07, 17 and 18 which do not have
a measured $\sigma_{\rm old}$.} and taking an equal weight for each galaxy,
we find that the best-fit FPs are
\begin{eqnarray}
{r_e \propto \sigma^{1.13\pm0.10} \Ie^{-0.62\pm0.07} 10^{-0.08\pm0.10}}\\
{r_e \propto \sigma^{1.09\pm0.08} \Ie^{-0.59\pm0.06} 10^{-0.11\pm0.07}}
\end{eqnarray}
for the \gunnr and Johnson $B$ bands, respectively. The errors in $\alpha$,
$\beta$, and $\gamma$ are estimated using the bootstrap method.
The rms scatter around these planes is 0.10 and 0.10 in $\Log r_e$ for
\gunnr and Johnson $B$, respectively. The parameters ($\alpha$, $\beta$,
$\gamma$) for \gunnr and Johnson $B$ agree within the uncertainties,
because the FP parameters are generally not highly sensitive to
the bandpass (J96). Therefore, we adopt the ($\alpha$, $\beta$,
$\gamma$) derived using the \gunnr data in the following discussion.
While $\alpha=1.13$ for the E+A FP agrees to within the errors with that
of the E/S0 FP ($\alpha = 1.24$) (the E+A FP is still viewed edge-on in
the projection in Figure \ref{fig:FP}$b$), $\beta = -0.62$ for the E+As
is $\sim 3\sigma$ discrepant from that of the E/S0s ($\beta = -0.82$).
In the context of interpreting differences between the FPs as primarily
due to {\sl M/L}, the tilt implies a mass dependence.  A similar tilt of the
FP of merger remnants is observed relative to the {\sl K}-band E/S0 FP
\cite[Fig.\,1;][]{Rothberg06}.

The tight correlation within the E+A FP suggests that \ml is correlated
with the structural parameters.  
The scatter around the plane appears to be comparable to the typical
scatter of E/S0 FPs ($\sim$ 0.1 dex).  Does this mean that E+As, many
of which are clear merger remnants, settle into a relaxed state on a
short timescale? Or is it possible that the correlation is the result
of a selection effect?  Our selection of galaxies with strong H$\delta$
absorption tends to choose galaxies with stronger bursts and/or the
younger ages; therefore it is possible that our sample represents only
a certain part of the FP, i.e., near the exclusion zone.  To address
this issue, a larger sample of E+As covering a wider range of H$\delta$
strengths is required.

Using the best fit FP parameters, \ml for E+As can be expressed as
  \begin{equation}
  {M/L} \propto M^{0.1} r_e^{0.4},
  \end{equation} 
compared with $M/L \propto M^{0.24} r_e^{0.02}$ for E/S0s.  Unlike in
the E/S0 FP, where \ml is a function mostly of mass, \ml for E+As
depends on the effective radius as well as the mass of the system.
E+As with smaller effective radii have smaller {\sl M/L}, possibly
resulting from a stronger burst and/or shorter time elapsed since the
starburst.  Because \ml is proportional to $(r_e \Ie)^{-1}$ at a fixed
velocity dispersion, one might expect that a smaller $r_e$ would lead to
a larger {\sl M/L}, but the observed trend is the opposite.  Therefore,
the surface brightness of E+As must increase faster than the mass or size
of the galaxy decreases. This inference is consistent with the observed KR
(Figure \ref{fig:FJR}), in which $\mue$ deviates further from the E/S0
relation for E+As with small effective radii.


\begin{figure}
\epsscale{1.20}
\plotone{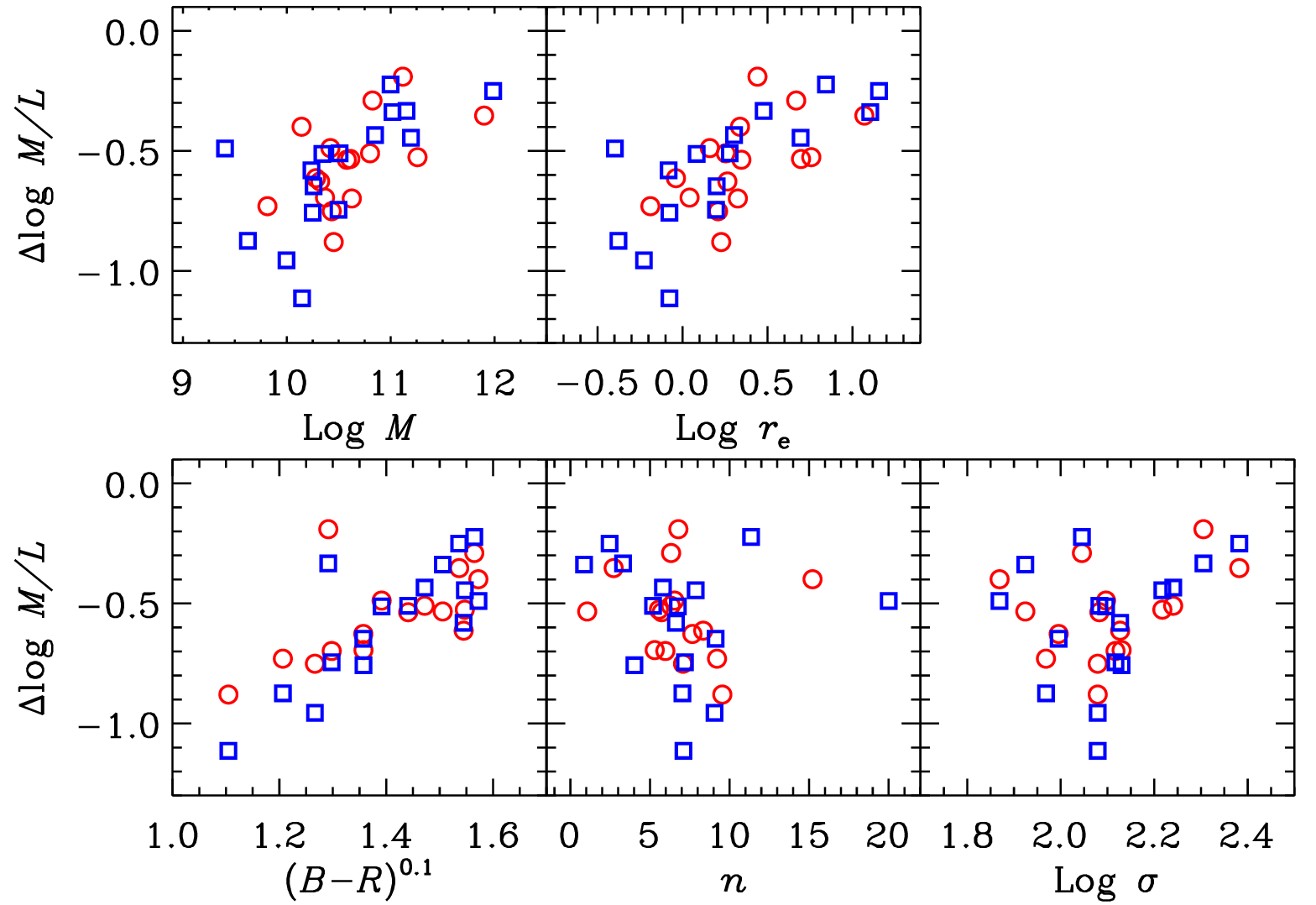}
\caption{
Correlations between the deviation of \ml from E/S0s for individual E+As
and the ($B-R$) color, dynamical mass, S\'ersic index $n$, velocity
dispersion of the old stellar population $\sigma_{\rm old}$, and the
effective radius $r_e$.  The squares and circles represent parameters from
the $B$ and $R$ bands, respectively.  $\Delta$\ml is tightly correlated
with ($B-R$) color, but not with $n$ or $\sigma$. We conclude that
the tilt, i.e. the variation of \ml along the E+A FP, is mostly due to
stellar population variations.  E+As with smaller effective radii have
younger stellar populations, on average, and therefore smaller {\sl M/L}.
\label{fig:fp_corr}}
\end{figure}

If \ml does indeed vary among E+As, then one might expect a correlation
between the deviation of \ml from E/S0s and the color of the galaxy.
In Figure \ref{fig:fp_corr}, we plot $\Delta$\ml against ($B-R$) color,
and against other parameters (mass, $r_e$, S\'ersic $n$, and $\sigma_{\rm
old}$).  We find a strong correlation between $\Delta$\ml and color,
but no significant correlation with $n$ or $\sigma$.  A Spearman rank
test shows that only ($B-R$) color correlates with $\Delta$\ml at a
significance level higher than 99.5\% for both the $r$ and $B$ bands. We
conclude that the tilt, i.e., the variation of \ml within the E+A FP,
does reflect a real variation of the stellar population.  E+As with
smaller effective radii have younger stellar populations, on average,
and therefore, smaller $M/L$.


The remaining question is what does the small $r_e$ physically represent?
Does it indicate that the intrinsic size of the galaxy is small or that
there was a strongly concentrated starburst?  We compare the effective
radii in $B$ and $r$ to address these questions.  In blue-core E+As
or E+As with the positive color gradients ($\sim$ 75\% of E+As), the
effective radius tends to be smaller in the $B$ band ($r_{e\mathrm{B}}$)
than in the $r$ band ($r_{e\mathrm{R}}$).  In these galaxies, it is likely
that $r_{e\mathrm{B}}$ will eventually become close to $r_{e\mathrm{R}}$
as the light from the young stars in the central region fades and
the color gradients flatten. However, most galaxies have fairly small
differences, $|\Log r_{e\mathrm{B}} - \Log r_{e\mathrm{R}}|$ $<$ 0.30,
which is clearly not enough to produce the tilt observed over two orders
of magnitude in effective radius.


To further test whether intrinsically small galaxies have smaller {\sl
M/L}'s, we repeat the orthogonal fitting, but this time exclude galaxies
with $|\Log r_{e\mathrm{B}} - \Log r_{e\mathrm{R}}|$ $>$ 0.15, i.e.,
for which the two effective radii differ by more than 25\%.\footnote{
We boost the statistics of this test by combining the $B$ and $R$
band sample assuming the same FPs parameters for both bands (see
\S\ref{sec:fp_tilt}).} For this subsample, the color gradients are
moderately flat and there is little dust, so strong evolution in the
effective radius due to stellar evolution is unlikely.  The fitting
of this subsample results in the same FP slope as for the full sample,
implying that intrinsically small and therefore less massive galaxies
have smaller \ml ratios.

Blue-core galaxies analogous to our blue core E+As (\S4.2) are found
among early-type galaxies at high-$z$ \citep{vanDokkum03,Treu05}, and
their rest-frame {\sl B}-band {\sl M/L} is estimated to be much smaller
(roughly $4 \times$) than that of normal E/S0 galaxies at the same epoch.
As a cautionary note, we examine how the difference between the blue
and red $r_e$ affects the measurement of \ml, especially for blue core
galaxies. When we use the blue $r_e$ to derive $M/L$ for our blue core
E+As, we find that \ml can be biased low by a factor of 1.25 to 5.5
relative to that derived using the red $r_e$.


\subsubsection{Will E+As Evolve into E/S0s?}

In summary, the scaling relations of E+As demonstrate that their
stellar populations are currently different from those of E/S0s.
This conclusion is drawn from the FP parameters ($\sigma$, $r_e$ and
$\mu_e$) in one photometric band, and therefore is derived independently
from our spectroscopic knowledge that E+A galaxies have significant A-type
populations.  The surprising finding is that among E+As, the variation of
stellar populations is closely tied to the structural parameters, i.e.,
E+As follow their own scaling relationships such that smaller or less
massive galaxies have a smaller {\sl M/L}.  Such a trend arises naturally
within a merger scenario, where low mass galaxies (the progenitors of
low-mass E+As) have higher gas fractions \citep{Young&Scoville91} and
could produce relatively larger populations of young stars.


Will E+As fade onto the fundamental plane of E/S0 galaxies after a
few Gyr?  This question is difficult to answer fully due to the complex
interplay between galaxy's dynamical evolution and the evolution of its
stellar populations.  However, if we ignore the dynamical evolution, we
can estimate the amount of fading using a stellar population synthesis
model \cite[][BC03]{BC03}.  We adopt a simple star formation history
for the E+As: a single instantaneous starburst on top of an underlying
10 Gyr old single burst population. Both populations have Salpeter IMFs
and solar metallicities.

\begin{figure}
\epsscale{1.0}
\plotone{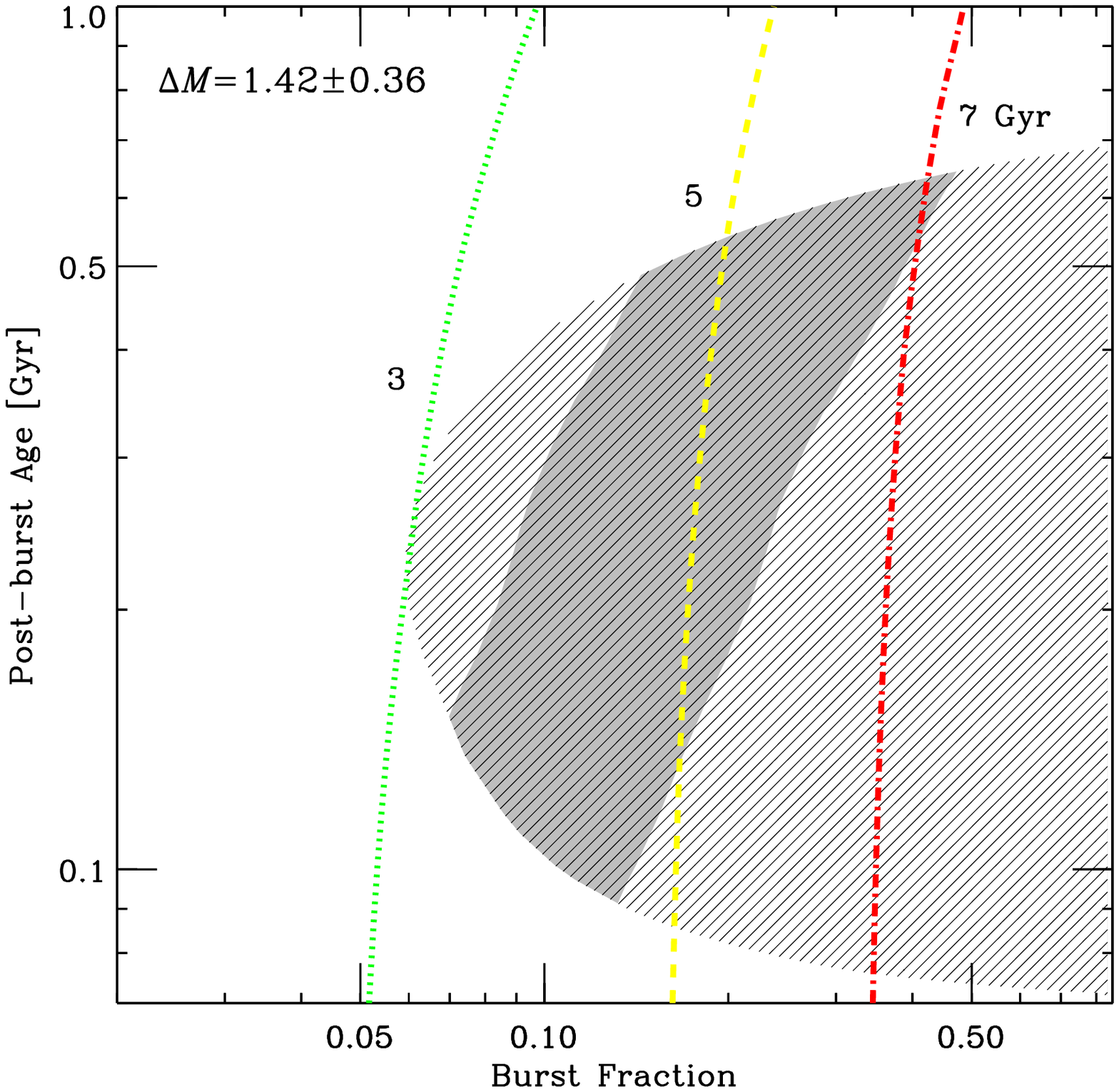}
\caption{
Time required for E+As to fade until their $(B-R)$ colors agree with that
of 10 Gyr old simple stellar population as a function of  burst fraction
and post-burst age.  From left to right, the contours represent fading
times of 3, 5 and 7 Gyr, respectively.  The lined region represents
galaxies that would be classified as E+As according to the Zabludoff
et al. (1996) selection criteria (average Balmer equivalent width $\lan
H\beta\gamma\delta \ran$ $>$ 5.5\AA).  The shaded region shows galaxies
that will fade by their observed offsets from the E/S0 fundamental plane
($\Delta M_r = 1.42 \pm 0.36$ mag) for the given burst fraction and
post-burst age.  This fading time lies between 3 and 7 Gyr, indicating
that E+As with reasonable post-burst ages (100 -- 700 Myr) and burst
fractions (7 -- 50\%) can evolve into E/S0s within $\sim$ 5 Gyr.
\label{fig:fading}}
\end{figure}

The largest uncertainties in determining the amount of fading and the
time required for settling onto the E/S0 FP are the unknown post-burst age
(the time elapsed since the burst) and burst strength (the fraction of the
stellar mass produced during the starburst).  Because we are unable to
break the degeneracy between post-burst age and burst-strength for this
sample (see, however, Yang et al. 2008, in prep.), we test whether the
required fading time is reasonable for the galaxies whose post-burst ages
and burst strengths are consistent with the FP offset (1.42 $\pm$ 0.36 mag
in Gunn $r$).  For various post-burst ages and burst fractions in Figure
\ref{fig:fading}, we calculate how much the galaxies should fade and
how long it will take from the given post-burst age until their ($B-R$)
colors agree with that of a 10 Gyr old simple stellar population, i.e.,
until they become E/S0s.  The lined region in Figure \ref{fig:fading}
shows the post-burst ages and burst fractions that would satisfy the E+A
selection criteria adopted by Zabludoff et al. (average Balmer equivalent
width $\lan H\beta\gamma\delta \ran$ $>$ 5.5\AA).  The shaded region
represents those galaxies that will fade by $\Delta M_r = $ 1.42 ($\pm
0.36$) mag until their ($B-R$) colors match the 10 Gyr old population
i.e., those E+As whose {\sl M/L}'s are consistent with the observed
FP offsets.  This shaded region illustrates the degeneracy between the
post-burst ages and burst strength: old/strong and young/weak bursts
both explain the observed FP offsets.  We also show the required fading
time (3, 5, and 7 Gyr) with contours.  The shaded region falls between 3
and 7 Gyr, indicating that E+A galaxies with reasonable post-burst ages
(100 -- 700 Myr) and burst fractions (7 -- 50\%) will settle onto the
E/S0 FP after $\sim$ 5 Gyr.

\section{Young Star Clusters}
\label{sec:cluster}

The properties of star cluster systems in early type galaxies, e.g.,
the color bimodality, provide important clues for how the host galaxies
formed and how they assembled their stellar content over time \cite[see
the recent review by ][]{Brodie06}.  If many or all E+A galaxies result
from galaxy-galaxy close interactions or mergers, as suggested here and
in previous work \citep{Zabludoff96, Yang04, Yang06, Blake04, Tran03,
Tran04, Goto05}, and evolve into E/S0s, as we argue in this paper, the
photometric properties of their cluster systems should be consistent, to
within evolutionary corrections, with those of early types. In the WFPC2
sample, we discovered young star clusters around four E+As\footnote{EA01A,
02, 03, 04} \citep{Yang04}.  These clusters are much brighter than
Galactic globular clusters and have blue colors consistent with ages
estimated roughly from the E+A galaxy spectra ($\lesssim$ Gyr).
Our new ACS observations of the remaining LCRS E+As also reveal a number
of point-like sources surrounding the galaxies. In this section, we
first present the evidence that these are newly-formed star clusters,
then investigate whether their colors and luminosities are consistent
with being created concurrently with the burst that generated the young
stars in the E+As, and finally test whether these cluster systems can
evolve into the globular cluster systems of present-day E/S0s.

\begin{figure}
\epsscale{1.2}
\plotone{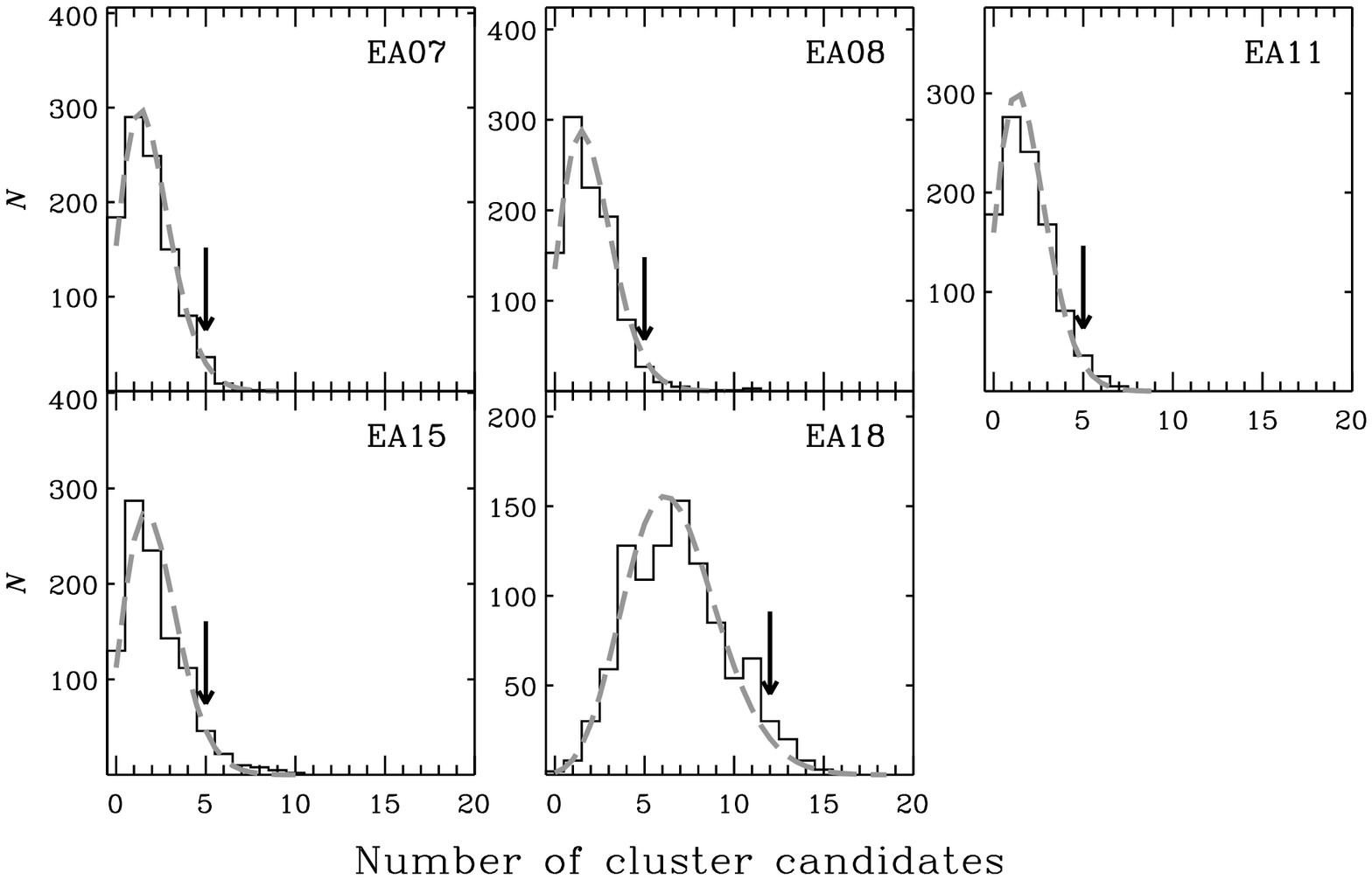}
\caption{
Distribution of the number of point sources in random beams sampling
the {\sl ACS} fields of the five E+As (EA07, 08, 11, 15, 18) with a
significant ($>90$\% confidence level) excess of surrounding sources.
In each field, the beam size corresponds to a 20 kpc radius at the
E+A's redshift.  In each panel, arrows represent the number of cluster
candidates within this radius centered on the E+A. The overlayed dashed
line is the Poisson distribution.
\label{fig:beam}}
\end{figure}

\subsection{Detection and Photometry of Cluster Candidates}

To detect cluster candidates, we use SExtractor \citep{Bertin&Arnouts96}
on both the original and model-subtracted residual $R$ band images (Fig.\,
\ref{fig:images_acs}). After testing various selection criteria and
visually inspecting the detected point sources, we require that cluster
candidates have at least 4 adjacent pixels whose flux in each pixel
is at least $3\sigma_{\rm sky}$ larger than the local sky value, where
$\sigma_{\rm sky}$ is the root mean square (rms) of the local sky values
estimated using a background mesh size of 4--8 pixels in SExtractor.
We use the residual images to find compact sources within the galaxies,
where the rapidly varying galaxy surface brightness prevents the reliable
detection of sources in the original images.  We choose to lower the
detection threshold to 2.5 $\sigma_{\rm sky}$ for the residual images,
because $\sigma_{\rm sky}$ is inflated by the residual galaxy light.
From this initial source list, objects with an apparent magnitude $m_R <
22$, elongation $\varepsilon > 1.7$, and isophotal area $A>36$ pixels
are rejected as possible foreground stars or background galaxies.

Because clusters are expected to be unresolved at the distance of our
E+A galaxies ($\sim$ 100 pc per pixel at $z\sim0.1$), contamination
from background galaxies, foreground stars, hot pixels, and residual
cosmic-rays could be problematic.  Therefore, we run a statistical test
to determine whether there is an excess of compact sources near each E+A.
First, we build a control sample of cluster-like sources from the entire
{\sl ACS} field ($\sim$200\arcsec$\times$200\arcsec) outside of a 20
kpc radius from the E+A using the same selection criteria.  Then we
calculate the distribution of the number of the sources within circular
beams of 20 kpc projected radius at random positions in the {\sl ACS}
field outside a 20 kpc radius from the E+As.  If there are significantly
more cluster candidates within the 20 kpc beam centered on the E+A than in
the random beams, then it is likely that these are true compact objects
associated with the E+A. We test various beam sizes, e.g., fixed angular
size versus physical size, and find that our conclusions are insensitive
to this choice.

Five E+As (EA07, 08, 11, 15, 18) have an excess of associated
point sources at a confidence level greater than 90\%.  We show
the number of cluster candidates found near the E+As relative to
the distribution found within random beams in Figure \ref{fig:beam}.
Because of the post-starburst nature of E+As and the merger signatures
clearly seen in EA07, 11, and 18, it is likely that these sources
are the newly formed star clusters often found in on-going mergers
\cite[The Antennae;][]{Whitmore_Schweizer95} and merger remnants
\cite[NGC7272;][]{Whitmore93}.

To test whether the colors and luminosities of these cluster candidates
are consistent with a merger origin and with evolution into the
globular cluster populations of E/S0s, we measure aperture magnitudes
in the $B$ and $R$ band images. Because many cluster candidates are
located within the galaxy, where there is a rapidly varying effective
background, obtaining reliable photometry is difficult.  We first
subtract the background galaxy light by fitting 7th order polynomials
to the background, excluding the cluster itself, within a 11$\times$11
pixel box centered on the point source.  We then add a flat background
that has the same mean as the fitted background to preserve the noise
properties. Finally, we measure magnitudes within a 2 pixel radius.
Because the uncertainties in the measured magnitudes and colors transfer
directly to those in the derived ages, we estimate photometric errors
using artificial cluster tests.  We place one thousand model star clusters
with known brightness into the relevant parts of the images and repeat
our procedure.  We find that if the model cluster is detected, the
magnitude error due to the background subtraction is nearly insensitive
to the underlying average sky value and depends mostly on the brightness
of the cluster.  Typically, the magnitude uncertainty is $\sim$0.25 mag
down to $m_R$ = 26.5.


\subsection{Luminosity Functions and Ages of the Star Clusters}
\label{sec:cluster_age_lf}

\begin{figure}
\epsscale{1.15}
\plotone{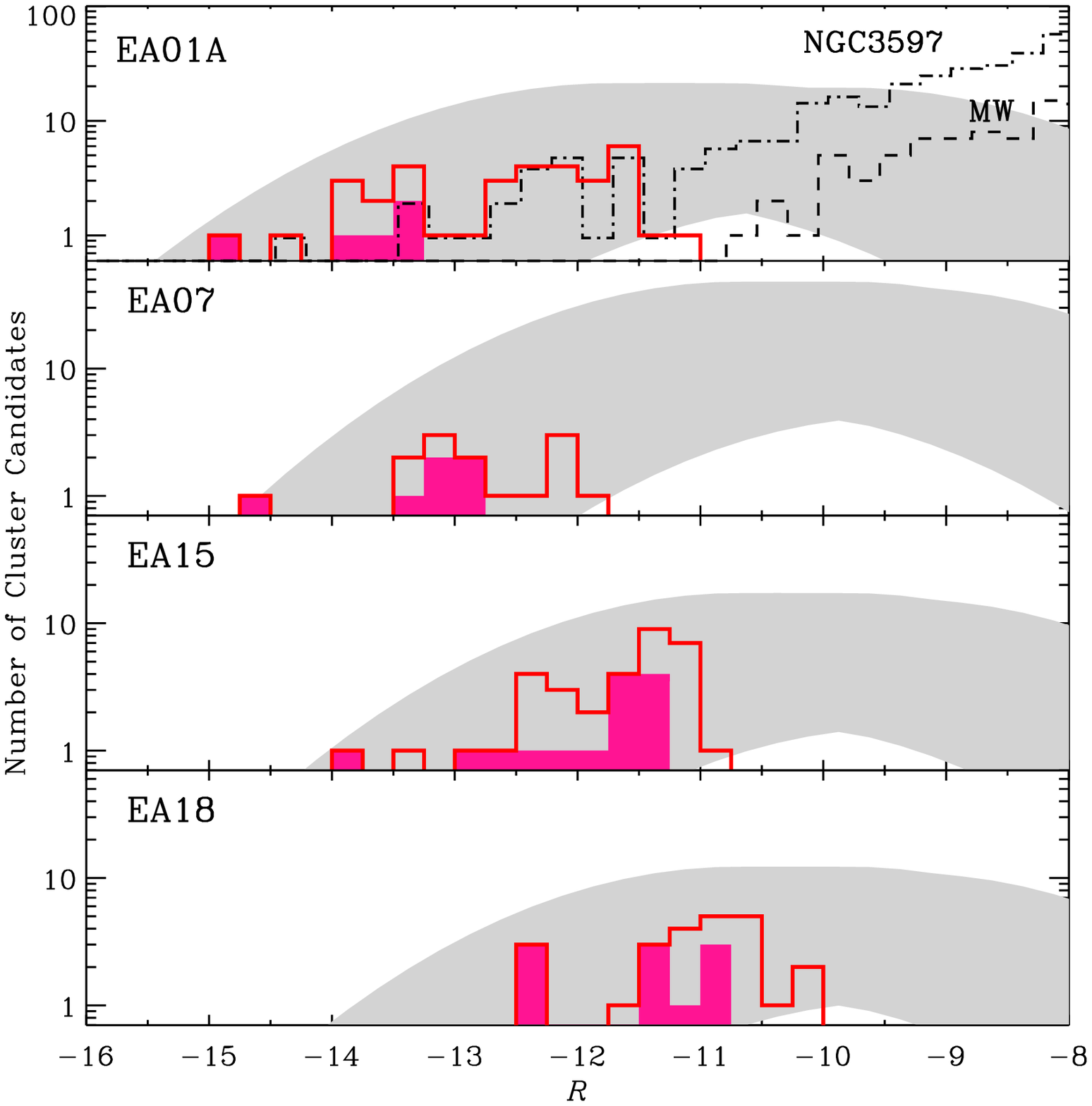}
\caption{
Cluster luminosity functions for the four E+As (EA01, 07, 15, 18) that
have more than five clusters with measurable colors.  The bluest (and
perhaps youngest) galaxy is at the top, the reddest galaxy at the bottom.
The solid lines are the LFs for the cluster candidates projected within
30 kpc of the host galaxy.  These LFs are incomplete fainter than about
the middle of their range.  The shaded histograms are for the clusters
with measured colors.  The LFs of Galactic GCs and the merging galaxy
NGC3597 are represented with dashed and dot-dashed lines, respectively.
The clusters in the E+As are much brighter than the Galactic globular
clusters ($M_R \lesssim -11$) but consistent with the bright end of the
NGC3597 cluster LF, suggesting that these compact point sources are the
population of star clusters formed during the starburst.  The luminosity
of the brightest clusters decreases as the color of the host galaxy
reddens, suggesting a fading trend with age.  The shaded regions are
the globular cluster LFs of E/S0s that are evolved backward to the epoch
of the E+A phase. The LFs of the cluster candidates fall in the shaded
region, indicating that they could evolve into the globular cluster
population of early-type galaxies.
\label{fig:cluster_lf}}
\end{figure}

In Figure \ref{fig:cluster_lf}, we present the luminosity functions
(LFs) of cluster candidates in the four E+As (EA01A, 07, 15, 18) that
have more than five clusters with measurable colors ($\sigma_{B-R} <
$ 0.35 mag) within a 30 kpc radius\footnote{We adopt a 20 kpc radius
for the random beam test to determine if cluster candidates exist, but
we analyze all cluster candidates within a 30 kpc radius to include as
many cluster candidates as possible.}.  The LFs of Milky Way globular
clusters \citep{Harris96} and clusters in a merging galaxy \cite[NGC
3597;][]{Carlson99} are also shown in the top panel for comparison. The
thick solid lines and the shaded histograms represent all E+A cluster
candidates within 30 kpc  and only those with measured colors,
respectively. The detected clusters in the four E+As are generally
much brighter than Galactic globular clusters ($M_R \lesssim -11$),
but consistent with the bright end of the NGC3597 cluster LF, suggesting
that these compact point sources are a population of star clusters formed
during the starburst.

More interesting is that the bright end of the LFs becomes fainter as the
EA number increases, i.e., as the host E+A becomes redder.  Furthermore,
we are able to detect a cluster population more often in E+As with smaller
EA number. We can reject with $\sim$90\% confidence that the distribution
of E+As with cluster populations (EA01A, 02, 03, 04, 07, 08, 11, 15, 18)
is random. All nine of these E+As exhibit tidal features or disturbed
morphologies (\S\ref{sec:tidal_feature}).  These trends suggest that the
newly-formed young star cluster systems fade or disrupt as the merger
remnant ages and morphologically relaxes. If E+As evolve into E/S0s,
we expect these cluster systems to become part of the globular cluster
systems of E/S0s.

\begin{figure}
\epsscale{1.0}
\plotone{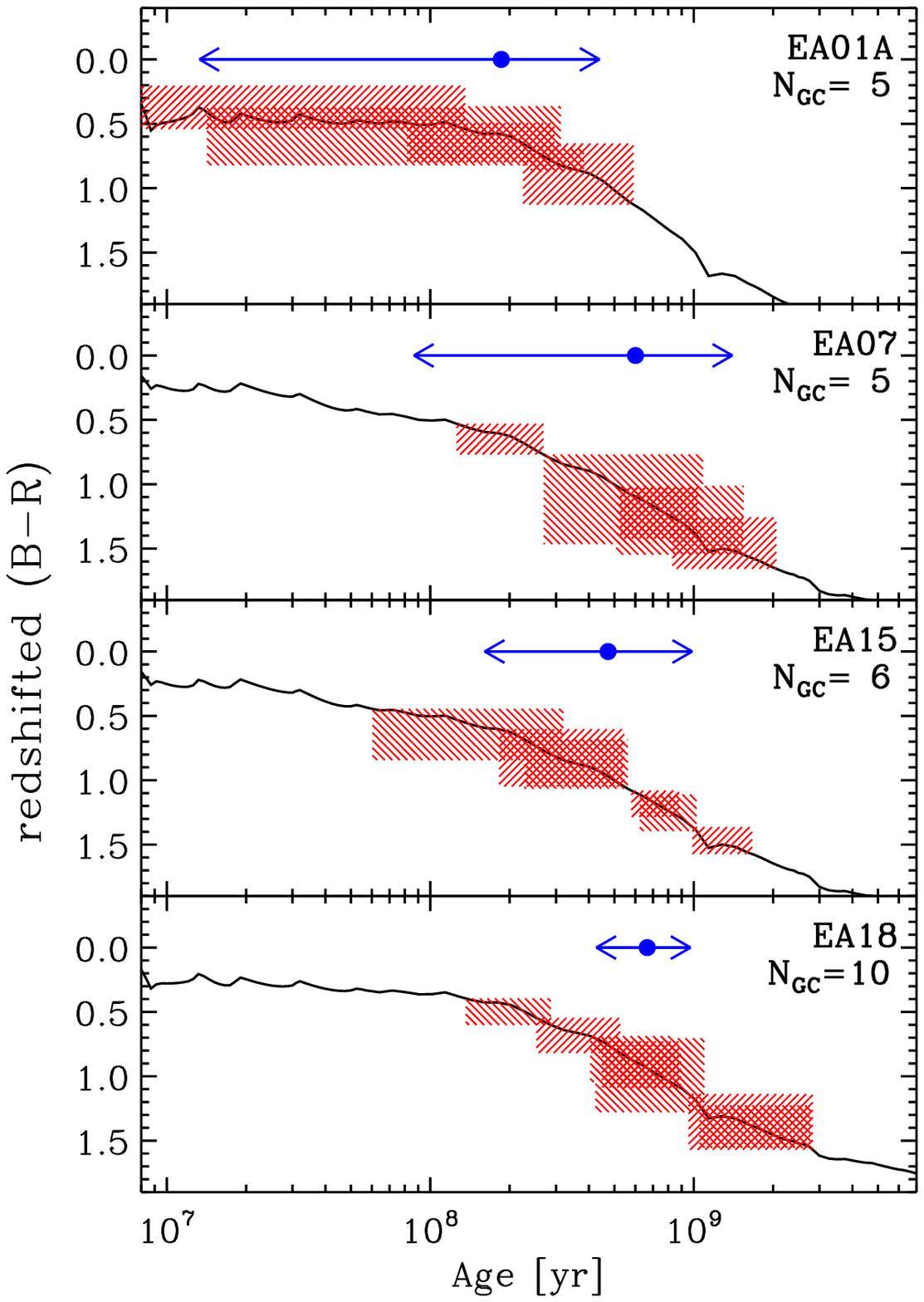}
\caption{
Evolution of the redshifted ($B-R$) colors of star clusters derived
from BC03 and the cluster colors from the four E+As (EA01, 07, 15,
18) in Figure \ref{fig:cluster_lf}.  For the model track, we assume an
instantaneous stellar population with a Salpeter IMF, masses ranging
from 1 to 100 $M_\sun$, and solar metallicity.  On the model track, we
show the measured cluster colors and the possible range of ages due to
the photometric errors (shaded region).  Only the clusters with color
uncertainties less than 0.35 mag are included in the age determination.
The ages of clusters derived from their colors are consistent with our
expectations from the E+A galaxy spectra: older than $\sim$10 Myr (the
lack of the emission lines due to the massive OB stars), but younger
than $\sim$ Gyr (the presence of strong Balmer lines due to the A stars).
The arrows represent the estimated post-burst age (time elapsed since the
starburst) using the methods described in \S \ref{sec:cluster_age_lf}.
Cluster ages become older as the EA number increases, i.e., as the host
E+A becomes redder.
\label{fig:cluster_age}}
\end{figure}

To pursue these suggestions, we determine the ages of individual clusters
by comparing the cluster colors with predictions from stellar population
synthesis models.  In Figure \ref{fig:cluster_age}, we show the ($B-R$)
evolution for star clusters derived from a BC03 model and the colors of
the star clusters found in three E+As (EA07, 15, 18). We also reproduce
Figure 10 from \citet{Yang04} for the clusters in EA01A for completeness.
To avoid uncertainties in the {\sl K}-correction and the magnitude
transformation due to the unknown SED, we calculate the redshifted
(\B{435}$-$\R{625}) color evolution using the spectra provided by the BC03
stellar synthesis model.  We assume an instantaneous stellar population
with a Salpeter initial mass function (IMF) over stellar masses ranging
from 1 to 100 $M_\sun$ and solar metallicity.  On the model tracks in
Figure \ref{fig:cluster_age}, we plot the observed cluster colors and
the possible age ranges due to the photometric errors (shaded region).
We include only clusters with color uncertainties less than 0.35 mag.
The ages of individual clusters are consistent with what one expects
from the E+A spectra: older than $\sim$10 Myr (the lack of emission
lines associated with massive OB stars) but younger than $\sim$ Gyr
(the presence of strong Balmer lines due to A stars).

To use the clusters to derive the time since the starburst, when the
majority of the clusters and the young galactic stars formed, we apply
a statistical test assuming that all star clusters formed during a
single burst and therefore that their spread in observed color arises
only from measurement errors.  Details of the post-burst age estimation
and statistical tests of the coeval assumption are described in Appendix
\ref{apdx:cluster_age_lf}.  We show the post-burst age inferred from the
clusters as horizontal bars in Figure \ref{fig:cluster_age}.  EA01A, 07,
15, and 18 have post-burst age ranges of [10\,Myr, 450\,Myr], [100\,Myr,
1.5\,Gyr], [150\,Myr, 1\,Gyr], and [400\,Myr, 1\,Gyr], respectively,
corresponding to the 95\% confidence level.  Note that it may be possible
to break the long-standing degeneracy between the burst strength and
the time elapsed since the starburst by using individual cluster ages,
because these systems are much simpler than the composite population of
the host galaxies. However, small number statistics and low S/N in the
color measurements do not allow us to put tighter constraints on the
post-burst ages at this time.



\begin{figure}
\epsscale{1.15}
\plotone{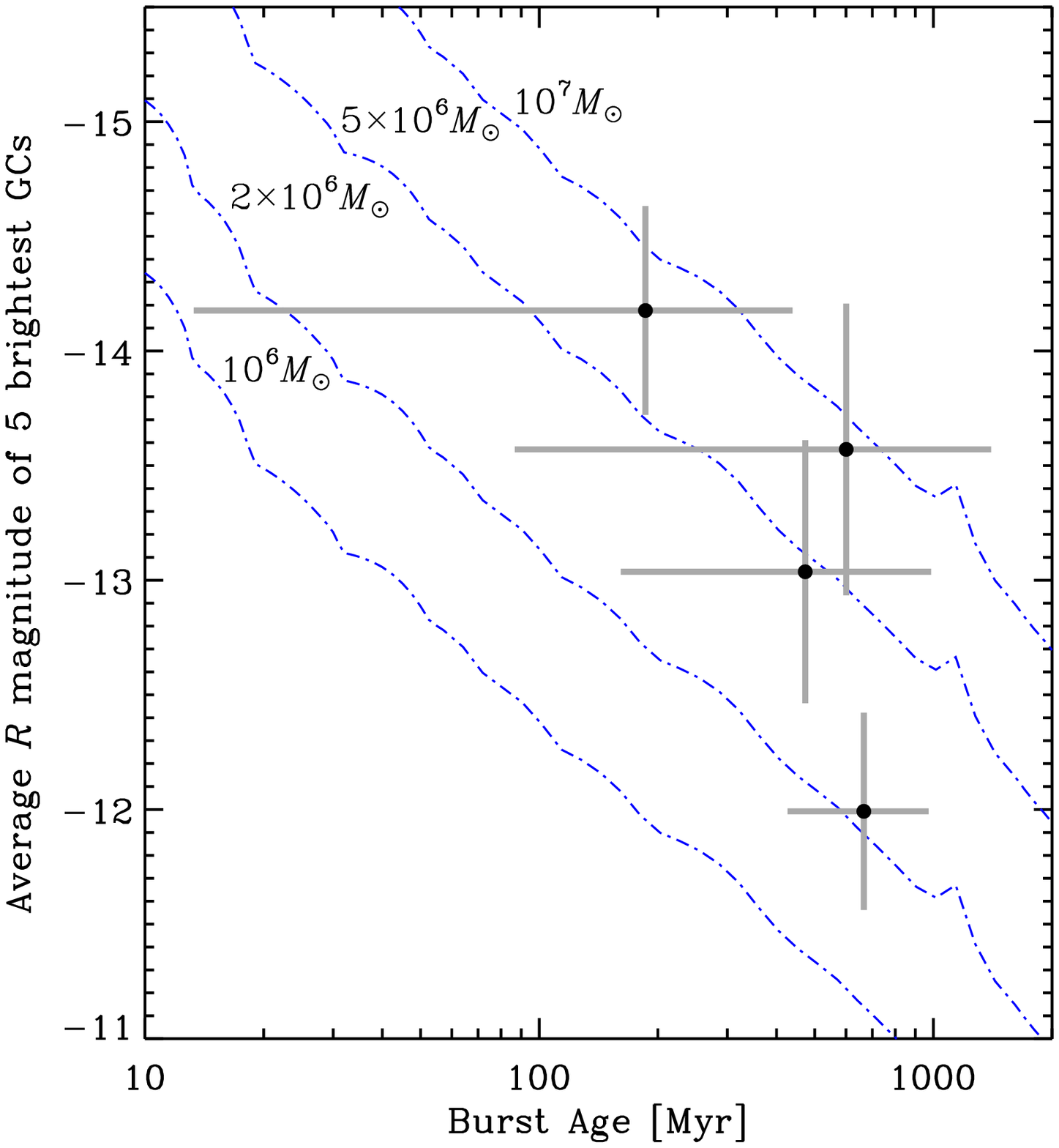}
\caption{
Evolution of the bright end of the star cluster LFs in the four
E+As (EA01A, 07, 15, 18) in Figures \ref{fig:cluster_lf} and
\ref{fig:cluster_age}.  We show the relation between the average
luminosity of the five brightest clusters in each E+A and the post-burst
age shown in Figure \ref{fig:cluster_age}.  Dot-dashed lines represent
BC03 model tracks for simple stellar populations (SSPs) with different
cluster masses (1, 2, 5 and 10 $\times$ $10^6\,M_\sun$).  The observed
fading of the cluster LFs is consistent with the color-derived post-burst
ages.
\label{fig:cluster_lf_evol}}
\end{figure}

The evolution of the cluster LFs is consistent with the model predictions.
While the uncertainties are fairly large, the derived post-burst ages
are consistent with the E+A spectra and tend to increase with EA number,
suggesting that the EA numbers {\it are} roughly consistent with a
sequence of post-burst ages. In Figure \ref{fig:cluster_lf_evol}, we
show that the evolution of the bright magnitude limit of the cluster LF
and the post-burst ages (derived from the cluster colors) are consistent
with the fading of a uniform cluster mass function.




\subsection{Evolution of the Cluster Population}

The remaining question is whether these young cluster candidates can
evolve into a significant part of the globular cluster systems of
present-day E/S0s.  Specifically, are the current numbers of clusters
in E+As consistent with the globular cluster luminosity function
(GCLFs) of E/S0s?  It has been proposed that the metal-rich GCs in
giant ellipticals formed during mergers of gas-rich spirals, producing
the bimodality of the GC subpopulations \citep{Ashman&Zepf92}.  Indeed,
many young GCs have been found in mergers \cite[e.g.,][]{Whitmore93}, and
\citet{Goudfrooij04} demonstrated that the LF of intermediate-age ($\sim$3
Gyr) GCs in NGC1316 can evolve dynamically into the red metal-rich
cluster populations that are ubiquitous in E/S0s.  Unfortunately, due
to low S/N in the color measurements and the small number statistics
of our cluster candidates, it is difficult to predict the evolution of
the individual cluster systems in our sample.  Instead, we ``predict''
the cluster LFs of E+As under the assumption that they are indeed the
progenitors of the E/S0 cluster population.


First, we assume the following form of the GCLFs, i.e., the number of
GCs per unit magnitude:
\begin{eqnarray}
\phi(M_R) = \frac{f_{\rm MR}}{\sqrt{2\pi}\sigma}  
\frac{S_N}{10^{0.4(M_{V}+15)}}
\exp[\frac{-(M_{R}-M_0)^2}{2\sigma^2}],
\end{eqnarray}
where $M_V$ is the total $V$-band magnitude of the host galaxy, $M_0$ and
$\sigma$ are the turnover magnitude and the dispersion of the Gaussian
function, $S_N$ is the specific frequency given by $S_N \equiv N_{\rm
GC}\cdot {10^{0.4(M_{V}+15)}}$ \citep{Harris&vandenBergh81}, and $f_{\rm
MR}$ is the fraction of metal-rich ``red'' clusters.  \citet{Harris01}
found that $\langle S_N \rangle$ = $3.6\pm0.5$, $M_0$ = $-7.4\pm0.15$
mag in the $V$-band and $\sigma$ = $1.4 \pm 0.03$ for local elliptical
galaxies ($M_V$ $\lesssim$ $-18$).

Second, to evolve this GCLF backward to the E+A epoch, we calculate the
fading of the turnover magnitude between an age corresponding to the
E+A clusters and that corresponding to E/S0 GCs, assuming that E/S0
GCs are 5 -- 10 Gyr old.  For solar metallicity and a Salpeter IMF,
a BC03 model predicts $\sim$ 2 mag of fading between 1 Gyr and 10 Gyr.
Note that these derived LFs will be lower limits because we consider
only photometric evolution.  For host galaxy luminosities, we adopt
$M_V$ = $M_V^{\rm E+A} + (1.3\pm0.73)$, as measured from the scaling
relationships.

The ``predicted" GCLFs taking into account all possible errors
mentioned above are consistent with the observed E+A cluster LFs (Figure
\ref{fig:cluster_lf}).  The most uncertain parameter is the age difference
between the clusters in E+As and in E/S0s due to the fast luminosity
evolution around 1 Gyr.  We use ages of 100 -- 500 Myr and 500 -- 1000 Myr
for EA01A and the other E+As, respectively.  We vary $f_{\rm MR}$ between
\case{1}{3} and \case{2}{3}.  Most of our observed LFs fall within the
uncertainties (shaded regions), indicating that the number of clusters
observed does not violate the constraint derived from the GCLFs of the
present-day E/S0s.  While it is highly uncertain how many clusters in
E+As will survive and how much they will contribute to the metal-rich
GCs in E/S0s, we conclude that it is at least possible that young star
clusters in E+As can evolve into the globular cluster systems of E/S0s.


\section{Conclusions}
\label{sec:conclusion}

We study the detailed morphologies of 21 E+A galaxies using high
resolution {\sl HST}/{\sl ACS} and {\sl WFPC2} images to investigate into
what E+A galaxies will evolve after their young stellar populations fade
away in a few Gyr.  Our findings are: \\

1. The morphologies of E+As are extremely diverse, ranging across
train-wrecks, barred galaxies, and blue-cores to relaxed disky galaxies.
Most of these galaxies lie in the field, well outside of rich clusters,
and at least 11 (55\%) have tidal or other disturbed features.  Our sample
includes one binary E+A system, in which both E+As are tidally disturbed
and interacting with each other.  These results support the picture in
which galaxy-galaxy tidal interactions or mergers are responsible for
triggering the E+A phase in many cases.

2. E+As are bulge-dominated systems (median bulge fraction $B/T$ = 0.59)
and their light is highly concentrated (S\'ersic index $n \gtrsim 5$).
When dust is negligible (at least 67\% of the time), E+As have high
concentration indices ($C \gtrsim 4.3$) consistent with those of
spheroids, but considerably larger asymmetry indices ($A \gtrsim 0.04$)
than ellipticals due to structures within a few $r_e$ that presumably
arise from the starburst and/or recent merger.  Thus E+As would be
morphologically classified as early-type galaxies once these disturbances
relax and the low surface brightness tidal features dissipate or fade.

3. The color morphologies of E+As are as diverse as their structural
morphologies. A large fraction (70\%) have positive color gradients
(bluer toward center), indicating that their young stellar populations
are more concentrated than their older populations.  We demonstrate that
evolution can invert these gradients into the negative gradients typical
of E/S0s if the inner parts of E+As have become more metal enriched than
the outer parts due to the centralized star formation.

4. Six E+As (30\%) exhibit compact (0.4--1.4 kpc) blue cores, which
might be the local analogs of the high-$z$ elliptical blue-cores
\citep{Menanteau01a}.  We discovered LINERs in three of these blue-core
E+As \citep{Yang06}, and the relationship between LINERs and blue-cores
could be an important clue to what stops the star formation in E+A
galaxies.

5. E+As stand apart from the E/S0 fundamental plane (FP) in the edge-on
projection, implying that the stellar populations of E+As are different
from that of E/S0s. E+As have, on average, a \ml that is 3.8 times
smaller than that of E/S0s.  The tilt of the E+A FP indicates that the
variation of the stellar populations among E+As is closely tied to their
structural parameters, i.e., E+As follow their own scaling relationships
such that smaller or less massive galaxies have smaller {\sl M/L}.
Such a trend arises naturally within a merger scenario, where low mass
galaxies (the progenitors of low-mass E+As) have higher gas fractions
\citep{Young&Scoville91} and could produce relatively larger populations
of young stars.

6. We find a population of unresolved compact sources in at least nine
E+A galaxies (45\%). The colors and luminosities of these young star
cluster candidates are consistent with the ages inferred from the E+A
spectra (0.01 -- 1 Gyr).  The bright end of the cluster luminosity
function fades as the host galaxy becomes redder, suggesting that the
newly-formed young star clusters age in parallel to their host. This
interpretation is confirmed by the color evolution of the cluster systems.

We have now examined the full set of E+As from the Las Campanas Redshift
Survey and so have representative results for local E+A galaxies.  We have
used high spatial resolution images to probe their detailed morphologies.
We find that their properties are either consistent with those of E/S0s
or, if left to evolve passively, will become like those of early-types.
The morphologies, color profiles, scaling relations, and young star
clusters suggest that E+As galaxies are caught in the act of transforming
from late-type to early-type galaxies.

\acknowledgements 

We thank an anonymous referee for his/her thorough reading of the
manuscript and helpful comments.  We thank Chien Peng for help in using
the GALFIT and useful discussions about galaxy morphology.  YY and AIZ
acknowledge support from NSF grant AST-02-06084, and {\sl HST} grant
GO-9781. Support for proposal GO-9781 was provided by NASA through a
grant from the Space Telescope Science Institute, which is operated by
the Association of Universities for Research in Astronomy, Incorporated,
under NASA contract NAS5-26555. \\

Facilities: \facility{HST(ACS,WFPC2)}


\clearpage
\begin{landscape}


\begin{deluxetable}{l rr cc l ll l}
\tablewidth{9in}
\tabletypesize{\small}
\tablecaption{Properties of E+A Galaxies \label{tab:basic}}
\tablehead{
\colhead{ID\tablenotemark{a}}&
\colhead{R.A.}&
\colhead{Dec.}&
\colhead{$z$\tablenotemark{a}}&
\colhead{Scale}&
\colhead{$\sigma_{\rm old}$\tablenotemark{b}}&
\colhead{$SFR$}&
\colhead{\ion{H}{1} mass}&
\colhead{Morphology}\\
\colhead{}&
\colhead{(J2000)}&
\colhead{(J2000)}&
\colhead{}&
\colhead{(kpc/\arcsec)}&
\colhead{(km s$^{-1}$)}&
\colhead{($M_{\sun}$ yr$^{-1}$)}&
\colhead{(10$^9$ $M_{\sun}$)}&
\colhead{}
}
\startdata
 EA01A&     11 01 19.05 &    $-$12 10  17.8 &  0.07463 &    1.42 &    $\phn35^{+ 39}_{- 35}$ &        $< 0.12\pm 0.07$\tablenotemark{c}&  $\phn\phn 7.7\pm 0.4$\tablenotemark{f}&   tidal-feature, companion, cluster \\
 EA01B&     11 01 18.39 &    $-$12 10  14.1 &  0.07463 &    1.42 &                   \nodata &        $< 0.07\pm 0.03$\tablenotemark{c}&                                 \nodata&   tidal-feature, companion, LINER \\
 EA02 &     02 17 39.88 &    $-$44 32  47.2 &  0.09873 &    1.82 &       $202^{+ 17}_{- 16}$ &             $< 4.4\phn$\tablenotemark{d}&                $< 3.4$\tablenotemark{f}&   tidal-feature, dust?, cluster \\
 EA03 &  12 09 \phn5.54 &    $-$12 22  35.8 &  0.08109 &    1.53 &       $120^{+ 22}_{- 20}$ &             $< 1.7\phn$\tablenotemark{e}&                $< 4.2$\tablenotemark{f}&   tidal-feature, cluster \\
 EA04 &  04 00 \phn0.34 &    $-$44 35  15.8 &  0.10122 &    1.86 &       $131^{+  9}_{-  9}$ &             $< 3.2\phn$\tablenotemark{d}&                $< 2.3$\tablenotemark{f}&   disturbed, dust?, cluster \\
 EA05 &  01 58 \phn1.52 &    $-$44 37  14.0 &  0.11731 &    2.12 &       $120^{+  7}_{-  8}$ &                                  \nodata&                                 \nodata&   bar, blue-core \\
 EA06 &     11 53 55.59 &    $-$03 10  36.3 &  0.08850 &    1.65 &    $\phn23^{+  1}_{- 23}$ &        $< 0.10\pm 0.03$\tablenotemark{c}&                                 \nodata&   dust, blue-core, LINER \\
 EA07 &  22 41 \phn9.75 &    $-$38 34  34.9 &  0.11413 &    2.07 &       $241^{+ 11}_{- 12}$ &                                  \nodata&                                 \nodata&   tidal-feature, dust, cluster \\
 EA08 &  14 32 \phn3.23 &    $-$12 57  33.1 &  0.11222 &    2.04 &    $\phn99^{+  8}_{- 11}$ &             $< 1.8\phn$\tablenotemark{e}&                                 \nodata&   tidal-feature, cluster \\
 EA09 &     01 17 38.28 &    $-$41 34  24.6 &  0.06513 &    1.25 &    $\phn74^{+134}_{-  5}$ &             $< 1.3\phn$\tablenotemark{e}&                                 \nodata&   blue-core \\
 EA10 &     02 11 42.96 &    $-$44 07  40.2 &  0.10493 &    1.92 &       $121^{+  7}_{-  8}$ &                                  \nodata&                                 \nodata&   tidal-feature, companion \\
 EA11 &     01 14 49.65 &    $-$41 22  31.8 &  0.12167 &    2.19 &       $174^{+ 12}_{- 11}$ &             $< 9.6\phn$\tablenotemark{d}&                $< 5.3$\tablenotemark{f}&   disturbed, companion, cluster \\
 EA12 &     12 05 59.79 &    $-$02 54  32.4 &  0.09713 &    1.80 &       $111^{+ 17}_{- 18}$ &      $\phn\phn 7.9\phn$\tablenotemark{e}&                                 \nodata&   tidal-feature, companion, dust \\
 EA13 &     11 19 52.44 &    $-$12 52  39.6 &  0.09572 &    1.77 &       $165^{+ 10}_{-  9}$ &             $< 2.1\phn$\tablenotemark{e}&                                 \nodata&   tidal-feature, companion, dust \\
 EA14 &  13 57 \phn1.68 &    $-$12 26  48.2 &  0.07046 &    1.34 &       $135^{+ 10}_{- 10}$ &             $< 1.5\phn$\tablenotemark{e}&                                 \nodata&   bar, blue-core \\
 EA15 &     14 40 44.26 &    $-$06 39  52.3 &  0.11381 &    2.07 &       $125^{+  9}_{- 10}$ &        $< 0.05\pm 0.02$\tablenotemark{c}&                                 \nodata&   cluster \\
 EA16 &     12 19 55.69 & $-$06 14 \phn 1.0 &  0.07642 &    1.45 &    $\phn93^{+  7}_{-  8}$ &        $< 0.05\pm 0.02$\tablenotemark{c}&                                 \nodata&   blue-core, LINER \\
 EA17 &     10 13 52.40 &    $-$02 55  47.2 &  0.06090 &    1.18 &    $\phn23^{+  1}_{- 23}$ &        $< 0.06\pm 0.03$\tablenotemark{c}&                $< 2.9$\tablenotemark{g}&   dust, blue-core, LINER \\
 EA18 &     00 22 46.79 &    $-$41 33  35.8 &  0.05985 &    1.16 &    $\phn23^{+  4}_{- 23}$ &             $< 1.2\phn$\tablenotemark{e}&  $\phn\phn 2.0\pm 0.3$\tablenotemark{g}&   disturbed, dust, cluster \\
 EA19 &     02 07 49.45 &    $-$45 20  51.1 &  0.06400 &    1.23 &    $\phn84^{+ 18}_{- 34}$ &      $\phn\phn 2.7\phn$\tablenotemark{e}&                $< 1.2$\tablenotemark{g}&   dust \\
 EA21 &     11 15 24.96 &    $-$06 45  13.8 &  0.09944 &    1.84 &       $134^{+  7}_{-  9}$ &             $< 1.7\phn$\tablenotemark{e}&                                 \nodata&   bar
\enddata

\clearpage
\tablenotetext{a}{E+A IDs and redshifts from \citet{Zabludoff96}.}

\tablenotetext{b}{Velocity dispersion of the old stellar population
from \citet{Norton01}.  The entries with $\sigma$ = 23 km s$^{-1}$
are the upper limits.}

\tablenotetext{c}{SFR from H$\alpha$ luminosity \citep{Yang06} using
the relation, SFR ($M_{\sun}$ yr$^{-1})$ = $L_{\rm H\alpha} $ $/\rm
1.27\times10^{41}ergs\ s^{-1}$ \citep{Kennicutt98}.}

\tablenotetext{d}{SFR from the radio-continuum \citep{Chang01} using
the relation, SFR ($M_{\sun}$ yr$^{-1})$ = $5.9\times10^{-22}$ $L_{\rm
1.4GHz}$ (W Hz$^{-1}$) \citep{Yun01}. Corrected for the cosmological
parameters adopted in this paper.  Note that we do not adopt the SFRs
derived from the gas surface densities, which are highly uncertain.}

\tablenotetext{e}{SFR from the radio-continuum \citep{Miller_Owen01}. Same
as footnote d.}

\tablenotetext{f}{\ion{H}{1} mass from \citet{Chang01}.  Corrected for
the cosmological parameters adopted in this paper.}

\tablenotetext{g}{\ion{H}{1} mass from \citet{Buyle06}.}
\end{deluxetable}

\clearpage
\end{landscape}




\begin{deluxetable}{cc rrr c}
\tablewidth{0pt}
\tabletypesize{\small}
\tablecaption{Magnitude Transformation \label{tab:transform}}
\tablehead{
\colhead{TCOL}&
\colhead{SCOL}&
\colhead{$c_0$}&
\colhead{$c_1$}&
\colhead{$c_2$}
}
\startdata
              \B{}$-$\B{439} &                    \B{439}$-$\R{702} &   $-$0.001 &   $-$0.029 &   $-$0.033 &  \\
              \B{}$-$\B{435} &                    \B{435}$-$\R{625} &      0.008 &   $-$0.048 &   $-$0.023 &  \\
               $r$$-$\R{702} &                    \B{439}$-$\R{702} &      0.496 &      0.014 &      0.023 &  \\
               $r$$-$\R{625} &                    \B{435}$-$\R{625} &      0.406 &   $-$0.084 &      0.016 &  \\
           \B{435}$-$\R{625} &                    \B{439}$-$\R{702} &   $-$0.107 &      0.938 &   $-$0.033 &  \\
\Bz{435}{0.1}$-$\Rz{625}{0.1}&        \Bz{439}{0.1}$-$\Rz{702}{0.1} &   $-$0.030 &      0.857 &   $-$0.009 &  \\
\enddata
\end{deluxetable}



\begin{deluxetable}{l ccccc}
\tablewidth{0pt}
\tablecaption{Concentrations and Asymmetries \label{tab:cas}}
\tablehead{
\colhead{ID}&
\colhead{$r(\eta = 0.2)$}&
\colhead{$C(R)$}&
\colhead{$A(R)$\tablenotemark{a}}&
\colhead{$m_B$}&
\colhead{$m_R$}  \\
\colhead{}&
\colhead{(arcsec)}&
\colhead{}&
\colhead{}&
\colhead{$(r<2r_p)$}&
\colhead{$(r<2r_p)$} 
}
\startdata

EA01A &    6.75 $\pm$ 0.06 &    3.10 $\pm$ 0.01 &  0.17 &   17.09 $\pm$ 0.30 &   16.77 $\pm$ 0.01  \\
EA01B &    1.28 $\pm$ 0.02 &    4.28 $\pm$ 0.02 &  0.08 &   17.44 $\pm$ 0.02 &   16.03 $\pm$ 0.01  \\
EA02  &    3.94 $\pm$ 0.05 &    4.96 $\pm$ 0.01 &  0.09 &   17.79 $\pm$ 0.04 &   16.24 $\pm$ 0.01  \\
EA03  &    2.35 $\pm$ 0.06 &    5.07 $\pm$ 0.02 &  0.07 &   16.76 $\pm$ 0.05 &   15.35 $\pm$ 0.01  \\
EA04  &    3.24 $\pm$ 0.06 &    4.81 $\pm$ 0.01 &  0.11 &   17.36 $\pm$ 0.01 &   15.77 $\pm$ 0.01  \\
EA05  &    1.31 $\pm$ 0.18 &    4.54 $\pm$ 0.17 &  0.05 &   18.65 $\pm$ 0.02 &   16.99 $\pm$ 0.06  \\
EA06  &    0.52 $\pm$ 0.01 &    3.40 $\pm$ 0.03 &  0.08 &   18.67 $\pm$ 0.01 &   17.41 $\pm$ 0.01  \\
EA07  &    6.93 $\pm$ 0.07 &    3.94 $\pm$ 0.01 &  0.15 &   16.91 $\pm$ 0.01 &   15.30 $\pm$ 0.01  \\
EA08  &    2.44 $\pm$ 0.05 &    4.71 $\pm$ 0.01 &  0.05 &   18.78 $\pm$ 0.01 &   17.27 $\pm$ 0.01  \\
EA09  &    2.56 $\pm$ 0.08 &    4.47 $\pm$ 0.01 &  0.11 &   19.01 $\pm$ 0.01 &   17.55 $\pm$ 0.01  \\
EA10  &    2.18 $\pm$ 0.06 &    4.57 $\pm$ 0.02 &  0.05 &   18.39 $\pm$ 0.01 &   16.91 $\pm$ 0.01  \\
EA11  &    1.94 $\pm$ 0.11 &    4.90 $\pm$ 0.04 &  0.07 &   18.79 $\pm$ 0.02 &   17.22 $\pm$ 0.01  \\
EA12  &    4.33 $\pm$ 0.04 &    4.09 $\pm$ 0.01 &  0.14 &   18.44 $\pm$ 0.01 &   16.84 $\pm$ 0.01  \\
EA13  &    6.45 $\pm$ 0.13 &    4.74 $\pm$ 0.01 &  0.27 &   17.62 $\pm$ 0.01 &   16.00 $\pm$ 0.01  \\
EA14  &    0.83 $\pm$ 0.03 &    3.64 $\pm$ 0.05 &  0.03 &   18.05 $\pm$ 0.01 &   16.69 $\pm$ 0.01  \\
EA15  &    1.67 $\pm$ 0.05 &    4.45 $\pm$ 0.02 &  0.04 &   19.02 $\pm$ 0.01 &   17.50 $\pm$ 0.01  \\
EA16  &    0.63 $\pm$ 0.06 &    3.84 $\pm$ 0.15 &  0.04 &   18.47 $\pm$ 0.02 &   17.30 $\pm$ 0.04  \\
EA17  &    2.84 $\pm$ 0.05 &    5.04 $\pm$ 0.01 &  0.08 &   18.58 $\pm$ 0.01 &   17.16 $\pm$ 0.01  \\
EA18  &    2.92 $\pm$ 0.05 &    3.73 $\pm$ 0.01 &  0.19 &   17.81 $\pm$ 0.01 &   16.37 $\pm$ 0.01  \\
EA19  &    4.43 $\pm$ 0.03 &    2.95 $\pm$ 0.01 &  0.10 &   18.06 $\pm$ 0.01 &   16.66 $\pm$ 0.01  \\
EA21  &    1.22 $\pm$ 0.17 &    4.64 $\pm$ 0.18 &  0.04 &   18.85 $\pm$ 0.04 &   17.26 $\pm$ 0.05 

\enddata
\tablenotetext{a}{Typical errors for $A$ are $\sigma(A)/A \simeq 0.015$.}
\end{deluxetable}



\renewcommand{\scr}{\small}

\begin{deluxetable}{l rcrrr c rcrrrr}
\tablewidth{0pt}
\tabletypesize{\small}
\tablecaption{\devauc and S\'ersic Profile Fit (\R{702} or \R{625} bands) \label{tab:devauc}}
\tablehead{
\colhead{}&
\multicolumn{5}{c}{$r^{1/4}$-Law}&
\colhead{}&
\multicolumn{6}{c}{S\'ersic $r^{1/n}$-Law} \\
\cline{2-6} \cline{8-13}
\colhead{ID}&
\colhead{$r_e$}&
\colhead{$\mu_e$}&
\colhead{$q$ \tablenotemark{a}}&
\colhead{$c$ \tablenotemark{a}}&
\colhead{$\chi_{\nu}^2$}&
\colhead{}&
\colhead{$r_e$}&
\colhead{$\mu_e$}&
\colhead{$n$}&
\colhead{$q$ \tablenotemark{a}}&
\colhead{$c$ \tablenotemark{a}}&
\colhead{$\chi_{\nu}^2$}\\
\colhead{}&
\colhead{(kpc)}&
\colhead{(mag arcsec$^{-2}$)}&
\colhead{($b/a$)}&
\colhead{}&
\colhead{}&
\colhead{}&
\colhead{(kpc)}&
\colhead{(mag arcsec$^{-2}$)}&
\colhead{}&
\colhead{($b/a$)}&
\colhead{}&
\colhead{}
}

\startdata
 EA01B& $    1.08$\scr$+ 0.07\atop - 0.01$& $   17.86$\scr$+ 0.12\atop - 0.02$& $ 0.70$&$-0.04$&2.88&& $    1.35$\scr$+ 0.40\atop - 0.01$& $   18.39$\scr$+ 0.55\atop - 0.04$& $    5.39$\scr$+ 0.99\atop - 0.05$& $ 0.68$&$-0.06$&2.53\\
 EA02 & $    2.76$\scr$+ 0.17\atop - 0.01$& $   19.83$\scr$+ 0.11\atop - 0.05$& $ 0.73$&$ 0.15$&1.51&& $    5.85$\scr$+ 2.01\atop - 0.01$& $   21.41$\scr$+ 0.57\atop - 0.35$& $    6.79$\scr$+ 0.76\atop - 0.48$& $ 0.75$&$ 0.34$&1.43\\
 EA03 & $    1.69$\scr$+ 0.09\atop - 0.01$& $   18.32$\scr$+ 0.10\atop - 0.05$& $ 0.97$&$-0.04$&2.26&& $    4.99$\scr$+ 1.03\atop - 0.01$& $   20.65$\scr$+ 0.37\atop - 0.22$& $    9.55$\scr$+ 0.54\atop - 0.37$& $ 0.98$&$-0.04$&1.50\\
 EA04 & $    2.12$\scr$+ 0.11\atop - 0.01$& $   18.92$\scr$+ 0.10\atop - 0.03$& $ 0.97$&$ 0.01$&1.42&& $    3.07$\scr$+ 0.37\atop - 0.01$& $   19.76$\scr$+ 0.23\atop - 0.11$& $    5.97$\scr$+ 0.29\atop - 0.17$& $ 0.98$&$-0.02$&1.31\\
 EA05 & $    1.62$\scr$+ 0.05\atop - 0.01$& $   18.57$\scr$+ 0.07\atop - 0.05$& $ 0.61$&$-0.10$&1.37&& $    2.93$\scr$+ 0.82\atop - 0.01$& $   19.91$\scr$+ 0.51\atop - 0.24$& $    7.08$\scr$+ 0.89\atop - 0.41$& $ 0.62$&$-0.06$&1.30\\
 EA06 & $    0.43$\scr$+ 0.01\atop - 0.01$& $   17.07$\scr$+ 0.01\atop - 0.04$& $ 0.86$&$-0.05$&0.70&& $    0.46$\scr$+ 0.01\atop - 0.01$& $   17.25$\scr$+ 0.05\atop - 0.07$& $    4.60$\scr$+ 0.12\atop - 0.21$& $ 0.86$&$-0.05$&0.69\\
 EA07 & $   11.75$\scr$+ 0.19\atop - 0.01$& $   21.27$\scr$+ 0.02\atop - 0.05$& $ 0.74$&$ 0.07$&1.27&& $    7.97$\scr$+ 0.35\atop - 0.01$& $   20.44$\scr$+ 0.08\atop - 0.08$& $    2.72$\scr$+ 0.11\atop - 0.09$& $ 0.75$&$ 0.07$&1.12\\
 EA08 & $    1.84$\scr$+ 0.05\atop - 0.01$& $   19.58$\scr$+ 0.04\atop - 0.02$& $ 0.89$&$ 0.07$&0.99&& $    4.00$\scr$+ 0.73\atop - 0.01$& $   21.25$\scr$+ 0.32\atop - 0.19$& $    7.67$\scr$+ 0.52\atop - 0.36$& $ 0.89$&$ 0.08$&0.85\\
 EA09 & $    2.18$\scr$+ 0.08\atop - 0.01$& $   20.86$\scr$+ 0.05\atop - 0.05$& $ 0.52$&$-0.37$&1.61&& $   20.92$\scr$+12.12\atop - 0.01$& $   25.58$\scr$+ 0.86\atop - 1.34$& $   15.22$\scr$+ 1.56\atop - 2.55$& $ 0.52$&$-0.37$&1.39\\
 EA10 & $    2.22$\scr$+ 0.03\atop - 0.01$& $   19.73$\scr$+ 0.02\atop - 0.05$& $ 0.81$&$ 0.05$&0.83&& $    3.28$\scr$+ 0.70\atop - 0.01$& $   20.57$\scr$+ 0.41\atop - 0.20$& $    5.74$\scr$+ 1.00\atop - 0.31$& $ 0.81$&$ 0.04$&0.75\\
 EA11 & $    1.80$\scr$+ 0.04\atop - 0.01$& $   18.92$\scr$+ 0.04\atop - 0.15$& $ 0.53$&$ 0.14$&1.16&& $    2.88$\scr$+ 0.44\atop - 0.01$& $   19.97$\scr$+ 0.29\atop - 0.15$& $    6.31$\scr$+ 0.94\atop - 0.24$& $ 0.53$&$ 0.13$&1.06\\
 EA12 & $    4.67$\scr$+ 0.25\atop - 0.01$& $   21.47$\scr$+ 0.07\atop - 0.10$& $ 0.81$&$ 1.13$&1.29&& $    9.14$\scr$+ 2.40\atop - 0.01$& $   22.83$\scr$+ 0.42\atop - 0.33$& $    6.33$\scr$+ 0.56\atop - 0.40$& $ 0.81$&$ 1.25$&1.22\\
 EA13 & $    5.72$\scr$+ 0.28\atop - 0.01$& $   20.50$\scr$+ 0.07\atop - 0.07$& $ 0.55$&$ 0.22$&0.94&& $    8.52$\scr$+ 2.41\atop - 0.01$& $   21.35$\scr$+ 0.47\atop - 0.35$& $    5.56$\scr$+ 0.72\atop - 0.48$& $ 0.54$&$ 0.17$&0.90\\
 EA14 & $    1.10$\scr$+ 0.01\atop - 0.01$& $   18.23$\scr$+ 0.00\atop - 0.03$& $ 0.57$&$ 0.18$&1.68&& $    1.44$\scr$+ 0.01\atop - 0.01$& $   18.83$\scr$+ 0.01\atop - 0.11$& $    5.30$\scr$+ 0.04\atop - 0.17$& $ 0.56$&$ 0.20$&1.58\\
 EA15 & $    1.45$\scr$+ 0.01\atop - 0.01$& $   19.23$\scr$+ 0.00\atop - 0.07$& $ 0.91$&$-0.07$&0.83&& $    2.38$\scr$+ 0.09\atop - 0.01$& $   20.34$\scr$+ 0.08\atop - 0.30$& $    6.54$\scr$+ 0.23\atop - 0.51$& $ 0.91$&$-0.07$&0.75\\
 EA16 & $    0.65$\scr$+ 0.01\atop - 0.01$& $   18.04$\scr$+ 0.01\atop - 0.02$& $ 0.83$&$ 0.05$&1.25&& $    1.71$\scr$+ 0.13\atop - 0.01$& $   20.18$\scr$+ 0.15\atop - 0.31$& $    9.22$\scr$+ 0.33\atop - 0.70$& $ 0.81$&$ 0.07$&0.84\\
 EA17 & $    1.35$\scr$+ 0.07\atop - 0.01$& $   20.02$\scr$+ 0.08\atop - 0.03$& $ 0.65$&$-0.13$&1.28&& $    5.16$\scr$+ 3.89\atop - 0.01$& $   22.82$\scr$+ 1.10\atop - 0.50$& $   10.44$\scr$+ 2.44\atop - 0.86$& $ 0.64$&$-0.11$&0.96\\
 EA18 & $    3.17$\scr$+ 0.03\atop - 0.01$& $   20.57$\scr$+ 0.01\atop - 0.06$& $ 0.50$&$-0.27$&1.66&& $    2.24$\scr$+ 0.04\atop - 0.01$& $   19.76$\scr$+ 0.04\atop - 0.08$& $    2.60$\scr$+ 0.06\atop - 0.09$& $ 0.49$&$-0.29$&1.42\\
 EA19 & $    4.98$\scr$+ 0.18\atop - 0.01$& $   21.41$\scr$+ 0.05\atop - 0.08$& $ 0.43$&$-0.14$&1.65&& $    3.22$\scr$+ 0.04\atop - 0.01$& $   20.16$\scr$+ 0.03\atop - 0.02$& $    1.06$\scr$+ 0.03\atop - 0.02$& $ 0.40$&$-0.10$&1.06\\
 EA21 & $    0.92$\scr$+ 0.01\atop - 0.01$& $   18.20$\scr$+ 0.02\atop - 0.03$& $ 0.72$&$ 0.21$&1.28&& $    2.20$\scr$+ 0.33\atop - 0.01$& $   20.09$\scr$+ 0.28\atop - 0.31$& $    8.35$\scr$+ 0.60\atop - 0.60$& $ 0.70$&$ 0.19$&1.04  
 
\enddata
\tablecomments{EA01A is excluded because it is so irregular.}
\tablenotetext{a}{
Typical errors for $q$ and $c$ are $\sim$ 0.01.
}
\end{deluxetable}

\renewcommand{\scr}{\small}

\begin{deluxetable}{l rcrr c rcrr rr c}
\tablewidth{0pt}
\tabletypesize{\small}
\tablecaption{Bulge-Disk Decomposition (\R{702} and \R{625} bands)\label{tab:decomp}}
\tablehead{
\colhead{}&
\multicolumn{4}{c}{$r^{1/4}$-Bulge}&
\colhead{}&
\multicolumn{4}{c}{Disk} \\
\cline{2-5} \cline{7-10}
\colhead{ID}&
\colhead{$r_e$}&
\colhead{$\mu_e$}&
\colhead{$q$ \tablenotemark{a}}&
\colhead{$c$ \tablenotemark{a}}&
\colhead{}&
\colhead{$r_d$}&
\colhead{$\mu_d$}&
\colhead{$q$ \tablenotemark{a}}&
\colhead{$c$ \tablenotemark{a}}&
\colhead{$B/T$}&
\colhead{$\chi_{\nu}^2$}\\
\colhead{}&
\colhead{(kpc)}&
\colhead{(mag arcsec$^{-2}$)}&
\colhead{($b/a$)}&
\colhead{}&
\colhead{}&
\colhead{(kpc)}&
\colhead{(mag arcsec$^{-2}$)}&
\colhead{($b/a$)}&
\colhead{}&
\colhead{}&
\colhead{}
}
\startdata
EA01B                 & $    0.73$\scr$+ 0.01\atop - 0.01$& $   17.35$\scr$+ 0.01\atop - 0.04$& $ 0.75$&$-0.06$&  & $    2.38$\scr$+ 0.23\atop - 0.01$& $   19.35$\scr$+ 0.13\atop - 0.07$& $ 0.39$&$ 1.21$&$    0.78$\scr$+ 0.01\atop - 0.03$& $ 2.40$&\\
EA03                  & $    0.71$\scr$+ 0.01\atop - 0.01$& $   16.94$\scr$+ 0.03\atop - 0.04$& $ 0.93$&$ 0.09$&  & $    3.60$\scr$+ 0.17\atop - 0.01$& $   19.19$\scr$+ 0.07\atop - 0.17$& $ 0.88$&$ 0.04$&$    0.54$\scr$+ 0.01\atop - 0.02$& $ 1.57$&\\
EA04                  & $    1.11$\scr$+ 0.01\atop - 0.01$& $   17.93$\scr$+ 0.02\atop - 0.08$& $ 0.87$&$ 0.08$&  & $    3.40$\scr$+ 0.11\atop - 0.01$& $   19.30$\scr$+ 0.06\atop - 0.19$& $ 0.72$&$ 0.32$&$    0.61$\scr$+ 0.01\atop - 0.03$& $ 1.24$&\\
EA05                  & $    1.12$\scr$+ 0.04\atop - 0.01$& $   17.96$\scr$+ 0.05\atop - 0.07$& $ 0.57$&$-0.15$&  & $    4.35$\scr$+ 0.98\atop - 0.01$& $   20.59$\scr$+ 0.32\atop - 0.30$& $ 0.77$&$-0.38$&$    0.68$\scr$+ 0.02\atop - 0.05$& $ 1.30$&\\
EA06                  & $    0.40$\scr$+ 0.01\atop - 0.01$& $   16.94$\scr$+ 0.05\atop - 0.03$& $ 0.85$&$-0.05$&  & $    4.82$\scr$+ 4.05\atop - 0.01$& $   22.49$\scr$+ 0.79\atop - 0.22$& $ 0.55$&$ 0.84$&$    0.85$\scr$+ 0.04\atop - 0.09$& $ 0.67$&\\
EA07\tablenotemark{b} & $   11.59$\scr$+ 0.48\atop - 0.01$& $   21.43$\scr$+ 0.03\atop - 0.07$& $ 0.69$&$ 0.17$&  & $    3.35$\scr$+ 0.63\atop - 0.01$& $   19.25$\scr$+ 0.03\atop - 0.24$& $ 0.82$&$-0.50$&$    0.85$\scr$+ 0.02\atop - 0.11$& $ 1.09$&\\
EA08                  & $    0.59$\scr$+ 0.03\atop - 0.01$& $   17.93$\scr$+ 0.08\atop - 0.00$& $ 0.89$&$-0.03$&  & $    2.16$\scr$+ 0.11\atop - 0.01$& $   19.24$\scr$+ 0.10\atop - 0.00$& $ 0.90$&$ 0.25$&$    0.45$\scr$+ 0.01\atop - 0.01$& $ 0.82$&\\
EA09                  & $    0.06$\scr$+ 0.01\atop - 0.01$& $   15.64$\scr$+ 0.09\atop - 0.04$& $ 0.98$&$-0.81$&  & $    1.45$\scr$+ 0.02\atop - 0.01$& $   18.71$\scr$+ 0.02\atop - 0.00$& $ 0.41$&$-0.21$&$    0.16$\scr$+ 0.01\atop - 0.01$& $ 0.84$&\\
EA10                  & $    1.16$\scr$+ 0.05\atop - 0.01$& $   18.90$\scr$+ 0.05\atop - 0.56$& $ 0.85$&$ 0.17$&  & $    3.13$\scr$+ 0.18\atop - 0.01$& $   19.98$\scr$+ 0.15\atop - 0.66$& $ 0.68$&$-0.12$&$    0.64$\scr$+ 0.01\atop - 0.12$& $ 0.74$&\\
EA11                  & $    1.06$\scr$+ 0.03\atop - 0.01$& $   18.00$\scr$+ 0.05\atop - 0.20$& $ 0.42$&$ 0.47$&  & $    3.24$\scr$+ 0.36\atop - 0.01$& $   20.10$\scr$+ 0.17\atop - 0.05$& $ 0.95$&$-0.68$&$    0.61$\scr$+ 0.01\atop - 0.02$& $ 0.92$&\\
EA12\tablenotemark{b} & $    1.49$\scr$+ 0.16\atop - 0.01$& $   19.74$\scr$+ 0.10\atop - 0.05$& $ 0.65$&$-0.48$&  & $    3.88$\scr$+ 0.56\atop - 0.01$& $   19.82$\scr$+ 0.19\atop - 0.03$& $ 0.69$&$-0.56$&$    0.36$\scr$+ 0.04\atop - 0.01$& $ 1.04$&\\
EA13                  & $    1.95$\scr$+ 0.33\atop - 0.01$& $   19.24$\scr$+ 0.19\atop - 0.25$& $ 0.61$&$ 0.10$&  & $    5.07$\scr$+ 0.61\atop - 0.01$& $   19.43$\scr$+ 0.27\atop - 0.24$& $ 0.43$&$ 0.07$&$    0.47$\scr$+ 0.05\atop - 0.05$& $ 0.83$&\\
EA14                  & $    0.75$\scr$+ 0.01\atop - 0.01$& $   17.62$\scr$+ 0.00\atop - 0.03$& $ 0.55$&$ 0.10$&  & $    2.93$\scr$+ 0.05\atop - 0.01$& $   19.88$\scr$+ 0.04\atop - 0.07$& $ 0.40$&$ 0.33$&$    0.71$\scr$+ 0.01\atop - 0.01$& $ 1.36$&\\
EA15                  & $    0.47$\scr$+ 0.01\atop - 0.01$& $   17.68$\scr$+ 0.01\atop - 0.15$& $ 0.92$&$ 0.02$&  & $    1.57$\scr$+ 0.01\atop - 0.01$& $   18.76$\scr$+ 0.01\atop - 0.12$& $ 0.90$&$-0.11$&$    0.48$\scr$+ 0.01\atop - 0.02$& $ 0.73$&\\
EA16                  & $    0.34$\scr$+ 0.01\atop - 0.01$& $   17.03$\scr$+ 0.05\atop - 0.02$& $ 0.81$&$ 0.02$&  & $    2.21$\scr$+ 0.09\atop - 0.01$& $   20.02$\scr$+ 0.08\atop - 0.05$& $ 0.81$&$ 0.14$&$    0.57$\scr$+ 0.01\atop - 0.01$& $ 0.72$&\\
EA17                  & $    0.25$\scr$+ 0.01\atop - 0.01$& $   17.48$\scr$+ 0.02\atop - 0.37$& $ 0.69$&$-0.15$&  & $    1.62$\scr$+ 0.03\atop - 0.01$& $   19.36$\scr$+ 0.03\atop - 0.11$& $ 0.58$&$ 0.06$&$    0.35$\scr$+ 0.01\atop - 0.03$& $ 0.78$&\\
EA18\tablenotemark{b} & $    2.93$\scr$+ 0.02\atop - 0.01$& $   20.59$\scr$+ 0.00\atop - 0.05$& $ 0.52$&$-0.30$&  & $    0.63$\scr$+ 0.01\atop - 0.01$& $   18.64$\scr$+ 0.06\atop - 0.00$& $ 1.00$&$-0.95$&$    0.91$\scr$+ 0.01\atop - 0.01$& $ 1.24$&\\
EA21                  & $    0.56$\scr$+ 0.01\atop - 0.01$& $   17.41$\scr$+ 0.03\atop - 0.03$& $ 0.70$&$ 0.33$&  & $    4.19$\scr$+ 0.32\atop - 0.01$& $   20.60$\scr$+ 0.10\atop - 0.10$& $ 0.65$&$-0.23$&$    0.59$\scr$+ 0.02\atop - 0.02$& $ 0.93$&\\
\enddata
\tablecomments{EA01A, EA02 and EA19 are excluded from the bulge-disk
decomposition because the GALFIT program fit does not converge.}
\tablenotetext{a}{Typical errors for $q$ and $c$ are $\sim$ 0.01.}
\tablenotetext{b}{E+As with unreliable bulge-disk decompositions.}
\end{deluxetable}




\newcommand{\fdcdr}{$\frac{d (B-R)}{d\Log r}$}
\begin{deluxetable}{l rrrr c}
\tablewidth{0pt}
\tabletypesize{\small}
\tablecaption{Color Gradients\label{tab:color_gradient}}
\tablehead{
\colhead{}&
\colhead{\fdcdr}&
\colhead{$\Log R_{\rm break}$}&
\colhead{\fdcdr$\big|_{\rm in}$}&
\colhead{\fdcdr$\big|_{\rm out}$}&
\colhead{}\\
\colhead{ID}&
\colhead{(mag dex$^{-1}$)}&
\colhead{(kpc)}&
\colhead{(mag dex$^{-1}$)}&
\colhead{(mag dex$^{-1}$)}&
\colhead{Morphology\tablenotemark{a}}
}
\startdata
  EA01B&  $   0.13\pm 0.01$ &              \nodata &              \nodata &              \nodata &       P  \\
  EA02 &  $   0.00\pm 0.03$ &  $   0.03\pm   0.03$ &  $  -0.25\pm   0.04$ &  $   0.38\pm   0.05$ &     N,D  \\
  EA03 &  $   0.20\pm 0.01$ &              \nodata &              \nodata &              \nodata &       P  \\
  EA04 &  $  -0.00\pm 0.01$ &  $   0.01\pm   0.05$ &  $   0.07\pm   0.01$ &  $  -0.20\pm   0.03$ &     N,D  \\
  EA05 &  $   0.44\pm 0.05$ &  $  -0.37\pm   0.02$ &  $   0.95\pm   0.07$ &  $   0.27\pm   0.02$ &    P,BC  \\
  EA06 &  $   0.44\pm 0.03$ &  $   0.13\pm   0.02$ &  $   0.60\pm   0.02$ &  $  -0.05\pm   0.05$ &  P,BC,D  \\
  EA07 &  $  -0.16\pm 0.02$ &  $   0.28\pm   0.03$ &  $  -0.46\pm   0.03$ &  $  -0.02\pm   0.01$ &     N,D  \\
  EA08 &  $   0.01\pm 0.01$ &  $   0.62\pm   0.03$ &  $  -0.04\pm   0.01$ &  $   0.48\pm   0.11$ &       P  \\
  EA09 &  $   0.06\pm 0.04$ &  $  -0.38\pm   0.01$ &  $   1.02\pm   0.05$ &  $  -0.18\pm   0.01$ &    P,BC  \\
  EA10 &  $   0.08\pm 0.01$ &  $   0.66\pm   0.03$ &  $   0.06\pm   0.01$ &  $   0.63\pm   0.09$ &       P  \\
  EA11 &  $   0.17\pm 0.04$ &  $   0.13\pm   0.03$ &  $  -0.09\pm   0.03$ &  $   0.43\pm   0.03$ &       F  \\
  EA12 &  $  -0.31\pm 0.05$ &  $   0.08\pm   0.02$ &  $   0.34\pm   0.04$ &  $  -0.68\pm   0.03$ &     N,D  \\
  EA13 &  $   0.08\pm 0.02$ &  $  -0.11\pm   0.03$ &  $   0.33\pm   0.02$ &  $  -0.02\pm   0.01$ &     F,D  \\
  EA14 &  $   0.29\pm 0.03$ &  $   0.13\pm   0.01$ &  $   0.75\pm   0.02$ &  $   0.08\pm   0.02$ &    P,BC  \\
  EA15 &  $   0.12\pm 0.01$ &              \nodata &              \nodata &              \nodata &       P  \\
  EA16 &  $   0.45\pm 0.02$ &  $   0.07\pm   0.02$ &  $   0.63\pm   0.02$ &  $   0.28\pm   0.02$ &    P,BC  \\
  EA17 &  $   0.18\pm 0.02$ &  $  -0.43\pm   0.03$ &  $   0.72\pm   0.06$ &  $   0.10\pm   0.01$ &  P,BC,D  \\
  EA18 &  $   0.06\pm 0.02$ &  $  -0.15\pm   0.04$ &  $   0.40\pm   0.04$ &  $  -0.10\pm   0.02$ &     F,D  \\
  EA19 &  $  -0.25\pm 0.02$ &  $   0.46\pm   0.03$ &  $  -0.37\pm   0.02$ &  $   0.23\pm   0.06$ &     N,D  \\
  EA21 &  $   0.09\pm 0.01$ &  $   0.13\pm   0.09$ &  $   0.15\pm   0.02$ &  $   0.04\pm   0.02$ &       F  
\enddata
\tablecomments{
Because a single linear fit is not the best representation of most
color profiles, the statistical uncertainties are always too small and
unrealistic. Therefore, we calculate bootstrap uncertainties.}
\tablenotetext{a}{
Morphology of the color profile: N(negative), P(positive), BC(blue-core),
F(flat or variable), D(dust)}
\end{deluxetable}


\clearpage
\appendix
\section{Detailed Morphologies of E+A Galaxies}
\label{apdx:qualitative_morphology}

High resolution {\sl HST} images enable us to identify a wealth of small
and large scale features.  For example, the low surface brightness tidal
features, small companion galaxies, blue cores, bars, and even point-like,
newly-formed star clusters are essential clues in understanding the
causes and end-products of the E+A phase.  In this appendix, we describe
the qualitative morphologies of 21 E+A galaxies.  Given that our E+A
sample was selected using uniform spectroscopic criteria (large
Balmer absorption and no [\ion{O}{2}] emission), it is striking that the
morphologies of E+A galaxies are so diverse.

\subsection{Tidal Features and Disturbed Morphologies}
\label{sec:tidal_feature}

Eight E+As (EA01AB, 02, 03, 07, 08, 10, 12, 13) show clear tidal features
indicative of recent galaxy-galaxy interactions. These features include
tidal tails that extend over a few tens of kpc and bridges that connect
apparent companions.  In addition to these dramatic large-scale tidal
features, another three (EA04, 11, 18) exhibit disturbed morphologies
such as shell-like structures (EA04 and EA11) and highly irregular
isophotes (EA18).  We conclude that the fraction of E+A galaxies with
readily identifiable, i.e., brighter than our detection limit
($\mu_R < 25.1\pm0.5$ mag arcsec$^{-2}$), merger/interaction signatures
is $\sim 55\%$ ($\pm 15$).

Since \citet{Zabludoff96} first claimed galaxy-galaxy interactions/mergers
as the main mechanism for E+A formation, other studies using larger
samples of E+As have reached similar conclusions.  For example,
\citet{Blake04} found that $\sim$13\% of E+As selected from the 2dFGRS,
using the same spectroscopic criteria as was done for the LCRS sample,
show tidal features or disturbed morphologies.  \citet{Goto05}
found that $\sim$30\% of his sample shows tidal features.  The higher
merger fraction that we report here is almost certainly due to the
improved sensitivity and higher resolution of {\sl HST} imaging rather
than intrinsic differences among E+A samples.  For example, the low
surface brightness features seen in EA08 and the shell-like structure
in EA04 would not be detected in typical ground-based imaging.

Do merger features correlate with increasing EA numbers (i.e., increasing
$D_{4000}$), which is likely to correlate with the mean stellar age of the
galaxy? We find tidal features more often among E+As with smaller EA
numbers. The distribution of EA numbers for E+As with tidal features
is not random at the $\sim$90\% confidence level. Because the duration
of the E+A phase ($\sim$ Gyr) is longer than the dynamical relaxation
time of the merger, we expect this correlation if EA numbers constitute
a rough age sequence.

\subsection{Companion Galaxies}

Among the 11 E+As with interaction/merger signatures, five  (EA01AB,
10, 11, 12, 13) have companion galaxies within $\sim$ 30 kpc that appear
to be interacting with the E+A. The EA01AB system is spectroscopically
confirmed \citep{Yang06}, but the others may only be projected companions.
For example, EA15's companion (R.A. = 14:40:45.2, DEC. = $-$06:39:53),
which we also observed in our slit, is at a different redshift.  The
properties of the projected companion galaxies are as diverse as those
of the E+As.  They range from almost 1:3 mergers (EA01AB) to apparent
1:75 minor mergers (EA10) as judged from the relative $R$ band fluxes.
Are the interactions with these companion galaxies responsible for the
starburst in E+As that ended $\lesssim$ Gyr ago?  If these apparent
on-going interactions produce the E+A spectra, then there was a close
interaction $\lesssim$ Gyr ago, and we are now witnessing a subsequent
passage.  Therefore, a wide range of merger configurations may cause
the E+A phenomenon.
Alternatively, another interaction/trigger produces the E+A spectra,
and the current interactions are solely by chance.  Larger surveys
including redshift measurements of the companion galaxies will help to
resolve this issue.  Conversely, not all interactions lead to E+A's.
\citet{Yagi06} present a spectroscopically confirmed interacting E+A
system where one galaxy is an E+A, but the other has neither current
star formation nor post-starburst signatures.



\subsection{Dust Features} 

Seven E+As (EA06, 07, 12, 13, 17, 18, 19) show dust features, such as
lanes and filamentary structures, in the two-color composites or the
residual images. In the WFPC2 sample, we also suspect that EA02 and
EA04 might have dust from their irregular residual images, but the
shallow \B{439} images prohibit us from confirming it.
Signatures of dust, although typically fairly minor, are present in over
a third (33--43\%) of our E+As.

There have been suggestions that star-forming galaxies could be disguised
as E+A galaxies due to obscuration by dust \citep{Smail99}. However,
this is not an issue for the E+As in this sample.  As discussed in
\S \ref{sec:color_profile}, only three E+As (EA07, 12, 19) have color
profiles seriously affected by dust lanes [$\Delta(B-R) \gtrsim 0.5$].
Furthermore, radio continuum emission is detected from only two of the
dusty E+As (EA12 and 19) with inferred star formation rates (SFR) of 5.9
and 2.2 $M_{\sun}$ yr$^{-1}$, respectively \citep{Miller_Owen01}. EA07
was not observed in the radio.  We conclude that the contamination rate
due to the E+A selection criteria adopted by \citet{Zabludoff96}, which
employs the average Balmer line strength and [\ion{O}{2}], is less than
15\% \cite[see also][]{Blake04,Goto04}.

\subsection{Barred Galaxies}

Only three E+As galaxies (EA05, 14, 21) exhibit clear signs of bars in
the residual images, while two other E+As (EA11 and EA19) have elongated
residuals that suggest the presence of a bar.  Because there is no
model profile for the bar component, we fit these galaxies as well as
possible using the various model GALFIT components. Because introducing
too many free components makes it impossible to interpret the components
physically \citep{Peng02},  we use only three-component models,
each with a combination of two S\'ersic profiles and one exponential disk
profile.  We show the best-fit model profiles in Figure \ref{fig:profile}
for the barred E+A galaxies.  The bars appear to be fit reasonably well
with a steep ($n < 0.8$) and flat ($q\sim0.5$) component. \\

\section{Post-Burst Age Determination}
\label{apdx:cluster_age_lf}

The ages of the young star cluster populations in E+A galaxies could
be used to break the degeneracy between the burst strength and the time
elapsed since the starburst, because a simple stellar population (SSP)
is an excellent approximation for the star clusters and therefore no
additional assumptions are required to model the underlying old stellar
populations.  To derive the time since the starburst (post-burst age),
when the majority of the detected clusters and young stars formed,
we use a statistical test and the simple assumption that all star
clusters formed during a single burst. Therefore, we assume that their
spread in colors is due entirely to color measurement errors.
For each post-burst age, we draw a thousand sets of $N$ cluster colors, each
scattered according to a Gaussian error distribution described by the
measurement error $\sigma_{err}^i$, where $N$ is the number of cluster
candidates found in each E+A.  These simulated colors are compared to the
distributions of observed colors. We determine the fraction of sets that
are statistically indistinguishable from the observations using the two
sample K-S test.  If the fraction of acceptable sets is less than 5\%,
we reject that model age as the time since the starburst.

\begin{figure}
\epsscale{0.6}
\plotone{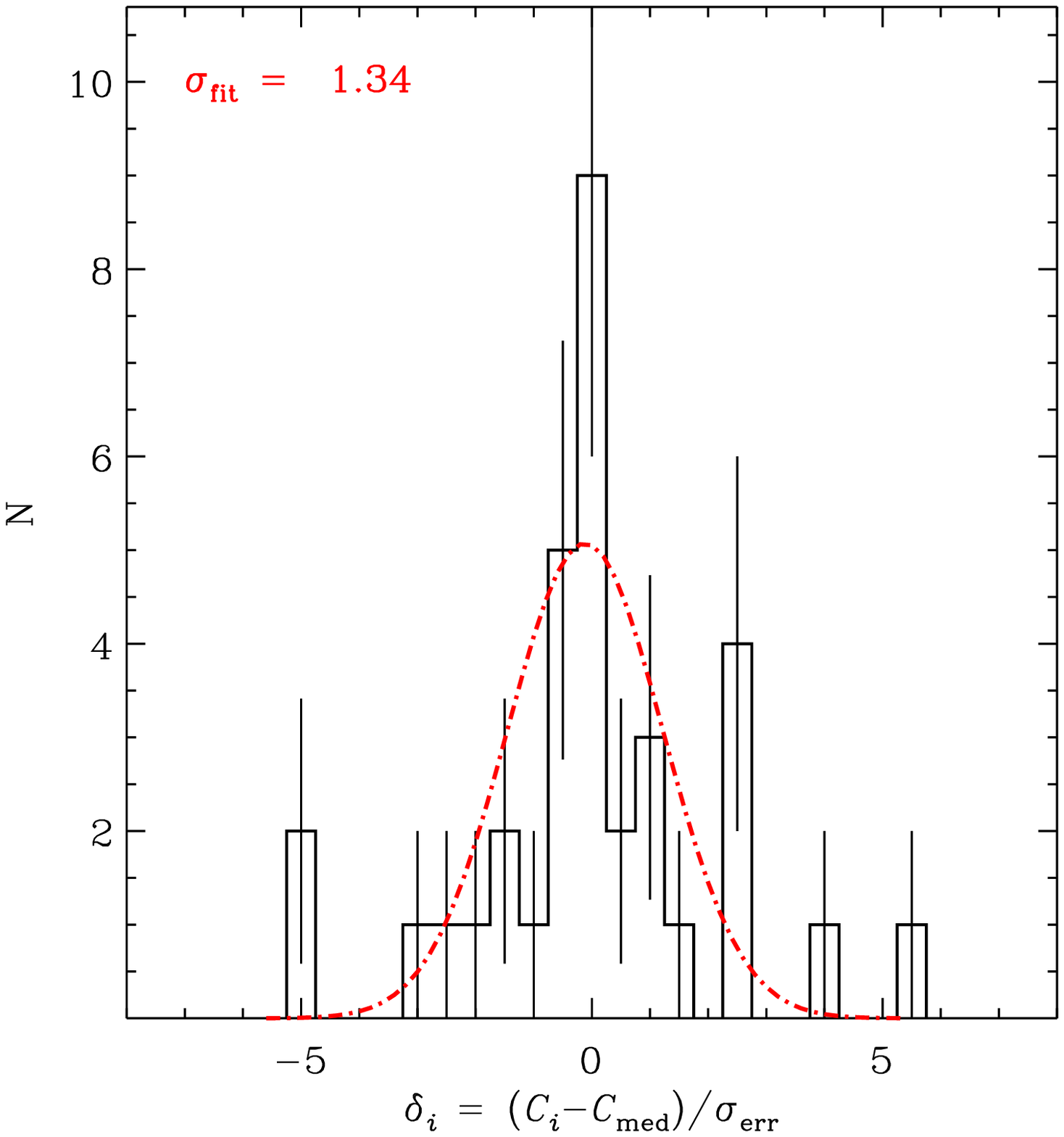}
\caption{
The composite distribution of $\delta_i$ = $(C_i-C_{med})/\sigma_{err}^i$
from five {\sl ACS} E+As with cluster candidates.  The histogram is
marginally consistent with a Gaussian with $\sigma = 1$.  There is
no strong indication against our assumption of Gaussian-distributed
uncertainties.
\label{fig:gaussianity}}
\end{figure}

Next we test whether the key assumptions in determining the post-burst ages ---
1) the Gaussian distribution of observed colors and 2) the single age
cluster population --- are statistically acceptable.  First, we consider
whether the assumption that the measured cluster colors are randomly
distributed according to a Gaussian is realistic.  Due to small number
statistics in each galaxy, it is not possible to test this assumption
for the clusters in each galaxy. Therefore, we apply the following test
to all clusters discovered in the five E+As (EA07, 08, 11, 15, 18).
For each E+A, we calculate the deviation of each cluster color from the
median color of the sample in terms of its measurement error, $\delta_i =
(C_i-C_{med})/\sigma_{err}^i$, where $C_i$ and $C_{med}$ are the measured
color of the cluster and the median color of clusters in the given
E+A, and $\sigma_{err}^i$ is the error in each cluster color including
uncertainties in background subtraction.  We examine the distribution
of $\delta_i$'s.  If the $\delta_i$'s follow a Gaussian distribution with
zero mean and a standard deviation of unity, our assumption of Gaussian
distributed errors can be justified. If not, e.g., if the $\delta_i$'s show
a skewed or flat distribution, their measured colors could be seriously
affected by the other factors.  In Figure \ref{fig:gaussianity}, we show
the distribution of $\delta_i$'s. The histogram is marginally consistent
with a Gaussian distribution with unit standard deviation ($\sigma_{\rm
fit} = 1.34$), therefore we conclude that there is no strong evidence
against this assumption.

Second, we test the validity of the assumption that all of the detected
star clusters formed during a single instantaneous burst.  We generate
a thousand sets of simulated cluster colors with an age spread of
$\Delta t$. Gaussian random errors with a dispersion corresponding
to the measurement errors $\sigma_{err}^i$ are added to the simulated
colors.  Now we calculate the $\delta_i$'s using these simulated sets
of cluster colors and compare the width $\sigma_{\rm fit}$ of the
$\delta_i$ distribution with the observed value $\sigma_{\rm fit} =
1.34$. If the observed $\sigma_{\rm fit}$ is significantly different
from the distribution of $\sigma_{\rm fit}$, then we reject the given
$\Delta t$. We find that $\Delta t < $ 0.1 Gyr is rejected at the 95\%
confidence level.  However, the small number statistics and low S/N of
the color measurements do not allow us to constrain the upper limits of
the burst durations.  Therefore, the instantaneous burst assumption might
not be strictly true, and the burst duration should be at least longer
than 100 Myr. On the other hand, using an E+A galaxy sample selected
from the Sloan Digital Sky Survey, \citet{Yang08} show that the timescale
over which star formation ends tends to be smaller than $\sim$ 200 Myr,
and larger burst durations (several hundred Myr) would wipe out the
correlations that we find among post-burst ages, global E+A colors (i.e.,
EA number), and the bright end of the star cluster luminosity function.
Therefore, we argue that the single burst age approximation is valid to
within a few hundred Myr.

\end{document}